\documentclass[11pt,a4paper]{article}
\usepackage{jheppub}
\usepackage{graphicx}
\usepackage{rotating}
\usepackage{color}
\usepackage{epsfig}
\usepackage{bm}
\usepackage{url}
\usepackage{slashed}
\usepackage{array}

\usepackage{tikz}

\newcommand*{\dittoclosing}{--- \raisebox{-0.5ex}{''} ---}

\newcommand{\figurefolder}{Figures}
\newcommand{\be}{\begin{equation}}
\newcommand{\ee}{\end{equation}}
\newcommand{\bea}{\begin{eqnarray}}
\newcommand{\eea}{\end{eqnarray}}
\newcommand{\bi}{\begin{itemize}}
\newcommand{\ei}{\end{itemize}}
\newcommand{\ben}{\begin{enumerate}}
\newcommand{\een}{\end{enumerate}}
\newcommand{\bt}{\begin{tabular}}
\newcommand{\et}{\end{tabular}}

\newcommand{\lc}{\left[}
\newcommand{\rc}{\right]}
\newcommand{\lp}{\left(}
\newcommand{\rp}{\right)}

\newcommand{\tree}{tree level}
\newcommand{\self}{self-energy}
\newcommand{\onel}{one loop}
\newcommand{\twol}{two loops}

\newcommand{\fig}{fig}
\newcommand{\annproc}{$\chi\chi\rightarrow f \bar f$}
\newcommand{\pr}{P_\text{R}}
\newcommand{\pl}{P_\text{L}}

\def\sl{\slashed }
\newcommand{\op}{\mathcal{O}}

\def\Tr{{\rm{Tr}}}

\def\Rea{{\rm{Re}\,}}

\def\g{\gamma }

\def\c{\chi }

\def\lg{{\mathchoice{~\raise.58ex\hbox{$<$}\mkern-14.8mu\lower.52ex\hbox{$>$}~}
                    {~\raise.58ex\hbox{$<$}\mkern-14.8mu\lower.52ex\hbox{$>$}~}
                    {\raise.59ex\hbox{{$\scriptscriptstyle <$}}\mkern-12.8mu%
                     \lower.01ex\hbox{{$\scriptscriptstyle >$}}}   {}   }} 
\def\gl{{\mathchoice{~\raise.58ex\hbox{$>$}\mkern-12.8mu\lower.52ex\hbox{$<$}~}
                    {~\raise.58ex\hbox{$>$}\mkern-12.8mu\lower.52ex\hbox{$<$}~}
                    {\raise.62ex\hbox{{$\scriptscriptstyle >$}}\mkern-12.0mu%
                     \lower.05ex\hbox{{$\scriptscriptstyle <$}}}  {}    }}  

\allowdisplaybreaks

\begin{document}

\begin{flushright}
TUM-HEP 953/14\\
arXiv:1409.3049v2 [hep-ph]\\
July 12, 2016
\end{flushright}

\vskip0.2in

\title{\huge 
Relic density computations at NLO: infrared finiteness and 
thermal correction}

\author[]{Martin Beneke,}
\author[]{Francesco Dighera,}
\author[]{Andrzej Hryczuk}

\affiliation[]{Physik Department T31, Technische Universit\"{a}t M\"{u}nchen, 
James-Franck-Str.~1, 85748 Garching, Germany}

\abstract{
There is an increasing interest in accurate dark matter relic density 
predictions, which requires next-to-leading order (NLO) calculations. 
The method applied up to now uses zero-temperature NLO calculations 
of annihilation cross sections in the standard Boltzmann equation 
for freeze-out, and is conceptually problematic, since it ignores the 
finite-temperature infrared (IR) divergences from soft and collinear 
radiation and virtual effects. We address this problem systematically by 
starting from non-equilibrium quantum field theory, and demonstrate on a 
realistic model that soft and collinear temperature-dependent divergences 
cancel in the collision term. Our analysis provides justification for the 
use of the freeze-out equation in its conventional form 
and determines the leading finite-temperature 
correction to the annihilation cross section. This turns out to 
have a remarkably simple structure. 
}


\maketitle


\vspace*{0.5cm}
\section{Introduction}
\label{sec:Intro}

The most widely studied and arguably most natural mechanism of generating 
the present-day abundance of dark matter (DM) is its thermal production 
in the early Universe followed by chemical decoupling (freeze-out) 
of the DM particles from the background plasma. 
For temperatures much higher than the mass of the DM particles the 
dark matter component remains in both chemical and kinetic equilibrium. 
When the temperature of the plasma drops, the interactions are not 
strong enough to keep the dark matter component in chemical equilibrium 
and the DM particle number freezes out. 

The precise moment when chemical decoupling happens is determined by 
two different physical processes: the expansion of the Universe governed by 
the Hubble rate $H$ and the annihilation rate $\Gamma$ 
of the DM particles. The precise description of the decoupling process is 
possible in kinetic theory, where the evolution of the system is given
by the transport equations for the phase-space distribution functions $f(p)$. 
If one assumes that \textit{i)} the Compton wavelength of 
DM particles is small with respect to inhomogeneity scale and 
\textit{ii)} one can adopt the quasi-particle approximation, one arrives 
at a semi-classical description. In this case the transport is governed by 
the Boltzmann equation and its solution can be used for the determination of 
the DM relic density for a given particle physics model.
The DM particle number density then follows the simple equation
\begin{equation}
\frac{dn_\c}{dt}+3Hn_\c = 
\langle \sigma_{\chi \bar\chi \rightarrow i j} v_{\rm rel} \rangle 
\lp n_\c^{\rm eq}n_{\bar \c}^{\rm eq} - n_\c n_{\bar \c}\rp ,
\label{eq:BoltzmannEq}
\end{equation}
where $H$ denotes the Hubble rate, and 
$\langle \sigma_{\chi \bar\chi \rightarrow i j} v_{\rm rel} \rangle$ 
the thermal average of the sum over all annihilation cross sections 
to two-particle final states $ij$. 

In recent years there has been an increasing interest in higher-order 
corrections to scattering and annihilation processes involving DM 
particles. The main phenomenological importance of such corrections 
is related to the modification of the annihilation spectra relevant for 
the indirect searches. Quite generally, 
the increasing precision of dark matter 
observations will require more accurate computations of the 
scattering and annihilation processes, in some cases at full 
next-to-leading order (NLO) in the coupling constant. In particular, 
it has also been noted recently that corrections to the 
annihilation rate can affect non-negligibly the relic density 
computation~\cite{Baro:2007em,Baro:2009na, Herrmann:2009wk, Herrmann:2009mp,Harz:2012fz,Chatterjee:2012db,Ciafaloni:2013hya,Herrmann:2014kma}. 
With this in mind the first numerical codes including the higher-order 
corrections are being developed, SloopS~\cite{Baro:2008bg, Baro:2009gn,SloopS} and DM@NLO \cite{Herrmann:2007ku,DMatNLO}. 
What is usually done is to compute the virtual and real 
radiation corrections to the 
two-particle processes $\chi\bar\chi\to ij$ using standard 
quantum field theory methods {\em at zero temperature}.

This procedure raises a number of questions, especially for 
relic density computations, since freeze-out occurs when the temperature 
of the Universe is small, but non-negligible compared to the DM 
particle mass.
\begin{itemize}
\item Why should the time evolution of $n_\chi$ be described 
by inclusive two-particle cross sections and a Boltzmann equation 
of the form applicable to $2\to 2$ reactions? The real radiation 
amplitude involves three-particle final states, 
typically containing an additional photon or gluon, which are 
themselves abundant in the plasma. Moreover, 
absorption processes exist, but are neglected in the computation. 
\item How do the soft and collinear infrared (IR) divergences cancel at 
finite temperature? It is well-known that IR divergences are more 
severe at finite temperature due to the enhancement from the 
Bose distribution at small momenta. Moreover, virtual and real 
scattering matrix elements, which are separately divergent, 
appear to be multiplied by different 
statistical factors in the full Boltzmann equation.
\item Assuming IR finiteness can be shown, what are the leading 
finite-temperature effects on the annihilation cross sections and 
the relic density?
\item Does the transport equation itself receive quantum corrections 
when it is derived from general principles of non-equilibrium 
quantum field theory (QFT) to NLO accuracy?
\end{itemize}
In this paper we address these questions. We demonstrate that 
the IR cancellation happens in the sum over ``cuts'' of individual 
self-energy diagrams similar to the situation at zero temperature, 
but involving the additional processes that occur in the plasma. 
The form of (\ref{eq:BoltzmannEq}) remains valid under the typical 
conditions of DM freeze-out, but the annihilation cross section is 
modified by a small and calculable finite-temperature correction. 
Remarkably, the final NLO finite-temperature correction has a 
very simple structure 
and can be computed directly from the zero-temperature tree
 level cross section.

The outline of the paper is as follows.
In section \ref{sec:IRBoltzmann} we discuss the IR problems that arise 
at NLO at finite temperature. Section \ref{sec:Bderivation} reviews the 
well-known derivation of the Boltzmann equation from non-equilibrium 
QFT, with emphasis on the application to the freeze-out process. 
Next we discuss the computation of the collision term in section 
\ref{sec:collision_term} and demonstrate the general procedure for a 
bino-like DM model. In section \ref{sec:results} we present and discuss 
the IR divergences cancellation and the result for the finite correction 
from thermal effects. We conclude in section~\ref{sec:conclusions}. 


\section{IR divergences and the Boltzmann equation}
\label{sec:IRBoltzmann}

In general the one-loop scattering amplitudes contain soft and 
collinear IR divergent terms. At zero temperature the Bloch-Nordsieck 
cancellations and Kinoshita-Lee-Nauenberg 
(KLN) theorem \cite{Bloch:1937pw,Kinoshita:1962ur,Lee:1964is} ensure that 
physical observables are free of both of these divergences, as they 
involve summation over initial and final degenerate states, in the sense 
of inclusiveness or experimental resolution in energy and angles. 
At finite temperature no general proof of such a theorem is known, 
nevertheless the cancellation was observed in all the particular cases 
studied in the literature (see e.g. \cite{Grandou:1991qr,Grandou:1990ir,Altherr:1991cq,Landshoff:1994ud,Baier:1989ub,Baier:1988xv,Altherr:1989yn,Gabellini:1989yk,Altherr:1988bg,Czarnecki:2011mr,Besak:2012qm,
Anisimov:2010gy,Salvio:2011sf,Laine:2011pq,Garbrecht:2013gd}). 
In both situations the main prerequisite for the cancellation is the 
inclusion of a soft or collinear gauge boson in the NLO computation. 

Whether and how the cancellation happens in NLO relic density computations 
has not yet been investigated.\footnote{
We note \cite{Wizansky:2006fm}, which, however, addresses the different 
question whether thermal corrections to mass and width parameters 
can have an effect on the relic density value in parameter regions 
where it depends sensitively on these parameters, such as in the 
co-annihilation or resonance region, or when the important decay 
channels are helicity-suppressed.} 
To formulate the problem, let us 
consider a typical freeze-out scenario of a weakly-interacting massive 
dark matter particle (WIMP), for which thermal and chemical 
equilibrium at temperatures larger than the freeze-out 
temperature $T_{f} \approx m_\chi/20$ is  maintained by 
$2\rightarrow 2$ scattering processes at leading order. 
In the  Friedmann-Robertson-Walker background the semi-classical 
Boltzmann equation for the evolution of the phase-space distribution 
function reads
\begin{equation}
\label{eq:Boltzmannf}
E\left(\partial_t-H \vec{p}\cdot \nabla_{\vec{p}} \right) f = \mathcal{C}[f].
\end{equation}
It can be rewritten as an equation for the number density 
$n_i(t)\equiv h_i \int \frac{d^3\vec p}{(2\pi)^3} f_i(p)$ of given 
species $i$ as
\begin{eqnarray}
\label{eq:Boltzmann}
\frac{dn_\c}{dt}+3Hn_\c=C_{\rm LO},
\end{eqnarray}
where $h_i$ is the number of internal degrees of freedom of particle $i$ 
and we assume that the thermal plasma is ``unpolarized'' with respect to 
the internal degrees of freedom. The integrated collision term for the 
leading-order (LO) annihilation (production) 
process $\c \bar{\c} \leftrightarrow i j$ 
is given by
\begin{eqnarray} 
\label{eq: collision_term_LO_pre}
C_{\rm LO} = \int d\Pi_{\c\bar \c ij}  \, 
|\mathcal{M}_{\c \bar{\c} \rightarrow i j}|^2
\left[ f_i f_j (1\pm f_\c)(1\pm f_{\bar \c}) - 
f_\c f_{\bar \c} (1\pm f_i)(1\pm f_j) \right],
\end{eqnarray}
where we defined 
\begin{eqnarray}
d\Pi_{\c\bar \c ijk\ldots} 
&=&  \,\frac{d^3 \vec p_\c}{(2\pi)^3 2 E_\c}
\frac{d^3 \vec p_{\bar \c}}{(2\pi)^3 2 E_{\bar\c}}
\frac{d^3\vec{p}_i}{(2\pi)^3 2E_i}
\frac{d^3\vec{p}_j}{(2\pi)^3 2E_j}
\frac{d^3\vec{p}_k}{(2\pi)^3 2E_k} \ldots 
\nonumber\\[0.2cm]
&&\times\,(2\pi)^4\delta^{(4)}(p_f- p_i),
\end{eqnarray}
with $p_f$ ($p_i$) denoting the sum of the final (initial) state momenta of 
the process, and moreover assumed CP invariance, which implies 
$|\mathcal{M}_{\c \bar{\c}\rightarrow i j}|^2=
|\mathcal{M}_{ij \rightarrow \c \bar{\c}}|^2$. The $\pm$ signs are 
chosen according to whether the particle is a boson (+) or fermion ($-$).
Note that here the squared matrix elements are defined 
to be \textit{summed} over the internal (spin) degrees of freedom of 
both the initial-state DM and final-state Standard Model (SM) particles. 
We shall make the standard assumption that the SM particles are kept in 
thermal equilibrium by frequent scatterings and that asymmetries 
are negligible. Therefore $f_{i,j} = f_{i,j}^{\rm eq}$, and also
the photon phase-space distribution $f_\gamma =
f_\gamma^{\rm eq}$ introduced below, are either
Bose-Einstein or Fermi-Dirac distributions with vanishing chemical 
potentials. For the DM particles we assume that kinetic equilibrium 
is maintained by 
frequent elastic scatterings with the particles from the thermal bath, 
resulting in $f_\chi \propto f_\chi^{\rm eq}$, 
where the factor of proportionality depends on temperature but not on 
the energy and momentum. Chemical equilibrium of the DM particles, however, 
is lost when the temperature falls below the freeze-out temperature $T_f$, and 
DM particle-number changing processes occur at insufficient rates.
Since in the $2 \to 2$ annihilation reaction all energies are 
of order $\mathcal{O}(m_\chi)$, and since $T_f\ll m_\chi$, all distribution 
functions are exponentially suppressed, and we can approximate 
$1\pm f \approx 1$. Under these assumptions the integrated collision 
term~(\ref{eq: collision_term_LO_pre}) takes the standard form 
\bea 
\label{eq: collision_term_LO}
C_{\rm LO} = \langle \sigma_{\chi \bar\chi \rightarrow i j} 
v_{\rm rel} \rangle \lp  n_\c^{\rm eq}n_{\bar \c}^{\rm eq} 
- n_\c n_{\bar \c}\rp,
\eea
where 
\begin{equation}
\langle \sigma_{\chi \bar\chi \rightarrow i j} v_{\rm rel} \rangle 
\equiv \frac{1}{n_\c^{\rm eq} n_{\bar \c}^{\rm eq}} 
\int\frac{d^3 \vec p_\c}{(2\pi)^3}\frac{d^3 \vec p_{\bar \c}}{(2\pi)^3} \ 
f^{\rm eq}_\chi f^{\rm eq}_{\bar \chi}
\,\sigma_{\chi \bar\chi \rightarrow i j} v_{\rm rel} ,
\end{equation}
denotes the thermally averaged cross section times velocity
\be 
\label{eq: sigmavT=0}
\sigma_{\chi \bar\chi \rightarrow i j} v_{\rm rel} \equiv 
\frac{1}{4 E_\c E_{\bar \c}} \int \frac{d^3 \vec p_i}{(2\pi)^3 2E_i}
\frac{d^3 \vec p_{j}}{(2\pi)^3 2E_j} \,
(2\pi)^4\delta^{(4)}(p_\c+p_{\bar{\c}} -p_i -p_j)\,
|\mathcal{M}_{\c \bar{\c} \rightarrow i j}|^2,
\ee
and we used $f_\chi = n_\chi/n_\chi^{\rm eq}\times f^{\rm eq}_\chi$.

As long as the amplitudes are computed at tree level, the 
cross section $\sigma_{\chi \bar\chi \rightarrow i j} v_{\rm rel}$ 
and hence the collision term is evidently IR finite. When the relic 
density computation described above is extended to NLO, what has been 
done up to now is to compute the {\em zero-temperature} 
annihilation cross section 
to NLO, while keeping the form of the remaining equations. This involves 
the one-loop correction to the $2\to 2$ annihilation processes 
$\c \bar{\c} \to i j$, and the tree-level radiation 
process $\c \bar{\c} \to i j\gamma$, assuming electromagnetic 
radiation for definiteness. The sum is IR finite by the usual 
zero-temperature IR cancellations.

This procedure is conceptually problematic. To see this, consider 
the NLO collision term 
\bea \label{eq: collision_term_NLO_phot}
C_{\rm NLO} &=& \int\! d\Pi_{\c \bar \c i j} \, 
\Bigl(
|\mathcal{M}^{\rm LO}_{\c \bar{\c} \rightarrow i j}|^2 
+ |\mathcal{M}^{\textrm{NLO }}_{\c \bar{\c} \rightarrow i j}|^2
\Bigr)
\lc f_i f_j  - f_\c f_{\bar \c}  \rc
\nonumber \\
& + & \int\! d\Pi_{\c \bar \c i j \g} \,
\biggl\{
|\mathcal{M}_{\c \bar{\c} \rightarrow i j \g}|^2 
\lc f_i f_j f_\g  - f_\c f_{\bar \c} (1 + f_\g) \rc	 \nonumber\\
&&\hspace{1.67cm}+\,|\mathcal{M}_{\c \bar{\c} \g\rightarrow i j}|^2 
\lc f_i f_j  (1 + f_\g) - f_\c f_{\bar \c} f_\g  \rc
\biggr\},
\eea
where again we used $1\pm f\approx 1$ except for the photon distribution 
function. The collision term \eqref{eq: collision_term_NLO_phot} contains both annihilation and production contributions,
which are however symmetric and can be described by the same 
thermally averaged cross section, as long as the theory is CP invariant and
the DM particles are in kinetic equilibrium. It can be most easily seen by making use of 
the detailed balance relation for the photon distribution function
\begin{equation}
f_\g = e^{-E_\g /T} (1+f_\g),
\end{equation}
the Maxwell approximation for the remaining ones and the energy conservation. 
It follows, that the collision term has the form analogous to \eqref{eq: collision_term_LO}
but with the thermally averaged cross section replaced by
\begin{eqnarray}
\langle \sigma^{\rm NLO} v_{\rm rel} \rangle &&
\equiv \frac{1}{n_\c^{\rm eq} n_{\bar \c}^{\rm eq}}  \int d\Pi_{\c \bar \c i j} \, f^{\rm eq}_\chi f^{\rm eq}_{\bar \chi} 
\nonumber\\
&&\hspace*{-2cm}\times\,
\biggl\{|\mathcal{M}^{\rm LO}_{\c \bar{\c} \rightarrow i j}|^2 
+ |\mathcal{M}^{\textrm{NLO }}_{\c \bar{\c} \rightarrow i j}|^2
+ \int d\Pi_\gamma \left[|\mathcal{M}_{\c \bar{\c} \rightarrow i j \g}|^2 
\,(1+f_\g)
+|\mathcal{M}_{ \c \bar{\c}\g \rightarrow ij}|^2  \,f_\g
\right]\biggr\},\qquad
\end{eqnarray}
with the interpretation $d\Pi_{\c\bar \c ij}d\Pi_\gamma = d\Pi_{\c\bar \c ij\gamma}$.
For this reason also when discussing the NLO corrections we will consider only annihilation processes.

The problematic approximation corresponds to setting $f_\g \to 0$, 
which amounts to computing the
thermal average of the \textit{zero-temperature} cross section. 
This step is not justified, since there are relevant 
regions of photon phase space $d\Pi_\g$, where the photon energy is 
small, in which case $f_\g \sim E_\g^{-1}$ is arbitrarily large. However,
if one simply keeps $f_\g$ in the expression for the collision term, 
the virtual one-loop and real terms, 
$|\mathcal{M}^{\textrm{NLO }}_{\c \bar{\c} \rightarrow i j}|^2$ 
and $\int d\Pi_\gamma \,
|\mathcal{M}_{\c \bar{\c} \g\rightarrow i j }|^2$, respectively, 
are multiplied by different factors, and the standard IR cancellation 
no longer occurs. Moreover, since  $f_\g \sim E_\g^{-1}$, an 
additional IR divergence is generated, which is more severe than 
the zero-temperature, logarithmic divergences.

It is now important to realize that the photons in the plasma contribute 
not only to the $2\to 3$ emission and $3\to 2$ absorption processes, 
but also to the virtual, one-loop two-body amplitude. Indeed, 
it has been shown in the special cases of muon decay \cite{Czarnecki:2011mr} 
and the right-handed neutrino production rate \cite{Besak:2012qm,
Anisimov:2010gy,Salvio:2011sf,Laine:2011pq,Garbrecht:2013gd} 
relevant to leptogenesis, that when finite-temperature Feynman 
rules are used in the computation of the decay or production rate, 
the additional 
IR divergence cancels. In particular, leptogenesis  
also involves a non-equilibrium situation. The proof of cancellation 
of all divergences in the general case does not seem to exist, 
though some partial results can be found 
in~\cite{Altherr:1988bg,Weldon:1991eg,Weldon:1993qh,Indumathi:1996ec}.  
Let us therefore add 
and make explicit the finite-temperature correction to 
the virtual correction by replacing 
\begin{equation}
|\mathcal{M}^{\textrm{NLO }}_{\c \bar{\c} \rightarrow i j }|^2
\to   |\mathcal{M}^{\textrm{NLO } T=0}_{\c \bar{\c} \rightarrow i j }|^2
+ |\mathcal{M}^{\textrm{NLO } T\neq0}_{\c \bar{\c} \rightarrow i j }|^2, 
\end{equation}
likewise for the inverse process.\footnote{Note that for the standard 
WIMP annihilation scenarios, there are no finite-temperature corrections 
to the tree-level amplitudes of $2\to2$ and $2\to3$ or $3\to 2$ processes, 
because at tree level annihilation occurs through $t$-channel exchange of 
a particle with mass larger than $m_\chi$, or through highly virtual 
$s$-channel particles.} 
Since the SM particles may have masses smaller or of order of $T_f$, we 
also abandon the assumption that  $f_{i,j}$ are exponentially small 
in the $2\to 3$ and $3\to 2$ processes, where particles $i,j$ need not 
have energy of order $m_\chi$. We can then extend and reorganize the NLO 
thermally averaged cross section into the expression 
\begin{eqnarray} 
\label{eq: collision_term_NLO_pre_1}
\langle \sigma^{\rm NLO} v_{\rm rel} \rangle_{T\neq 0}\!
&=& \!\frac{1}{n_\c^{\rm eq} n_{\bar \c}^{\rm eq}}  \int \!\! d\Pi_{\c \bar \c i j} \, f^{\rm eq}_\chi f^{\rm eq}_{\bar \chi}  \nonumber\\
\,
&&\hspace*{-1cm}
\biggl\{\Bigl(|\mathcal{M}^{\rm LO}_{\c \bar{\c} \rightarrow i j}|^2 
+ |\mathcal{M}^{\textrm{NLO }T=0}_{\c \bar{\c} \rightarrow i j}|^2
+\! \int\!\! d\Pi_\gamma \,|\mathcal{M}_{\c \bar{\c} \rightarrow i j \g}|^2 
\Bigr)
\nonumber\\[0.2cm]
&&\hspace*{-1cm}+\,
|\mathcal{M}^{\textrm{NLO } T\neq0}_{\c \bar{\c} 
\rightarrow i j}|^2 
+ \int\!\! d\Pi_\gamma \bigg[
 f_\g \lp |\mathcal{M}_{\c \bar{\c} \rightarrow i j \g}|^2 + 
|\mathcal{M}_{\c \bar{\c} \g \rightarrow i j}|^2 \rp
\nonumber\\
&& \hspace{-1cm} 
+ \,f_i \lp |\mathcal{M}_{\c \bar{\c} i \rightarrow j \g}|^2\pm|\mathcal{M}_{\c \bar{\c} \rightarrow i j \g}|^2  \rp
+ f_j \lp |\mathcal{M}_{\c \bar{\c} j \rightarrow i \g}|^2\pm|\mathcal{M}_{\c \bar{\c} \rightarrow i j \g}|^2 \rp
\bigg]\biggr\}.\qquad \quad
\end{eqnarray}
Note that we have neglected terms with more than three distribution 
functions, as they are necessarily exponentially suppressed relative to 
those given, since the kinematics of $2 \leftrightarrow 3$ processes 
allows only one particle to be soft. The NLO collision term also 
includes the processes $\c \bar \c j\leftrightarrow i\gamma$, 
$\c \bar \c i\leftrightarrow j\gamma$, which appear first at this order.

In the cross section above there are both $T$-independent and $T$-dependent 
IR divergences. The former are present in the second line 
on the right-hand side of (\ref{eq: collision_term_NLO_pre_1}). However, 
the expression in the parentheses is IR finite by the standard 
$T=0$ KLN cancellations, and we will not discuss it further in this 
work. 
Our main interest is in the remaining two lines which contain 
the finite-temperature correction to the one-loop virtual amplitude 
and emission and absorption processes multiplied by additional 
phase-space distribution functions. Our aim is to show that 
these terms are IR finite and to evaluate the leading 
correction.
Indeed, our main result will be that the relic density can be obtained
by solving the equation analogous to \eqref{eq: collision_term_LO} 
with collision term
\bea 
\label{eq: collision_term_NLO}
C_{\rm NLO} = \langle \sigma^{\rm NLO} 
v_{\rm rel} \rangle_{T\neq 0} \lp  n_\c^{\rm eq}n_{\bar \c}^{\rm eq} 
- n_\c n_{\bar \c}\rp,
\eea
and the NLO thermally averaged cross section replaced 
by \eqref{eq: collision_term_NLO_pre_1}, 
which now depends also on $T$ through a finite-temperature correction. 

Owing to the presence of the 
$\mathcal{M}^{\textrm{NLO }T\neq 0}_{\c \bar{\c} \leftrightarrow i j}$ term, 
in order to obtain a meaningful result at NLO, one needs to perform the 
computation of the amplitudes in the thermal field theory formalism.
The starting point for a systematic treatment is non-equilibrium 
quantum field theory and the closed time-path (CTP) 
formalism~\cite{Schwinger:1960qe,Keldysh:1964ud}. 
In the next section we review the derivation of the Boltzmann equation 
from the Kadanoff-Baym equations \cite{Baym:1961zz} with application 
to relic density computations. The general strategy of this derivation 
is well-known and gives a prescription for the computation of the collision 
term, which consistently takes into account all the thermal corrections. 
We recapitulate it here to set up the notation for the concrete 
calculations to follow. These are performed in an example model 
for DM annihilation, where we can directly observe the cancellation of 
both soft and collinear divergences. As we will show, the IR finiteness 
of the collision term is related to the finiteness of DM particle 
self-energy diagrams in the thermal background. The formalism allows us to 
compute the finite-temperature correction and we find that the 
naive zero-temperature NLO relic density computations are accurate 
up to corrections of order $\mathcal{O}(\alpha \tau^2)$, 
where  $\tau\equiv T/m_\c \ll 1$ and $\alpha$ is 
the fine structure constant. The correction 
has a remarkably simple form.


\section{Derivation of the Boltzmann equation}
\label{sec:Bderivation}

\begin{figure}[t] 
 \centering
 \includegraphics[width=0.35\textwidth]{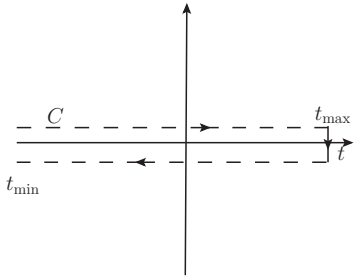}
 \caption{The contour $C$ in the complex time plane. The value 
$t_{\rm{max}}$ can be taken to be $+\infty$ for practical computations.}
 \label{fig:contour}
\end{figure}

In this section we briefly review the derivation of the kinetic equation 
for non-equilibrium propagators, and from this the Boltzmann equation 
for the phase space density functions by performing the Wigner 
transformation and gradient expansion (see, for example, 
\cite{KB,kainulainen:2001cn,prokopec:2003pj,Calzetta:1986cq,Carrington:2004tm}). 
We start with the closed time-path (CTP) formulation of 
non-equilibrium quantum field theory (reviewed, e.g., in the 
book~\cite{LeBellac}), where all correlation functions are defined on a 
complex time plane along the contour $C$, see fig.\ref{fig:contour}. 
The contour Green function for a fermion is defined as
\begin{equation}
iS_{\alpha\beta}(x,y)\equiv\langle T_C \psi_\alpha(x)\bar\psi_\beta(y) 
\rangle,
\end{equation}
where $T_C$ denotes the time ordering operation along the contour. 
It corresponds to four Green functions with real-time arguments:
\begin{eqnarray}
iS_{\alpha\beta}^>(x,y)& \equiv& \langle \psi_\alpha(x)\bar\psi_\beta(y) 
\rangle \qquad iS_{\alpha\beta}^<(x,y)\equiv  
-\langle \bar\psi_\beta(y)\psi_\alpha(x) \rangle \\
iS_{\alpha\beta}^c(x,y)& \equiv& \langle T^c \psi_\alpha(x)\bar\psi_\beta(y) 
\rangle \quad  iS_{\alpha\beta}^a(x,y)\equiv  
\langle T^a \psi_\alpha(x)\bar\psi_\beta(y) \rangle,
\end{eqnarray}
where $T^c(T^a)$ denotes chronological (anti-chronological) 
time ordering in real time.\footnote{Often the upper branch is called `1' 
and the lower `2', and the propagators are denoted as $S^>=S^{21}$, 
$S^<=S^{12}$, $S^c=S^{11}$ and $S^a=S^{22}$.} 
The brackets $\langle\ldots\rangle$ imply 
averaging over an ensemble at time $t_{\rm{min}}$. The free-field 
momentum-space propagators are summarized in appendix~\ref{sec:FeynRules}.  
The formalism describes a general non-equilibrium system, where the 
physical macroscopic observables are averages over an ensemble. The CTP 
formulation originates from the need to describe the time evolution of 
an operator expectation value (``in-in formalism'') rather than a 
scattering matrix element (``in-out formalism'').

In an interacting non-equilibrium system the two-point Green functions 
depend on both space-time coordinates, which may be chosen as relative 
coordinate $r=x-y$, and averaged (macroscopic) coordinate 
$X=\frac{x+y}{2}$. In equilibrium the system can depend only on 
the relative coordinate due to translation invariance. Therefore, for 
systems not far from equilibrium it is useful to perform the Wigner 
transform and define the Green functions
\begin{equation}
G(X,p)\equiv\int_{t_{\rm{min}}}^{t_{\rm{max}}} d^4u \,e^{ipu} \,
G\left(X-u/2,X+u/2\right),
\end{equation}
(and similarly the self-energies). 
The dependence on $p$ describes the fluctuations on the microscopic scale of 
particle interactions, while the coordinate $X$ describes the macroscopic 
space-time variations. In equilibrium, the Wigner-space Green 
functions  depend only on the momentum $p$.

The contour Green functions obey the Dyson-Schwinger equation
\begin{eqnarray}
S_{\alpha\beta}(x,y)&=&S^0_{\alpha\beta}(x,y)-\int_C d^4z \int_C d^4z' 
S^0_{\alpha\gamma}(x,z) \Sigma_{\gamma\rho}(z,z') S_{\rho\beta}(z',y),
\end{eqnarray}
where the superscript `0' denotes the free propagators, and $\Sigma$ 
is the self-energy. The Dyson-Schwinger equations lead to the 
Kadanoff-Baym equations \cite{Baym:1961zz}
\begin{eqnarray}
\label{eq:KBf}
&& (i\sl\partial_x-m_{\c})S^{\lg}(x,y)-\int d^4z \left( \Sigma_h(x,z) 
S^\lg (z,y) + \Sigma^\lg(x,z) S_h (z,y) \right)= \mathcal{C_\c}\,,
\end{eqnarray}
where the collision is defined as
\begin{eqnarray}
\mathcal{C_\c}&\equiv& \frac{1}{2}\int d^4z\left( \Sigma^>(x,z) S^< (z,y) -  
\Sigma^<(x,z) S^> (z,y) \right).
\end{eqnarray}
Here the subscript $h$ denotes the hermitian part, 
$\Sigma_h=\Sigma^c-\frac{1}{2}\left( \Sigma^> + \Sigma^<\right)$ and 
analogously for the Green functions.\footnote{Note that 
it is actually $\g^0 \Sigma_h$ which is hermitian, since 
$(\g^0 \Sigma_h)^\dag = \g^0 \Sigma_h$, and the same for $S_h$.} 
These equations for the Green functions 
are exact functional equations, but too difficult to solve. At this point 
we apply the approximations described in the introduction. First, we 
transform to Wigner space and take $t_{\rm{min}}=-\infty$. 
Then we perform the gradient expansion up to the first order in gradients, 
upon which \eqref{eq:KBf} becomes
\begin{eqnarray}
&& \left(\sl p+\frac{i}{2}\sl \partial - m_\c \right) S^\lg - 
\Sigma_h S^\lg - \Sigma^\lg S_h +\frac{i}{2}\lbrace \Sigma_h ,
S^\lg \rbrace +\frac{i}{2}\lbrace \Sigma^\lg ,S_h \rbrace 
+\mathcal{O}(\nabla^2) = \mathcal{C_\c}\,,\qquad
\label{eq:KBfW}
\end{eqnarray}
where the Green functions now depend on $X$ and $p$, 
$\partial_\mu \equiv \frac{\partial}{\partial X^\mu},$
and 
\begin{equation}
\lbrace A, B\rbrace = \frac{\partial A}{\partial p_\mu}
\frac{\partial B}{\partial X^\mu} - \frac{\partial A}{\partial X_\mu}
\frac{\partial B}{\partial p^\mu},
\end{equation}
denotes an analogue of the Poisson bracket with respect to 
the coordinates $X$ and $p$. 
The expanded collision term in Wigner space reads
\begin{eqnarray}
\label{eq:collisionWf}
\mathcal{C_\c}&=& \frac{1}{2}\left( \Sigma^> S^<  -  \Sigma^< S^> \right) 
- \frac{i}{4}\left( \lbrace\Sigma^>, S^<\rbrace  - 
\lbrace \Sigma^< ,S^>\rbrace \right) +\mathcal{O}(\nabla^2) .
\end{eqnarray}

By performing the gradient expansion we assume that the variations of  
physical quantities in the coordinate $X$ are small with respect to the 
typical inverse momenta of the plasma excitations. The latter are of the 
order of plasma temperature $T$. As concerns the former, in the homogeneous 
Universe, the macroscopic variation of dark matter particle number density 
is set by the expansion rate $H$ and the annihilation rate 
$\Gamma \sim n\, \alpha^2/m_\c^2 \sim \alpha^2 T^{3/2} 
m_\c^{-1/2} e^{-m_\c/T}$, both of which are of the same order, when the 
number density freezes out at $m_\c/T \sim 20$. 
Thus gradients $\nabla\sim \Gamma$
are exponentially suppressed and it is a good approximation to 
keep only the zeroth order in the gradient expansion, which corresponds 
to neglecting all terms with Poisson brackets in~\eqref{eq:KBfW}. 
In addition there is 
also an expansion in the coupling constants of the interactions, as long 
as they are weak. As we show below, when the collision term is 
evaluated at lowest non-vanishing order (and in zeroth order of the 
gradient expansion), one recovers the standard freeze-out equation 
for the DM number density. Since in the cases of interest 
\begin{equation}
\frac{\nabla}{T} \ll \alpha \ll 1,
\end{equation}
the next order in the coupling expansion is much more important than 
higher-order gradient terms, but still allows for a perturbative 
expansion in $\alpha$. Thus, in the following, when we consider 
relic density computations at NLO, we mean next-to-leading order in 
the coupling constants in the collision term, but leading order in 
the gradient expansion. The relevant equation is then 
\begin{eqnarray}
&& \left(\sl p+\frac{i}{2}\sl \partial - m_\c \right) S^\lg - \left[ 
\Sigma_h S^\lg + \Sigma^\lg S_h\right]_{|\rm NLO}
= \frac{1}{2}\left( \Sigma^> S^<  -  \Sigma^< S^> \right)_{|\rm NLO}.
\label{eq:KBfWLO}
\end{eqnarray}
Separating the hermitian and anti-hermitian parts leads to constraint and 
kinetic equations
 \begin{eqnarray}
2p^0 i \gamma^0 S^\lg - \bigl\lbrace\vec{p}\cdot\vec{\gamma} \gamma^0
+m_\c\gamma^0+\Sigma_h \g^0, i\gamma^0 S^\lg\bigr\rbrace 
- \lbrace i\Sigma^\lg \g^0,\g^0 S_h\rbrace &=& 
i\mathcal{C}_{\chi}- i\mathcal{C}_{\chi}^\dagger, \quad
\label{eq:KBconstr} \\[0.2cm]
 i\partial_t i \gamma^0 S^\lg - 
\left[\vec{p}\cdot\vec{\gamma}\gamma^0 +m_\c\gamma^0+\Sigma_h \g^0, 
i\gamma^0 S^\lg\right] 
- \left[ i\Sigma^\lg \g^0,\g^0 S_h\right]
&=& i\mathcal{C}_{\chi}+ i\mathcal{C}_{\chi}^\dagger, \quad
\label{eq:KBkin}
\end{eqnarray}
where here  $\lbrace\cdot,\cdot\rbrace$ and  $\left[\cdot,\cdot\right]$ 
denote the anti-commutator and commutator, respectively.

The constraint equation (\ref{eq:KBconstr}) to zeroth order takes the 
simple form
\begin{equation}
\label{eq:KBconstr0}
\bigl\lbrace (\sl p - m_\c)\g^0 , i \g^0 S^\lg\bigr\rbrace = 0.
\end{equation}
It describes the spectral properties of the quasi-particles and in 
particular puts constraints on the structure of the Green function. 
Inserting the most general parameterization of the Dirac matrix 
structure compatible with spatial isotropy,
\begin{equation}
iS^\lg = m_\c \left(g_s^\lg + g_p^\lg \gamma^5 \right) + 
g_{v0}^\lg\, p^0\gamma^0 - g_{v3}^\lg \,\vec{p}\cdot\vec{\gamma} 
+ g_{a0}^\lg\, p^0\gamma^0\gamma^5  - 
g_{a3}^\lg \,\vec{p}\cdot\vec{\gamma} \gamma^5 
+ g_{t}^\lg [\gamma^0,\vec{p}\cdot\vec{\gamma}],\
\end{equation}
the constraint equation (\ref{eq:KBconstr0}) leads to the conditions
\begin{equation}
g^\lg_s=g^\lg_{v0}=g^\lg_{v3}\equiv g^\lg, \qquad 
g^\lg_p=g^\lg_{a0}=g^\lg_{a3}=g^\lg_t=0.
\end{equation}
Hence the Green function must be of the form
\begin{equation}
\label{eq:constraint}
iS^\lg=(\sl p +m_\c)g^\lg.
\end{equation}
The general solution of the constraint equation (\ref{eq:KBconstr0}) can 
also be written in the form of the Kadanoff-Baym ansatz
\begin{eqnarray} \label{eq: KBansatz1}
iS^< &=& -A_0\left[\Theta(p^0)f_\c(\vec{p}\,) + 
\Theta(-p^0)(1-f_{\bar\c}(-\vec{p}\,))  \right ], \\[0.2cm]
\label{eq: KBansatz2}
iS^> &=& -A_0\left[-\Theta(p^0)(1-f_\c(\vec{p}\,)) - 
\Theta(-p^0)f_{\bar\c}(-\vec{p}\,)  \right ].
\end{eqnarray}
where $A_0(X,p)=\langle \left[\psi(x),\bar\psi(y)\right]\rangle_0$ describes 
the spectral properties of the quasi-particles in zeroth order, and 
is given by
\begin{equation}
A_0(X,p)= 2\pi\delta(p^2-m_\c^2)(\sl p +m_\c)\varepsilon(p^0),
\end{equation}
with $\varepsilon$ the 
sign function $\varepsilon (p^0)=\Theta ( p^0) -\Theta( -p^0)$. 
By comparing with (\ref{eq:constraint}) we identify
\begin{eqnarray}
\label{eq:g12}
g^<(X,p) &=& -2\pi\delta(p^2-m_\c^2)
\left[\Theta(p^0)f_\c(\vec{p}\,) - \Theta(-p^0)(1-f_{\bar\c}(-\vec{p}\,))  
\right ],\\[0.2cm]
\label{eq:g21}
g^>(X,p) &=& -2\pi\delta(p^2-m_\c^2)
\left[-\Theta(p^0)(1-f_\c(\vec{p}\,)) 
+ \Theta(-p^0)f_{\bar\c}(-\vec{p}\,)\right ] .
\end{eqnarray}

Finally, the Boltzmann equation follows from combining the 
kinetic equation (\ref{eq:KBkin}) in zeroth order of the gradient 
expansion with the quasi-particle approximation and the solution 
(\ref{eq:constraint}), (\ref{eq:g12}-\ref{eq:g21}) of the zeroth-order 
constraint equation. We first note that the term 
$\bigl[\vec{p}\cdot\vec{\gamma}\gamma^0 +m_\c\gamma^0, i\gamma^0 S^\lg\bigr]$ 
vanishes with the above ansatz for $S^\lg$. Next we examine the terms
containing commutators with self-energies. 
We assume that the deviation of the
DM particle distribution from thermal equilibrium is sufficiently small 
that the self-energy can be computed with propagators (\ref{eq: KBansatz1}-\ref{eq: KBansatz2}).
Then at one-loop we can use parametrizations
\begin{equation}
\Sigma_h=\alpha p^0\g^0-\beta\vec{p}\cdot \vec{\g} + \sigma m_\c, 
\qquad  \Sigma^\lg=a^\lg p^0\g^0-b^\lg\vec{p}\cdot \vec{\g} + c^\lg m_\c,
 \end{equation}
where $\alpha$,$\beta$, $\sigma$, $a^\lg$, $b^\lg$ and $c^\lg$ are 
scalar functions of the momentum. With this ansatz one can check that 
both $\bigl[i\Sigma^\lg\gamma^0, \gamma^0 S_h\bigr]$  and
$\bigl[\Sigma_h\gamma^0, i\gamma^0 S^\lg\bigr]$ are proportional 
to $\vec{p}\cdot\vec{\g}$ and for this reason, after taking the trace 
over spinor indices, will not contribute to the Boltzmann equation.

Then, multiplying (\ref{eq:KBkin}) by $2\Theta(p^0)$,  taking the trace,  
and integrating over $p^0$, we obtain, after using 
(\ref{eq:constraint}) with (\ref{eq:g12}-\ref{eq:g21}):
\begin{equation}
\label{eq:collBEf}
E_\chi \partial_t f_\c = -E_\chi \int \frac{dp^0}{(2\pi)}\,\Theta(p^0) 
\,\frac{1}{2}\,\Tr\{\mathcal{C_\c} +\mathcal{C}_\c^\dagger \}.
\end{equation}
In the FRW background, the time derivative needs to be replaced by the 
covariant one, from which we recover~\eqref{eq:Boltzmannf} 
with the collision term $\mathcal{C}[f]$ given by the right-hand side of 
(\ref{eq:collBEf}). 
The Boltzmann equation 
for the number density~(\ref{eq:Boltzmann}) has the integrated collision 
term given by
\be \label{eq: collision_term_matching}
C = -h_\chi \int\frac{d^4 p}{(2\pi)^4} \,\Theta(p^0) \,\frac{1}{2}
\,\Tr\{\mathcal{C_\c} +\mathcal{C}_\c^\dagger \}.
\ee
In the next section we will demonstrate the consistency between the above 
equation and~(\ref{eq: collision_term_LO_pre}), and compute the 
collision term to NLO.


\section{The collision term} 
\label{sec:collision_term}

In this section we present the computation of the collision term at 
zeroth order in the gradient expansion in a  ``bino-like'' DM model. 
In this model the DM Majorana fermion $\chi$ annihilates at tree-level 
into SM fermions via a $2\rightarrow 2$ process, mediated by $t$- and 
$u$-channel exchange of a heavy scalar particle $\phi$.
The collision term is computed including the {\em thermal} NLO 
contributions to the annihilation process.

After introducing the model, we illustrate explicitly the calculation of 
the right-hand side of (\ref{eq:KBfWLO}) at tree-level in the CTP formalism 
to show how the collision term can be expressed in terms of 
annihilation cross sections and phase-space distributions, as in the standard 
expression (\ref{eq: collision_term_LO_pre}). We then proceed to 
the main part of the paper and compute the thermal NLO corrections.
As we are primarily interested in the infrared divergence cancellation
at finite temperature and the leading finite-temperature correction, we 
drop the terms that can be associated with the $T=0$ NLO correction to 
the annihilation cross section, 
even though its finite part is parametrically larger than the finite-$T$ 
correction. The computation of the zero-temperature cross section is 
already well understood and could be 
straightforwardly included in the formalism.
We note that while we focus on a particular model to perform
explicit calculations, the procedure itself is much more general
and can be applied to a variety of different DM scenarios.


\subsection{The model}
\label{sec:model}

We consider the extension of the Standard Model by an $SU(2)\times U(1)$ 
singlet Majorana fermion and a scalar doublet $\phi=(\phi^+,\phi^0)^T$. The 
relevant terms in the Lagrangian read
\begin{eqnarray}
\mathcal{L} &=& - \frac{1}{4} F^{\mu\nu} F_{\mu\nu} + \bar f 
\left(i\sl D-m_f \right) f 
+\frac{1}{2}\,\bar\chi\left(i\sl \partial-m_\c \right)\chi\nonumber\\
&&  + \,
(D_\mu \phi)^\dagger (D^\mu \phi)-m_{\phi}^2 \phi^\dagger \phi +
\left(\lambda\bar\chi P_L f^- \phi^+ 
+h.c.\right)\,,
\end{eqnarray}
where the SM fermions form a left-handed doublet $f=(f^0,f^-)^T$. 
In this model the only interaction involving the DM particle $\c$ is the 
Yukawa interaction with the ``sfermion'' $\phi$ and SM (light) fermion 
doublet $f$, of which we include only the charged component. 
The neutral component would affect the inclusive tree-level cross section 
through the $\lambda \bar\chi P_L f^0 \phi^0$ interaction, 
which allows $\c\bar \c\to f^0\bar{f}^0$, however this process receives 
no radiative corrections since it contains only electrically neutral 
particles. 
The scenario we have in mind, realized in the minimal 
supersymmetric SM (MSSM) if 
the dark matter is the bino, is an electroweak or TeV scale DM particle, 
and a scalar (sfermion) with mass $m_\phi > m_\chi \approx 
\mathcal{O}(0.1 {\rm -} 1\,\textrm{TeV})$. 
In this situation the freeze-out occurs after the electroweak phase 
transition. In the covariant derivative $D_\mu=\partial_\mu - i e A_\mu$ 
we therefore keep only the electromagnetic term.

The motivation for studying this particular scenario follows  
from its relevance to the dark matter phenomenology of the MSSM, 
and from its relative simplicity. Moreover, in such a model 
the zero-temperature NLO corrections have been shown to be 
significant \cite{Ciafaloni:2011oq}, since they lift the helicity 
suppression of the LO annihilation process. 


\subsection{Calculation of the collision term at LO}
\label{sec:collision_term_calculation_LO}

In the CTP formalism the fermion collision term~(\ref{eq:collisionWf})
to leading order in gradient expansion is given by 
\begin{eqnarray}
\mathcal{C_\c} &=& \frac{1}{2}\left( \Sigma^> S^<  -  \Sigma^< S^> \right).
\label{eq:LOcollision}
\end{eqnarray}
In the calculation of the self-energies the phase-space distribution 
functions of all the interacting particle species appear in the 
finite-temperature propagators, see appendix~\ref{sec:FeynRules}. 
The two terms $\Sigma^<$ and $\Sigma^>$ account for all 
possible processes involving the interacting species, which includes 
annihilation, production and scattering processes for $\chi$, 
as well as absorption processes characteristic of the 
finite-temperature plasma. In the kinetic equation for the 
particle number density, the contributions from particle-number 
preserving scattering 
processes $\chi f \rightarrow \chi f$ cancel out after summing over 
the two terms on the right-hand side of (\ref{eq:LOcollision}) and 
after taking the trace and 
performing the integral over the particle four-momentum 
in (\ref{eq: collision_term_matching}). These terms will therefore 
be omitted right away.

\begin{figure}[t] 
 \centering
 \includegraphics[width=0.225\textwidth]{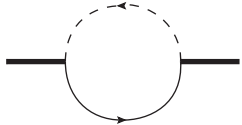}
 \caption{The DM \self\ at \onel. The same diagram topology with 
reversed arrows is not shown.}
 \label{fig: 1loop}
\end{figure}

\begin{figure}[t] 
 \centering
 \includegraphics[width=0.75\textwidth]{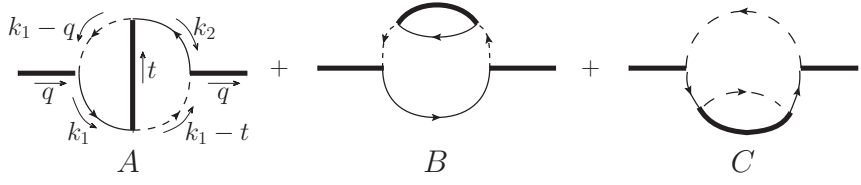}
 \caption{The DM \self\ at \twol. The same diagram topologies 
with reversed arrows are not shown for simplicity. 
In the following they are into account and denoted by a 
superscript \textit{rev}. 
}
 \label{fig: 2loop}
\end{figure}

We start from the calculation at leading order in the coupling (loop) 
expansion to show the correspondence between the self-energy diagrams 
and annihilation processes.
The one-loop self energy, shown in fig.~\ref{fig: 1loop}, describes 
$1\leftrightarrow 2$ processes, which are not relevant 
for the relic-density computation, because they are kinematically 
forbidden or exponentially 
suppressed.\footnote{The $\phi$ co-annihilation scenario can 
be straightforwardly included in the presented formalism, but is 
beyond the scope of this work. In this scenario the diagram of 
fig.~\ref{fig: 1loop} describes $\phi\leftrightarrow \chi f$ 
(inverse) decays.}
Therefore, the LO annihilation process $\chi\chi \leftrightarrow f\bar f$ 
must be contained in the two-loop 
self-energy diagrams of \fig.~\ref{fig: 2loop}.
The self energies $\Sigma^{<,>}$ are computed from the diagrams 
discussed above by applying the Feynman rules of the CTP formalism 
with the propagators of appendix~\ref{sec:FeynRules}, and the proper 
treatment of the fermion-number violating interactions
for Majorana fermions~\cite{Denner1992}.
We denote the propagator of the charge-conjugate fermion field as
$S^{' ab}(p) \equiv C\lp S^{ab}(p) \rp^T C^{-1}$, where $C$ is
the charge-conjugation matrix and the transpose is with respect 
to the spinor indices only.

%
\begin{figure}[t] 
 \centering
 \includegraphics[width=1\textwidth]{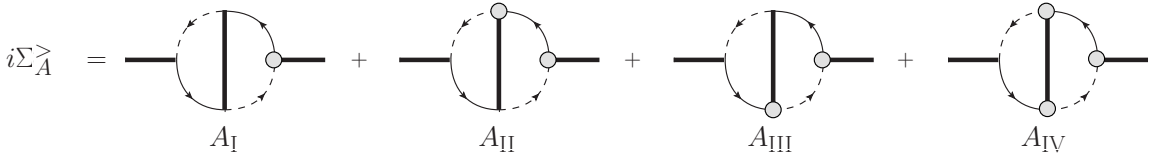}
 \caption{$i\Sigma^>_A$ as given by the CTP Feynman rules.
 Uncircled and circled vertices denote type `1' and type `2'  
 vertices,  respectively.}
 \label{fig: 2loop_A}
\end{figure}

\begin{figure}[t]
\centering
\includegraphics[width=1\textwidth]{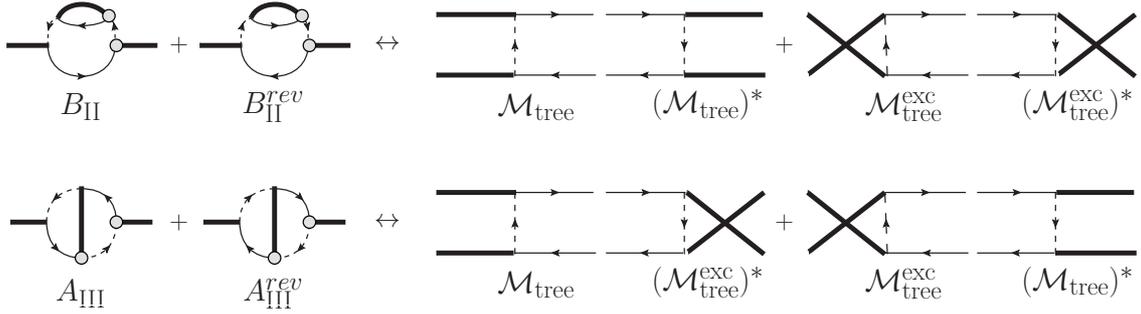}
\caption{Tree-level annihilation diagrams for a Majorana 
fermion and their matching with the two-loop self-energies. 
Note the correspondence between reversing the fermion flow arrows 
and crossing the external legs.
}
\label{fig:diagtree}
\end{figure}

Let us consider the contribution to $\Sigma^>(q)$ from diagram $A$ in 
\fig.~\ref{fig: 2loop}.  Since $\Sigma^> = \Sigma^{21}$, 
the left vertex is of the type `1' and the right of type `2', while one 
has to sum over both types of internal vertices. 
This leads to the sum of the four diagrams in fig.~\ref{fig: 2loop_A}, 
where uncircled and circled vertices denote type `1' and type `2' 
vertices, respectively. Fixing the fermion flow and assigning the momenta 
as in fig.~\ref{fig: 2loop} the whole expression appearing in the 
collision term reads
\begin{eqnarray}
i\Sigma^>_{A}\lp q\rp iS^< \lp q\rp&=&\!\!\!
\sum_{a,b=1,2}\!\! \!-(-1)^{a+1}(-1)^{b+1}\lambda^4\int\frac{d^4t}{\lp2\pi\rp^4}
\frac{d^4k_1}{\lp2\pi\rp^4}\frac{d^4k_2}{\lp2\pi\rp^4}\,\lp2\pi\rp^4
\delta^{(4)}(q+t-k_1-k_2) 
\nonumber\\
&& \hspace{-1.75cm}\times i\Delta^{1a}\lp k_1-q\rp i\Delta^{2b}
\lp k_1-t\rp\pr iS_f^{'a2}\lp -k_2\rp\pl iS^{ab}\lp t\rp\pl iS_f^{b1}
\lp k_1\rp\pr iS^{12}\lp q\rp. 
\label{eq: AIIIsum}
\end{eqnarray}
At this point we note that the thermal part of the sfermion propagator
is exponentially suppressed, since  $m_\phi>m_\chi \gg T_f$. Dropping this 
part implies that only the 11 and 22 components of $\Delta^{ab}$ are 
non-vanishing, so the ends of a scalar (dashed) line must either both 
be circled or not. Hence the only diagram in fig.~\ref{fig: 2loop_A} 
that we have to compute is $A_\text{III}$.
Taking the trace over the spinor indices, 
which accounts for the 
polarization sum 
in the number density equation, 
the previous equation simplifies to
\bea
\Tr\{\Sigma^>_{A_{\rm III}}\lp q\rp S^< \lp q\rp \}&=&
-\lambda^4\int\frac{d^4t}{\lp2\pi\rp^4}\frac{d^4k_1}{\lp2\pi\rp^4}
\frac{d^4k_2}{\lp2\pi\rp^4}\lp2\pi\rp^4\delta^{(4)}(q+t-k_1-k_2) 
\nonumber\\
&& \hspace{-3.6cm} \times \underbrace{i\Delta^{11}\lp k_1-q\rp i\Delta^{22}
\lp k_1-t\rp}_{\equiv\mathcal{S}} \,
\underbrace{\Tr\{\pr iS_f^{'12}\lp -k_2\rp\pl iS^{12}\lp t\rp\pl iS_f^{21}
\lp k_1\rp\pr iS^{12}\lp q\rp\}}_{\equiv\mathcal{F}}. \qquad
\label{eq: AIII}
\eea
Since in the scalar part $\mathcal{S}$ we need only the $T=0$ part 
of the propagators, we have 
\be
\mathcal{S}=\frac{i}{\lp k_1-q\rp^2-m_\phi^2}\,
\frac{-i}{\lp k_1-t\rp^2-m_\phi^2}.
\ee
In the fermion part $\mathcal{F}$ both the $T=0$ and the thermal parts 
contribute, in principle. However, the expression involves only the purely 
thermal off-diagonal CTP propagator, 
leaving
\bea
\mathcal{F}&=&\textrm{Tr}\lbrace \pr \lp \sl k_2+m_f\rp 
\pl \lp\sl t+m_\chi\rp \pl \lp\sl k_1+m_f\rp \pr \lp\sl q+m_\chi\rp \rbrace 
\nonumber\\[0.1cm]
&& \hspace{-0.4cm} \times(2\pi)^4\delta(q^2-m_\chi^2)\,\delta(t^2-m_\chi^2)\,
\delta(k_1^2-m_f^2)\,\delta(k_2^2-m_f^2)
\nonumber\\[0.1cm]
&& \hspace{-0.4cm} \times \!
\lc \Theta(-k_2^0)f_f(-\vec{k}_2)-\Theta(k_2^0)\bigl( 1-f_{\bar{f}}(\vec{k}_2) 
\bigr)\rc  
\lc \Theta(t^0)f_\chi(\vec{t}\,)-\Theta(-t^0)\lp 1-f_\chi(-\vec{t}\,) \rp\rc 
\nonumber\\
&& \hspace{-0.4cm} \times \!
\lc -\Theta(k_1^0)\bigl( 1-f_f(\vec{k}_1) \bigr)+ 
\Theta(-k_1^0)f_{\bar{f}}(-\vec{k}_1)\rc  
\lc \Theta(q^0)f_\chi(\vec{q}\,)-\Theta(-q^0)\lp 1-f_\chi(-\vec{q}\,) \rp\rc\!.
\qquad\
\label{eq:F}
\eea
The last two lines of (\ref{eq:F}) lead to 16 distinct terms describing 
different processes in the thermal plasma. Half of them vanish after 
multiplying by $\Theta(q^0)$ as needed for (\ref{eq: collision_term_matching}). Out of 
the remaining 8 terms 5 are kinematically forbidden, since they refer
to $4\leftrightarrow 0$ and $1\leftrightarrow 3$ processes. One is left 
with two terms corresponding to scatterings $\chi f \rightarrow \chi f$ 
and $\chi \bar{f} \rightarrow \chi \bar{f}$, which do not contribute to 
the number-changing processes and cancel out after including the 
$\Sigma^<S^>$ contribution, and one term describing the annihilation 
process $\chi\chi \rightarrow f\bar{f}$. Only this last term contributes 
to the integrated collision term.
As explained in the introduction, we assume the 
background plasma to be in thermal equilibrium with zero chemical
potential and therefore the SM fermion distribution function is the Fermi-Dirac one
for both particle and antiparticle, $f_{\bar{f}} = f_f = f^{\rm{eq}}_f $ and
\be \label{eq: Fermi-Dirac}
\Theta(p^0)f^{\rm eq}_f(\vec p\,) = \frac{1}{e^{\beta p^0}+1} \equiv f_F(p^0) 
\qquad \textrm{with } p^0\equiv \sqrt{\vec p^{\,2}+m_f^2}\ .
\ee
Finally we get
\bea
\Tr\{\Sigma^>_{A_{\rm III}}\lp q\rp S^< \lp q\rp\}&=& 
\frac{1}{2E_{\chi_1}}\lp2\pi\rp\delta\lp q^0-E_{\chi_1}\rp 
\int\frac{d^3\vec{t}}{\lp2\pi\rp^3 2E_{\chi_2}}
\nonumber\\
&& \hspace*{-2cm} 
\times \int \frac{d^3\vec{k}_1}{\lp2\pi\rp^3 2 E_{f_1}} 
\frac{d^3\vec{k}_2}{\lp2\pi\rp^3 2 E_{f_2}} 
\lp2\pi\rp^4 \delta^{(4)}\lp q+t-k_1-k_2\rp 
\,|\mathcal{M}_{A_{\rm III}}|^2
\nonumber\\[0.1cm]
&&  \hspace*{-2cm}  \times 
\lc f_\c(\vec{q}\,)  f_\c(\vec{t}\,) \bigl( 1- 
f_f^{\text{eq}}(\vec{k_1})\bigr) \bigl( 1- f_f^{\text{eq}}(\vec{k_2})\bigr) \rc, 
\label{eq: AIII_fin}
\eea
where all the momenta are on-shell. Adding the hermitian conjugate and integrating this expression with 
$-h_\chi d^4q/(2\pi)^4\, \frac{1}{2}\Theta(q^0)$ as appropriate to the 
collision term for the $\chi$ number 
density (\ref{eq: collision_term_matching}) and accounting for the 
factor $1/2$ in (\ref{eq:LOcollision}), 
the structure of the result is now manifestly as in 
(\ref{eq: collision_term_LO_pre}), with a zero-temperature annihilation   
cross section times velocity multiplied by the statistical factors 
corresponding to the process 
$\chi_1(q)\chi_2(t)\rightarrow f(k_1)\bar f(k_2)$.
The matrix element squared can be recognized as the interference 
term between the two tree-level diagrams for the annihilation process
\annproc, as shown in fig.~\ref{fig:diagtree}. Specifically, 
\bea
|\mathcal{M}_{A_{\rm III}}|^2 = 
-\lambda^4\ \mathcal{S}\ \text{Tr}\{\cdots\} 
=\mathcal{M}_\textrm{tree} \lp\mathcal{M}_\textrm{tree}^\textrm{exc}\rp^*,
\label{eq: M_AIII} 
\eea
where the trace refers to the first line of (\ref{eq:F}).
The same procedure applied to the diagram $B$ in \fig.~\ref{fig: 2loop}
and to the corresponding diagrams with reversed arrows leads to 
the identifications
\bea
|\mathcal{M}_{B_{\rm II}}|^2 &=& |\mathcal{M}_\textrm{tree}|^2,
\nonumber\\
|\mathcal{M}_{A_{\rm III}}^{rev}|^2 &=& 
\mathcal{M}_\textrm{tree}^\textrm{exc} \lp\mathcal{M}_\textrm{tree}\rp^*,
\nonumber\\
|\mathcal{M}_{B_{\rm II}}^{rev}|^2 &=& 
|\mathcal{M}_\textrm{tree}^\textrm{exc}|^2.
\eea
Diagram $C$ of \fig.~\ref{fig: 2loop} does not contribute, since as 
discussed above we can ignore any contribution with an off-diagonal 
sfermion CTP propagator.

The calculation of $\Sigma^< S^>$ is analogous and reproduces 
the first term in (\ref{eq: collision_term_LO_pre}), which corresponds to
the production process $f \bar f\rightarrow\chi\chi$.
We therefore conclude that -- as anticipated -- at LO in the CTP formalism, 
that is, inserting the DM \self\ at \twol\ into 
(\ref{eq:LOcollision}) and~(\ref{eq: collision_term_matching}) 
for the integrated collision term, leads to the standard Boltzmann 
equation~(\ref{eq:Boltzmann}). This is true under the assumptions of the gradient 
expansion and quasi-particle approximation, which are well-satisfied 
for the standard scenario of freeze-out of an initially thermal 
DM particle population. At LO in the coupling expansion the  
integrated collision term is, provided the 
\tree\ $2\rightarrow 2$ processes are $\chi\chi\leftrightarrow f\bar f$, as in 
(\ref{eq: collision_term_LO_pre}).

\subsection{Tree-level annihilation cross section}

For later reference, we give the tree-level $\chi\chi\to f\bar f$ 
cross section in our model. More precisely, we give the cross section 
times velocity expanded in the small velocity (partial waves) 
in the non-relativistic regime,\footnote{We recall that this  
cross section is \textit{summed} over both initial and 
final state polarizations.}
\be
\label{eq:sigmaab}
4 E_\c^2 \ \sigma_{\chi\chi \rightarrow f \bar f\ } v_{\rm rel} = 
a_{\rm tree} + b_{\rm tree} v^2,
\ee
where $v$ is the CM velocity of one DM particle and we extracted the 
flux factor $4E_\c^2$ so that $a_{\rm tree}$ and $b_{\rm tree}$ 
are dimensionless, and correspond to the s- and p-wave contributions,
respectively.
It proves useful to adopt variables rescaled by the DM mass, i.e. 
\begin{equation}
\tau\equiv\frac{T}{m_\c}, \qquad 
\xi\equiv\frac{m_{\phi}}{m_\c}, \qquad
\epsilon\equiv\frac{m_f}{2m_{\c}}.
\end{equation}
Then 
\bea \label{eq:abtree}
a_{\rm tree} &=& \frac{2\lambda^4}{\pi} \epsilon ^2 
\frac{\sqrt{1-4\epsilon ^2}}{\left(1+\xi ^2-4\epsilon ^2\right)^2}, 
\\[0.2cm]
b_{\rm tree} &=&\frac{4\lambda^4}{3\pi}
\frac{
1+\xi ^4 -\epsilon ^2 \left(9 +8 \xi^2+5 \xi ^4 \right) 
+\epsilon ^4 \left(31 + 46\xi^2 + 7\xi^4\right) 
-8 \epsilon ^6 \left(9+7 \xi ^2\right) +112 \epsilon ^8}
{\sqrt{1-4 \epsilon ^2} \left(1+\xi ^2-4\epsilon ^2\right)^4}.
\nonumber \\
\eea
Note that  always  $\xi \geq 1$ and $\epsilon \leq \frac{1}{2}$, 
but typically $\epsilon \ll 1$. In the first term the appearance of 
the $\epsilon^2$ factor implies the well-known helicity suppression of 
s-wave annihilation of a Majorana fermion into SM fermions.


\subsection{Collision term at NLO}
\label{sec:collision_term_calculation_NLO}

Now that we understand how to match the collision term in the CTP 
formalism to the form of the freeze-out equation with 
the standard computation of annihilation cross sections, we are ready 
to consider the NLO thermal corrections.
They are encoded in the three-loop DM self-energy diagrams obtained
by adding a photon line to the diagrams 
$A$ and $B$ in \fig.~\ref{fig: 2loop} in all possible ways.\footnote{
Diagrams leading to s-channel photon exchange via a 
loop-induced $\c \bar\c \gamma$  coupling of 
DM  to the photon do not contribute to the \textit{thermal} correction. 
The virtual process with the thermal photon is kinematically forbidden, while 
with thermal fermion is suppressed by additional power of momentum of the 
thermal particle, leading to vanishing correction at the 
order $\mathcal{O}(\tau^2)$.
} From the annihilation amplitude 
point of view they can be arranged into three classes: 
\textit{i)} processes corresponding to thermal emission and absorption, 
\textit{ii)} thermal internal virtual corrections
and \textit{iii)} thermal corrections to mass and 
wave-function renormalization on the external legs. 
We use this classification for organizing the discussion of the computation, 
even though it is somewhat artificial from the self-energy diagram 
point of view. The reason is that we want to show a clear connection 
between the usual way of doing calculations and the 
quantities appearing in the collision 
term as derived from CTP formalism. When showing the results for 
IR divergence cancellation and leading thermal correction 
we revert to the more natural classification 
based on different self-energy diagrams.

At NLO there are 20 three-loop self-energy diagrams contributing to 
$\Sigma^>$  of a Majorana fermion.\footnote{For a Dirac fermion 
there are only half that number, as no clashing arrows are allowed and 
hence the diagrams of type $A$ vanish.} 
They are given in tables~\ref{tab:seA} and~\ref{tab:seB}, together with the 
corresponding processes they describe after associating the terms 
in the CTP sums with matrix elements squared.  
Since the thermal part of the propagators always contains the on-shell 
delta function $\delta(p^2-m^2)$ we refer to these contributions 
as ``cuts'' of the self-energy diagrams.

In the remainder of this section we describe the method of performing the 
calculations emphasizing the differences with respect to the $T=0$ case. 
The results and their discussion will follow in section \ref{sec:results}.

\begin{figure}[t]
\centering
\vskip-0.8cm
\includegraphics[width=1\textwidth]{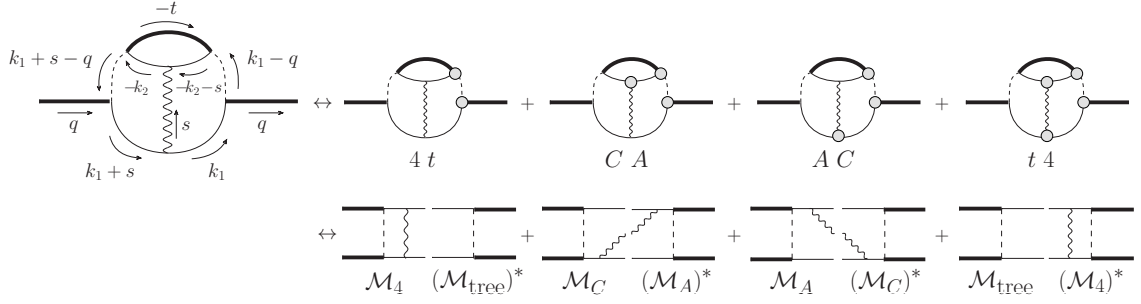}
\caption{An example three-loop self-energy contribution to  $i\Sigma^>$
decomposed into 
a sum over ``cuts'' and the interpreation of the cuts as scattering 
processes. $i\Sigma^>$ is obtained by taking the sum over all possible 
diagrams in which the vertex attached to the external line on the left 
(right) is of type `1' (`2'). 
The correspondence between reversing the charge flow arrows and crossing 
the external fermion legs 
is the same as displayed in fig.~\ref{fig:diagtree}. For
simplicity, from this figure on, we will denote with a single diagram 
with no arrows the sum of the two diagrams with and without reversed arrows.}
\label{fig: 3_matched}
\end{figure}

\setlength\tabcolsep{14pt}
\begin{table}[ht]
\centering
\begin{tabular}{c|ll}
\multicolumn{1}{c|}{CTP diagram} & \multicolumn{1}{c}{\hspace{-1.2cm} Real} & \multicolumn{1}{c}{\hspace{-1.2cm}  Virtual} \\
\hline 
\includegraphics[scale=0.4]{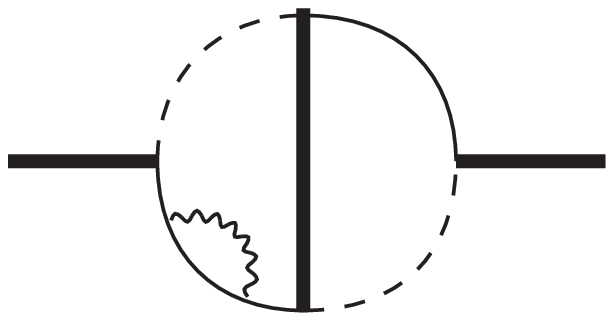} & \includegraphics[scale=0.45]{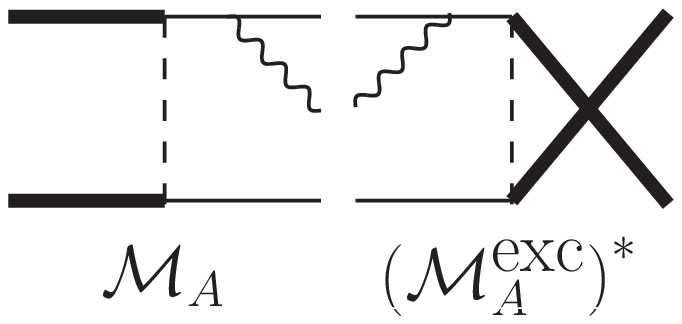} &  \includegraphics[scale=0.45]{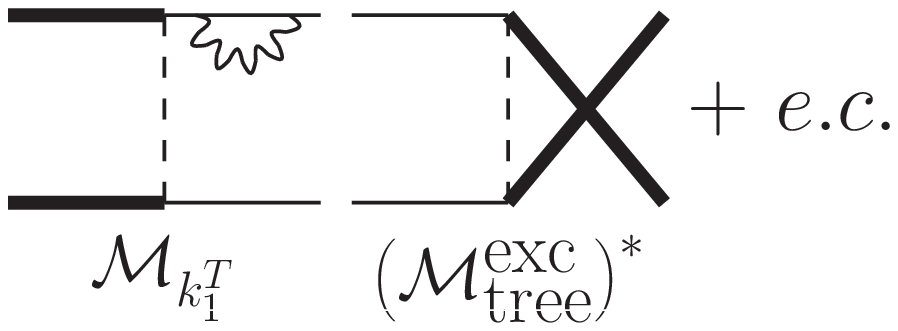} \\
 
\includegraphics[scale=0.4]{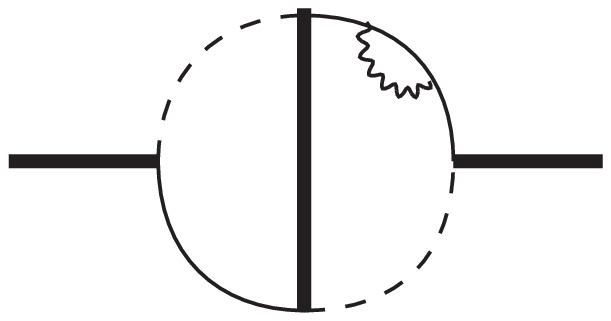} & \includegraphics[scale=0.45]{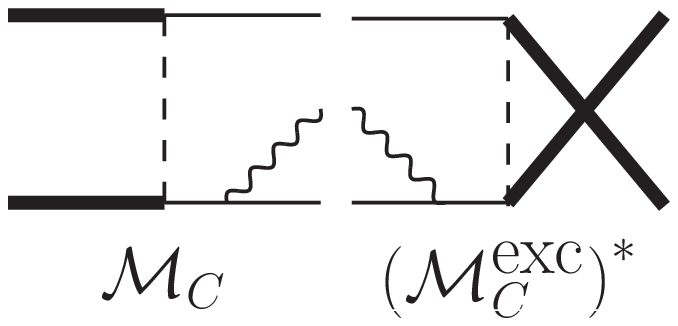} & \includegraphics[scale=0.45]{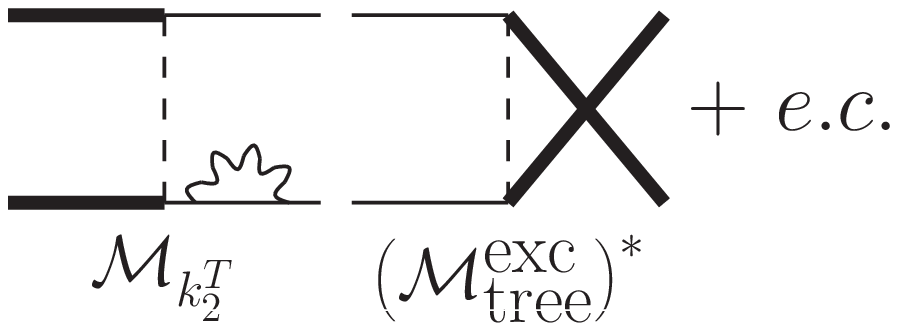} \\

\includegraphics[scale=0.4]{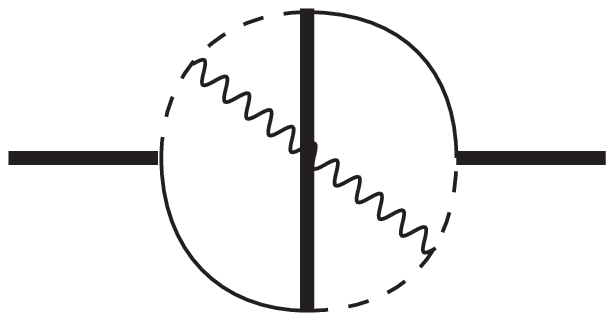} & \includegraphics[scale=0.45]{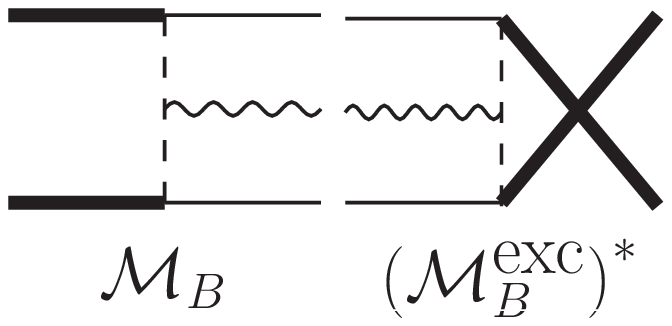} & \\

\includegraphics[scale=0.4]{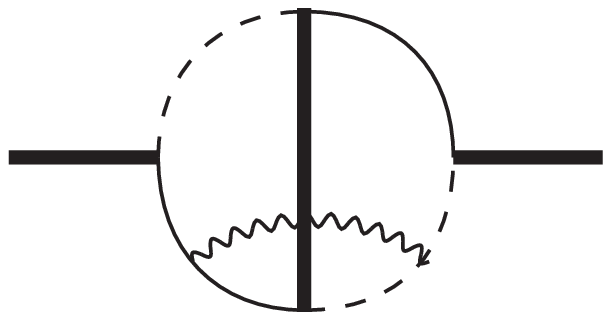} & \includegraphics[scale=0.45]{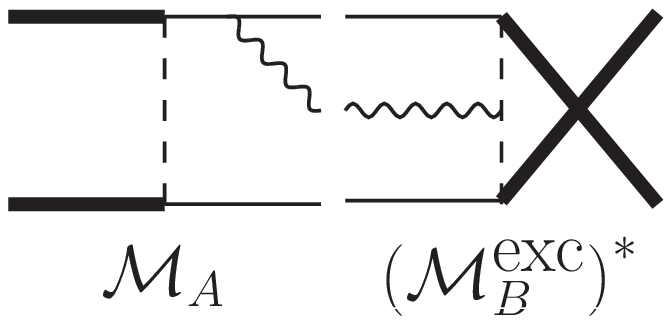} & \includegraphics[scale=0.45]{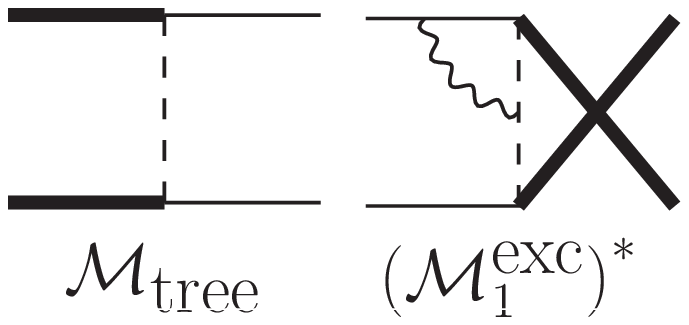} \\

\includegraphics[scale=0.4]{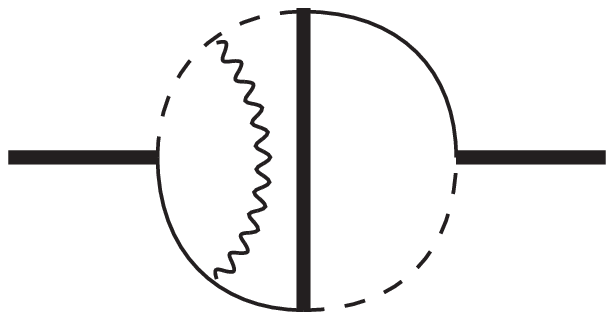} & \includegraphics[scale=0.45]{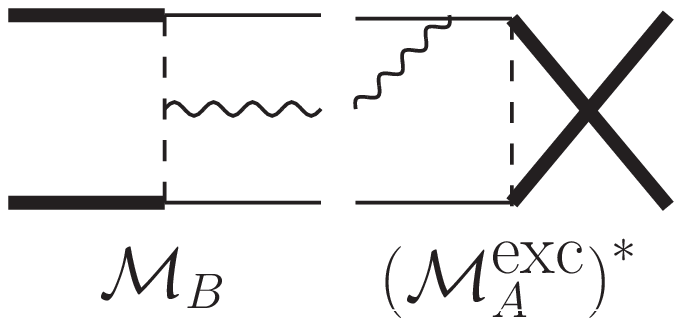} & \includegraphics[scale=0.45]{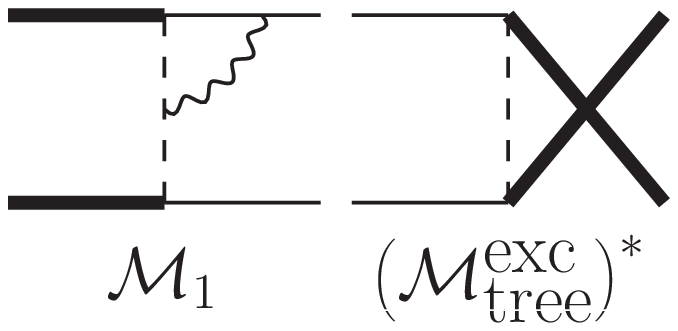}  \\

\includegraphics[scale=0.4]{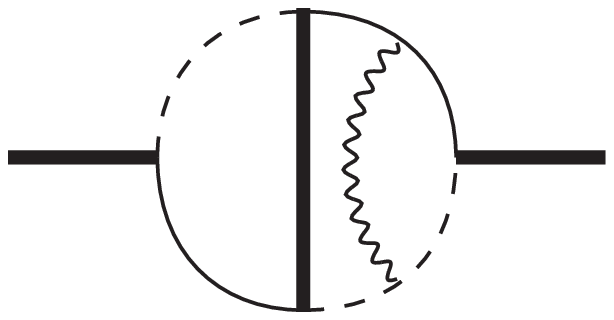} & \includegraphics[scale=0.45]{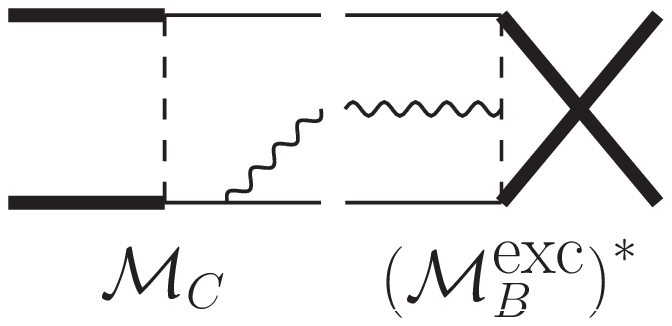} &  \includegraphics[scale=0.45]{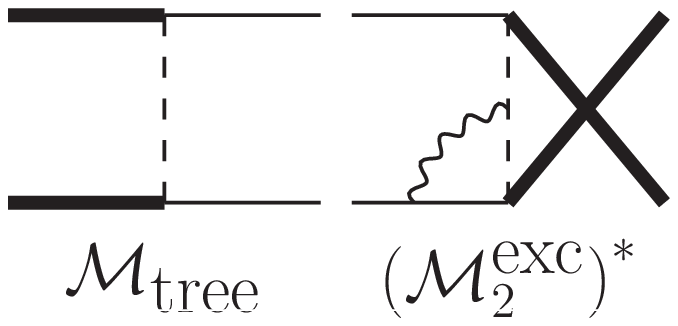} \\

\includegraphics[scale=0.4]{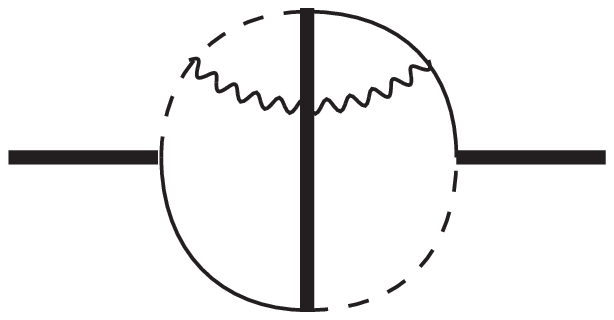} & \includegraphics[scale=0.45]{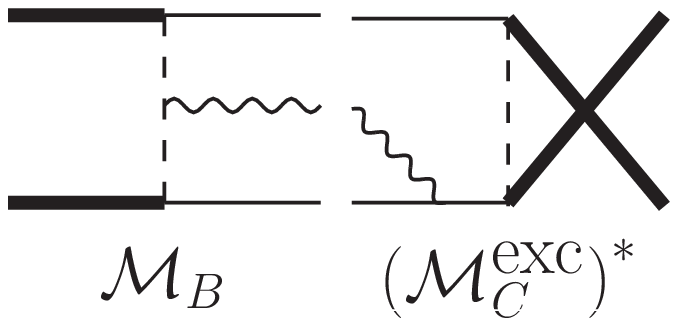} & \includegraphics[scale=0.45]{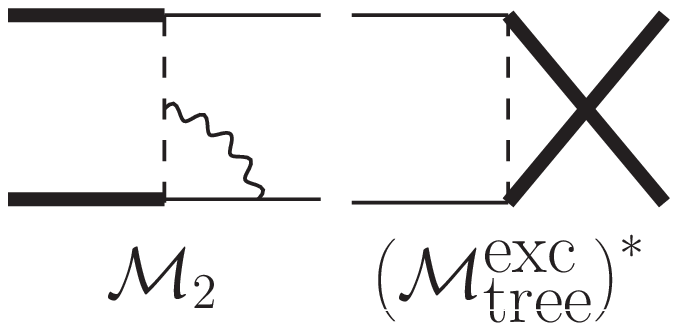}  \\

\includegraphics[scale=0.4]{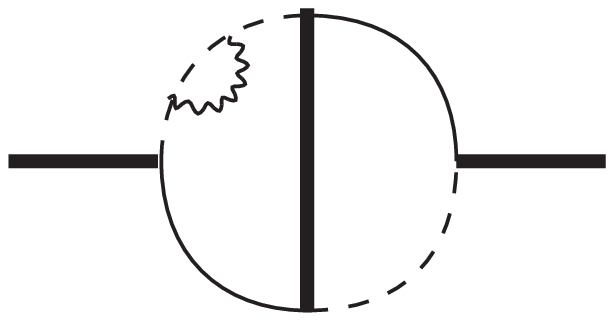} & & \includegraphics[scale=0.45]{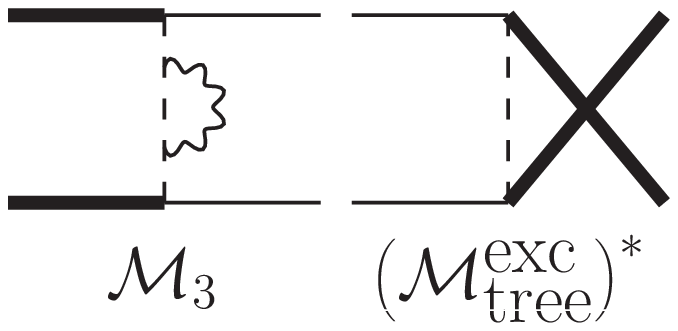}  \\

\includegraphics[scale=0.4]{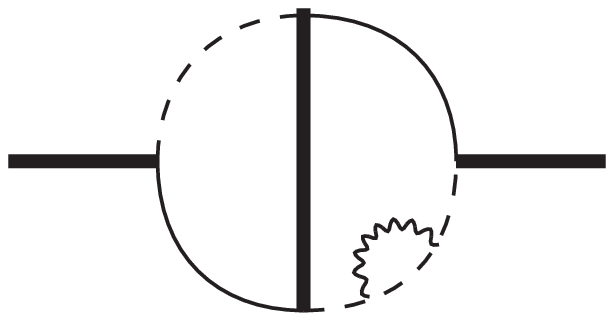} & & \includegraphics[scale=0.45]{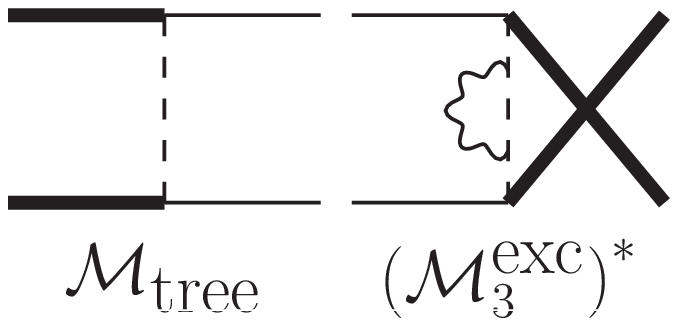}   \\

\includegraphics[scale=0.4]{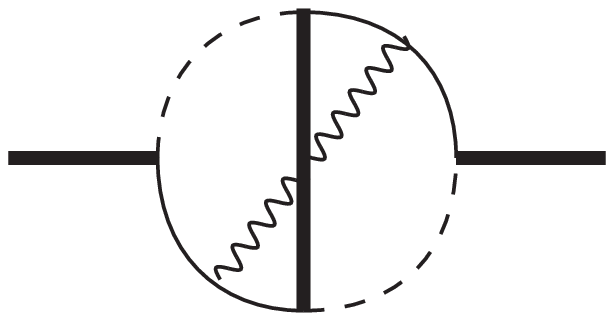} & \includegraphics[scale=0.45]{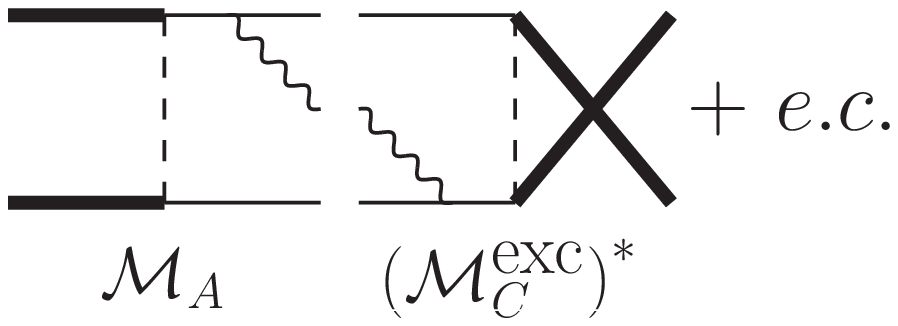}  & \includegraphics[scale=0.45]{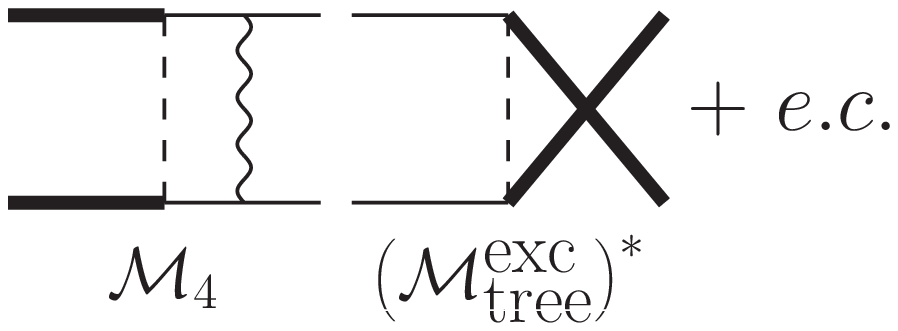} \\

\end{tabular}
\caption{The self-energy diagrams of type A and the correspondence to the diagrams leading to real emission and absorption, virtual corrections and the correction to the external SM fermion legs. The \textit{e.c.} stands for exchanging the DM fermion legs in both parts of the amplitude and complex conjugation. 
}
\label{tab:seA}
\end{table}

\setlength\tabcolsep{14pt}
\begin{table}[ht]
\centering
\begin{tabular}{c|lll}
\multicolumn{1}{c|}{CTP diagram} & \multicolumn{1}{c}{\hspace{-1.2cm} Real} & \multicolumn{1}{c}{\hspace{-1.2cm}  Virtual}  \\
 \hline 
 \includegraphics[scale=0.4]{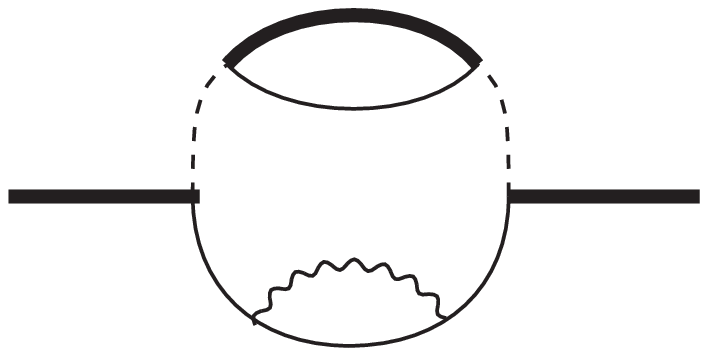} & \includegraphics[scale=0.45]{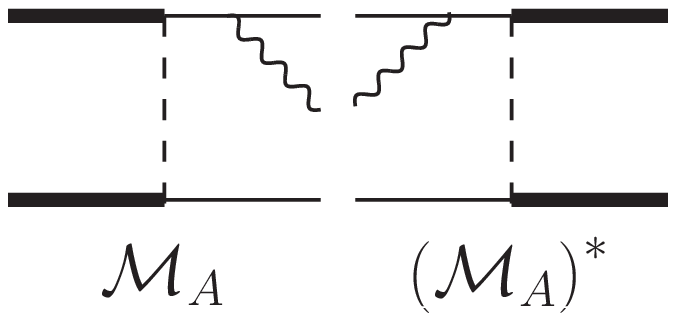} & \includegraphics[scale=0.45]{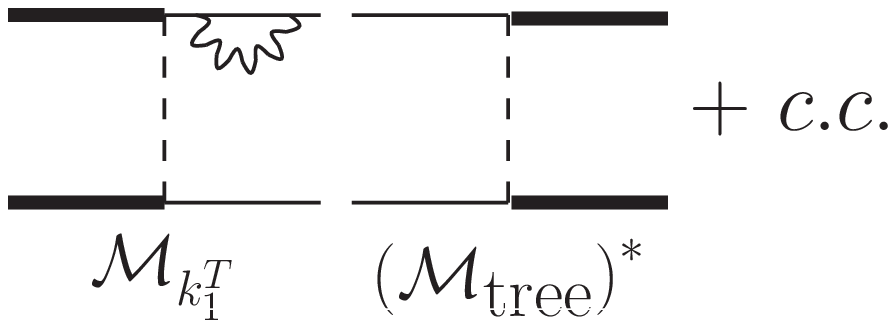} \\
  
 \includegraphics[scale=0.4]{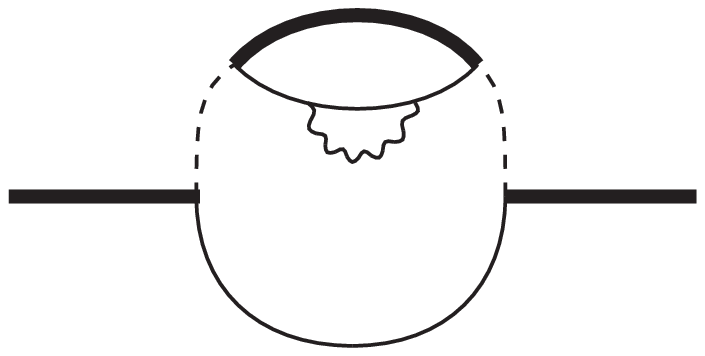} & \includegraphics[scale=0.45]{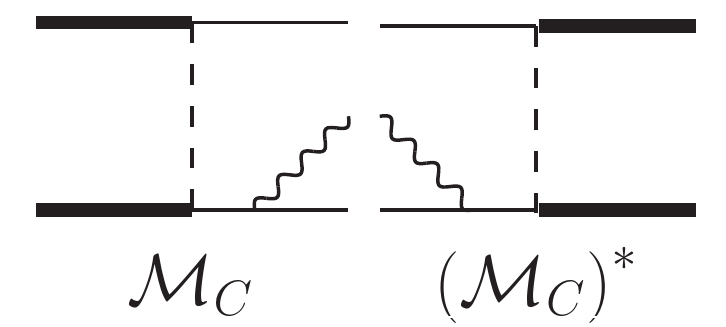} & \includegraphics[scale=0.45]{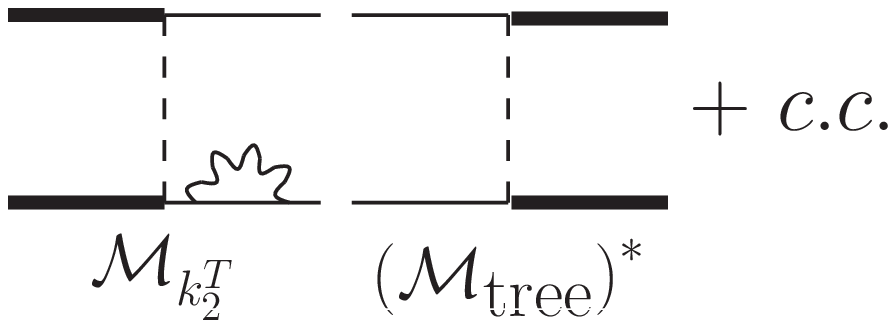} \\
 
 \includegraphics[scale=0.4]{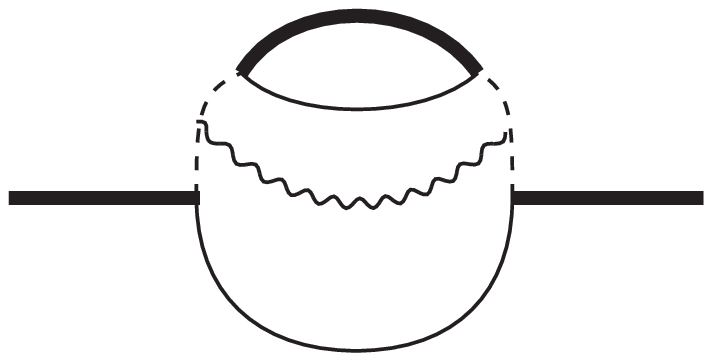} & \includegraphics[scale=0.45]{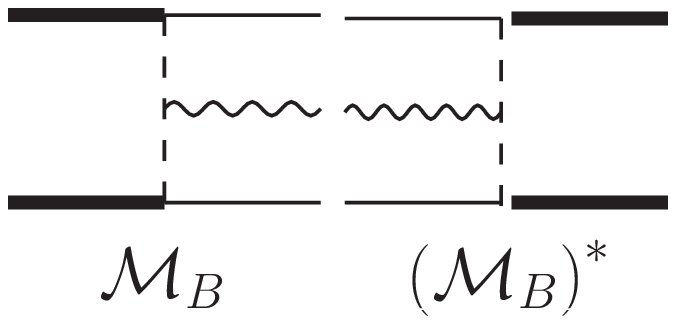} & \\
 
 \includegraphics[scale=0.4]{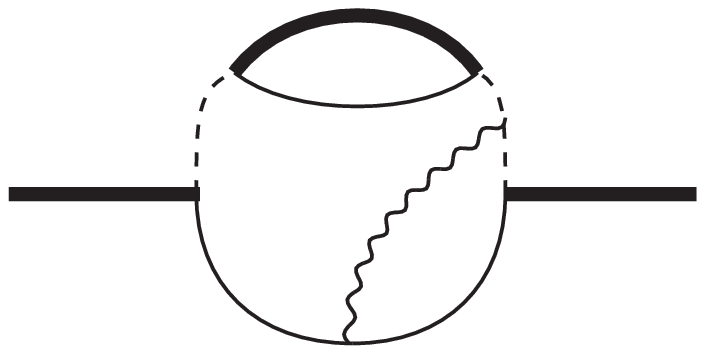} & \includegraphics[scale=0.45]{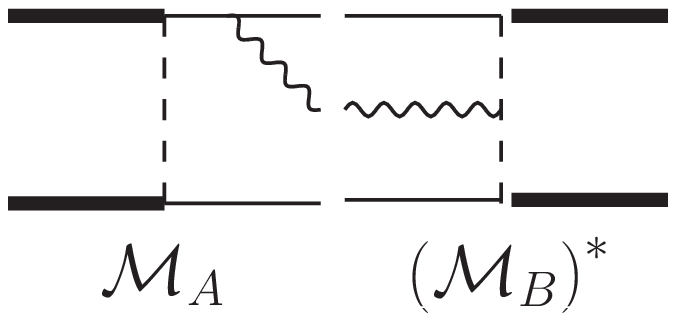} & \includegraphics[scale=0.45]{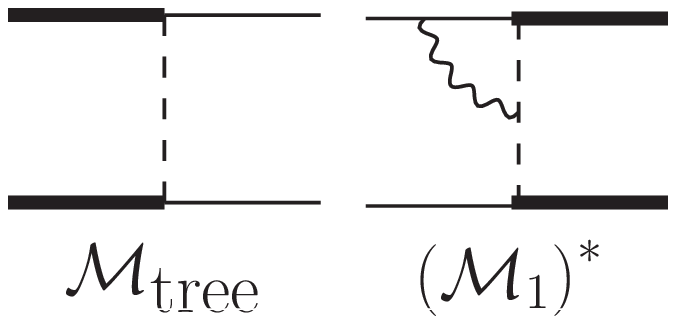} \\
 
 \includegraphics[scale=0.4]{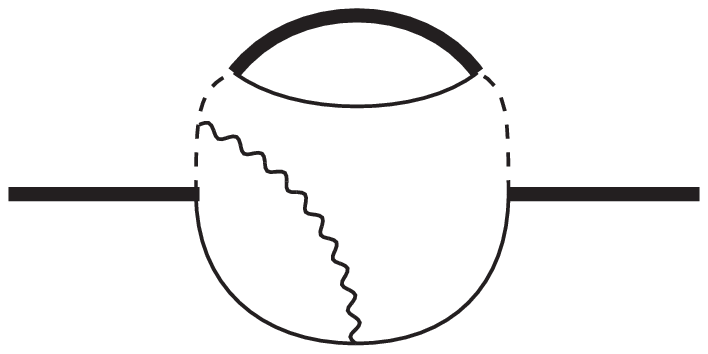} & \includegraphics[scale=0.45]{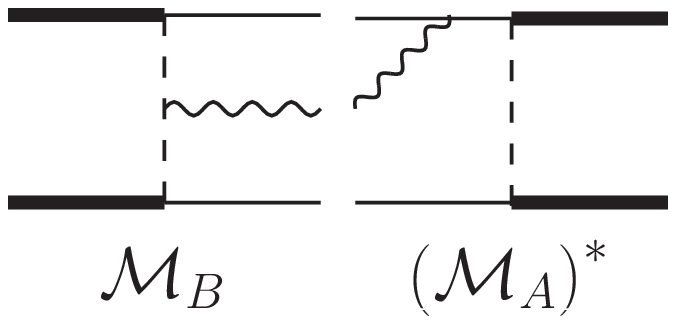} & \includegraphics[scale=0.45]{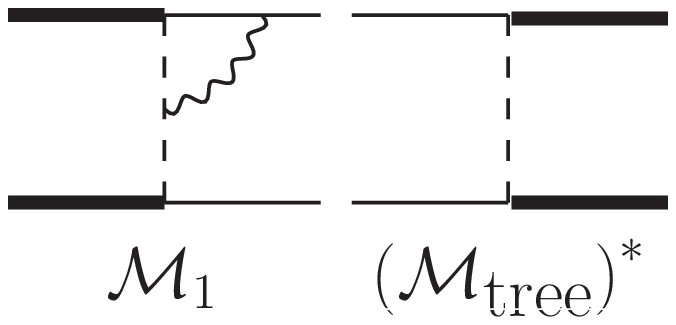}  \\
 
 \includegraphics[scale=0.4]{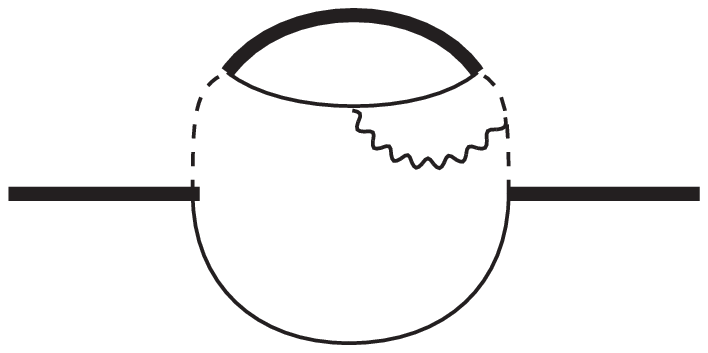} & \includegraphics[scale=0.45]{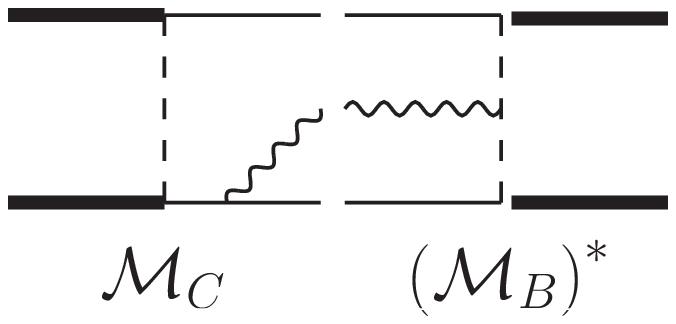} &  \includegraphics[scale=0.45]{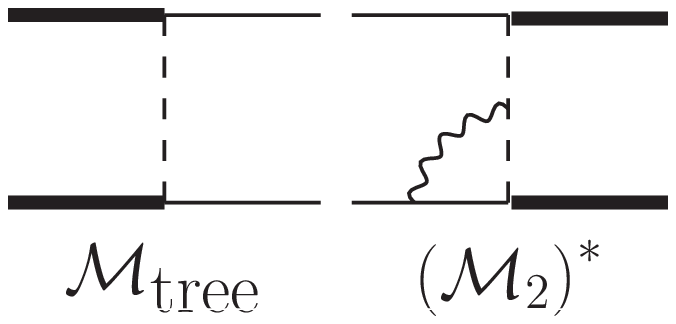} \\

 \includegraphics[scale=0.4]{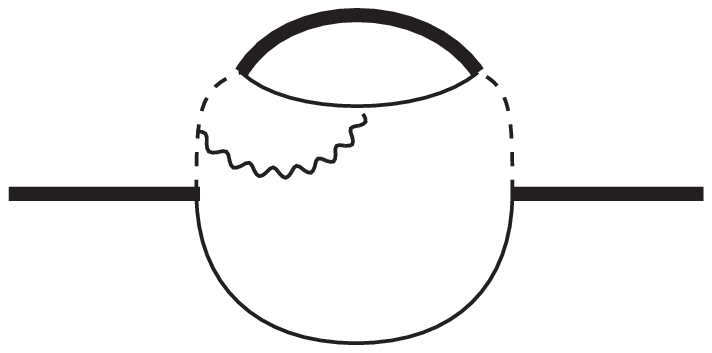} & \includegraphics[scale=0.45]{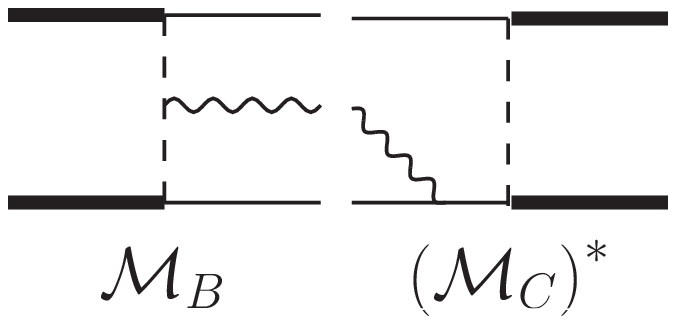} & \includegraphics[scale=0.45]{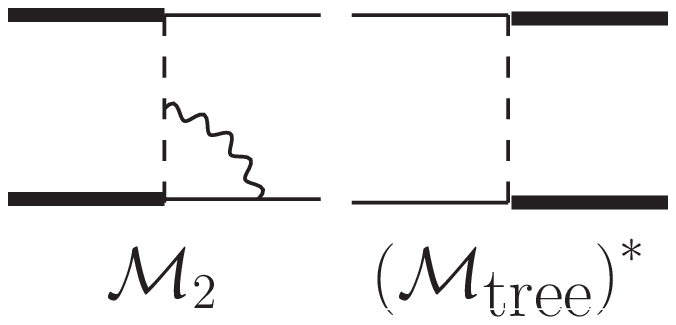}  \\
 
 \includegraphics[scale=0.4]{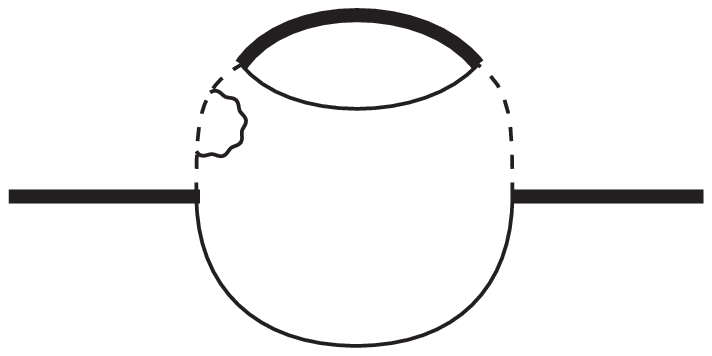} & & \includegraphics[scale=0.45]{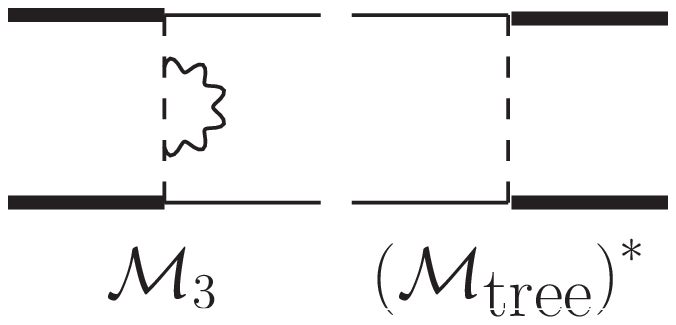}  \\
 
 \includegraphics[scale=0.4]{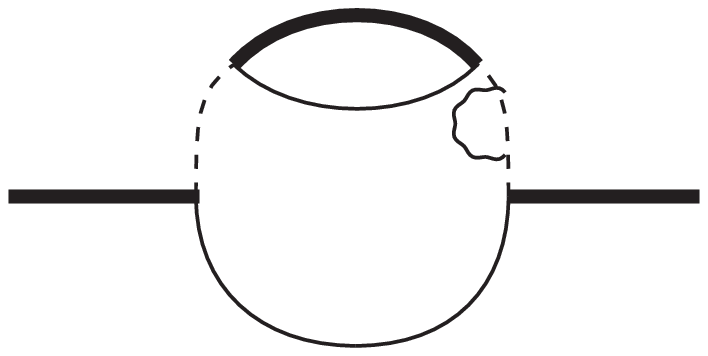} & & \includegraphics[scale=0.45]{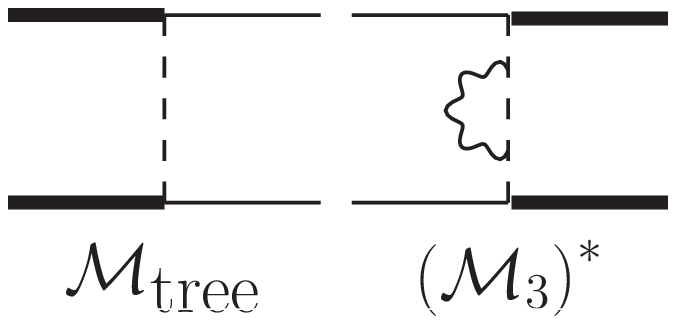}   \\
 
 \includegraphics[scale=0.4]{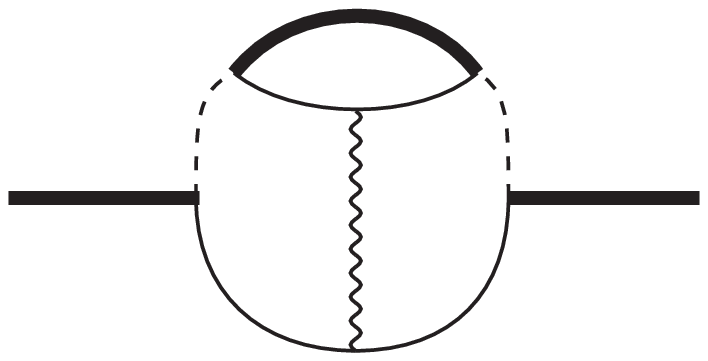} & \includegraphics[scale=0.45]{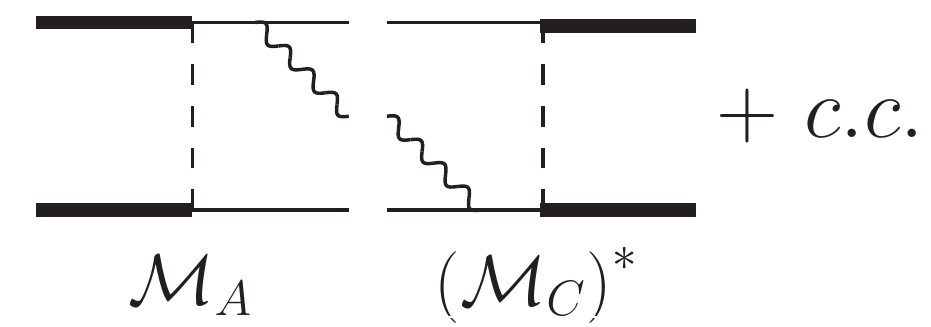}  & \includegraphics[scale=0.45]{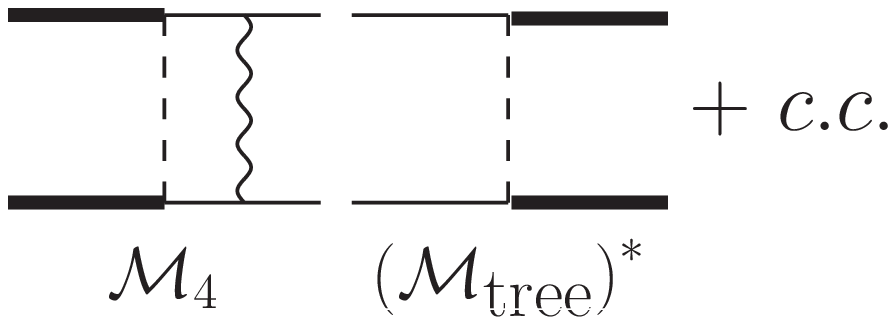} \\

\end{tabular}
\caption{The self-energy diagrams of type B and the correspondence to the diagrams leading to real emission and absorption, virtual corrections and the correction to the external SM fermion legs. The $c.c$ stands for complex conjugation.}
\label{tab:seB}
\end{table}

As an example we consider the self-energy diagram in 
fig.~\ref{fig: 3_matched}, which corresponds to the last diagram in 
table~\ref{tab:seB}. Following the rules described in 
section~\ref{sec:collision_term_calculation_LO}, we note that every 
three-loop CTP self-energy contribution to $\Sigma^{<,>}$ contains $2^4=16$ 
different terms from the CTP sum over circled and uncircled vertices. 
Most of them, however, do not contribute as they involve the 
exponentially suppressed thermal part of the sfermion. In case of 
the example shown in fig.~\ref{fig: 3_matched} only four terms remain, 
which we may associate with virtual and real photon NLO corrections. 
To confirm this interpretation, we consider the second cut of the diagram
in fig.~\ref{fig: 3_matched}, labelled $CA$, and show that it corresponds 
to the interference term of the two real photon emission amplitudes 
from the different final state legs multiplied by the associated 
Bose enhancement factors. Proceeding as in 
section~\ref{sec:collision_term_calculation_LO} we obtain for this 
contribution the expression
\bea
\Tr\{i\Sigma^>_{C A}\lp q\rp iS^< \lp q\rp \}&=&
 -\lambda^4 e^2 \int \frac{d^4t}{\lp2\pi\rp^4} \frac{d^4k_1}{\lp2\pi\rp^4} 
\frac{d^4k_2}{\lp2\pi\rp^4} \frac{d^4s}{\lp2\pi\rp^4} \lp2\pi\rp^4 
\delta^{(4)}(q+t-k_1-k_2-s)
\nonumber \\[0.2cm]
&& \hspace{-3cm} \times\,
\underbrace{i\Delta^{11}\lp k_1+s-q\rp 
i\Delta^{22}\lp k_1-q\rp}_{\equiv\mathcal{S}} \
   \underbrace{\Tr\{\pr iS^{21}\lp -t\rp \pl iS^{12}\lp -k_2\rp\gamma^\mu   
iS^{22}\lp -k_2-s\rp \}}_{\equiv\mathcal{F}_1}
\nonumber \\
&& \hspace{-3cm} \times\, \underbrace{iD^{21}_{\mu\nu}\lp s\rp}_{\equiv\mathcal{V}} \ \underbrace{\Tr\{\pl iS^{21}\lp k_1\rp 
\gamma^\nu iS^{11}\lp k_1+s\rp \pr iS^{12}\lp q\rp \}}_{\equiv\mathcal{F}_2}.
\label{eq:cnlof1f2}
\eea
In the scalar part $\mathcal{S}$ it is again sufficient to keep only 
the $T=0$ part of the propagators, while the photon propagator 
$\mathcal{V}$ contains only the thermal part.
Omitting for brevity the traces over the numerator Dirac matrices we get
\bea
\mathcal{V} &=& -g_{\mu\nu}\ 2\pi\,\delta(s^2)  
\lc \Theta(s^0) \lp 1 + f_\g(\vec{s}\,) \rp +  
\Theta(-s^0) f_\g(-\vec{s}\,) \rc,
\\[0.4cm]
\mathcal{F}_1 &\propto& 
2\pi \delta\lp t^2-m_\c^2\rp
\lc -\Theta(-t^0) \lp 1-f_\chi(-\vec{t}\;)\rp +  
\Theta(t^0) f_\chi(\vec{t}\;)\rc
\nonumber\\
&&
\times
2\pi \delta\lp k_2^2-m_f^2\rp 
\lc \Theta(-k_2^0) f_f(-\vec{k}_2) -  \Theta(k_2^0) ( 1-f_{\bar{f}}( \vec{k}_2))\rc 
\nonumber\\
&&
\times
\bigg[ \frac{-i}{(k_2+s)^2-m_f^2} - 2\pi \delta\lp (k_2+s)^2 -m_f^2 \rp 
\nonumber\\
&&
\hspace{0.5cm}\times
\lc \Theta(-k_2^0-s^0) f_f(-\vec{k}_2-\vec{s}\,) +  \Theta(k_2^0+s^0) f_{\bar{f}}( \vec{k}_2+\vec{s}\, )\rc \bigg],
\\[0.2cm]
\mathcal{F}_2 &\propto& 
2\pi \delta\lp k_1^2-m_f^2\rp \lc -\Theta(k_1^0) (1-f_f(\vec{k}_1) ) +  
\Theta(-k_1^0) f_{\bar{f}}( -\vec{k}_1)\rc
\nonumber\\
&&
 \times\,
\bigg[\frac{i}{(k_1+s)^2-m_f^2} - 2\pi \delta\lp (k_1+s)^2 -m_f^2 \rp 
\nonumber\\
&&
\hspace{0.5cm}\times
\lc \Theta(k_1^0+s^0)f_f(\vec{k}_1+\vec{s}\,) 
+ \Theta(-k_1^0-s^0) f_{\bar{f}}( -\vec{k}_1-\vec{s}\,) \rc \bigg]
\nonumber\\
&&
 \times\, 
(2\pi) \,\delta\lp q^2-m_\c^2\rp 
\lc \Theta(q^0) f_\c(\vec{q}\,) -  \Theta(-q^0) \lp 1-f_\c(-\vec{q}\,)\rp\rc .
\eea
From the above expressions we see that the contributions from the thermal 
parts of the $S^{11}(k_1+s)$ and $S^{22}(-k_2-s)$ are vanishing, since the 
combination of $\delta$-functions multiplying those terms has no support.
We are then left with 32 terms, from which again half vanishes 
after multiplying by $\Theta(q^0)$. Out of the remaining terms 8 describe
emission and 8 absorption of a photon attached to the tree-level diagram 
of type B. 
Among those terms there are 6 that correspond to processes 
which are kinematically forbidden and from the remaining ones 6 describe 
scatterings and 4 annihilation.

Only the annihilation terms eventually contribute to the Boltzmann 
equation for $\chi$ particle number, hence (\ref{eq:cnlof1f2}) simplifies to
\bea
\label{eq:trssnlo}
\Tr\{\Sigma^>_{ C A}\lp q\rp S^< \lp q\rp\}&=& 
\frac{1}{2E_{\chi_1}}\,(2\pi)\delta\lp q^0-E_{\chi_1}\rp 
\int\frac{d^3\vec{t}}{\lp2\pi\rp^3 2E_{\chi_2}} 
\nonumber\\
&& \hspace{-3.4cm} 
\times\,\int \frac{d^3\vec{k}_1}{\lp2\pi\rp^3 2 E_{f_1}} 
\frac{d^3\vec{k}_2}{\lp2\pi\rp^3 2 E_{f_2}} 
\frac{d^3\vec{s}}{\lp2\pi\rp^3 2 E_\g} \lp2\pi\rp^4 
 f_\c(\vec q\,)  f_\c(\vec t\,)
 \nonumber\\[0.2cm]
&& \hspace{-3.4cm}  \times\,
\Big[
\delta^{(4)}(q\!+\!t\!-\!k_1\!-\!k_2\!-\!s) \,
|\mathcal{M}_{CA}(k_1,k_2,s)|^2
  \bigl(1- f_f^{\text{eq}}(\vec k_1)\bigr) \bigl( 1- f_f^{\text{eq}}(\vec k_2)\bigr) \bigl(1+ f_\g^{\text{eq}}(\vec s\,)\bigr)
 \nonumber\\
&&  \hspace{-3.4cm} +\,
 \delta^{(4)}(q\!+\!t\!-\!k_1\!-\!k_2\!+\!s) \,
|\mathcal{M}_{CA}(k_1,k_2,-s)|^2
f_\g^{\text{eq}}(\vec s\,)  \bigl(1-f_f^{\text{eq}}(\vec k_1)\bigr) \bigl( 1- f_f^{\text{eq}}(\vec k_2)\bigr) 
 \nonumber\\
&&  \hspace{-3.4cm} -\,
 \delta^{(4)}(q\!+\!t\!-\!k_1\!+\!k_2\!-\!s) \,
|\mathcal{M}_{CA}(k_1,-k_2,s)|^2
 f_f^{\text{eq}}(\vec k_2) \bigl(1-f_f^{\text{eq}}(\vec k_1)\bigr)  \bigl(1+ f_\g^{\text{eq}}(\vec s\,)\bigr)
 \nonumber\\
&&  \hspace{-3.4cm} -\,
 \delta^{(4)}(q\!+\!t\!+\!k_1\!-\!k_2\!-\!s) \,
|\mathcal{M}_{CA}(-k_1,k_2,s)|^2
 f_f^{\text{eq}}(\vec k_1) \bigl( 1- f_f^{\text{eq}}(\vec k_2)\bigr) \bigl(1+ f_\g^{\text{eq}}(\vec s\,)\bigr)
\Big],\qquad\quad
\eea
where the equilibrium distribution function for the photon is
given by the Bose-Einstein statistics
\be \label{eq: Bose-Einstein}
\Theta(p^0)f^{\rm eq}_\g(\vec p\,) = \frac{1}{e^{\beta p^0}-1} 
\equiv f_B(p^0) \qquad \textrm{with } p^0\equiv|\vec p\,|.
\ee
The factors $|\mathcal{M}_{CA}|^2 $ collect the traces 
contained in the definition of $\mathcal{F}_1$ and $\mathcal{F}_2$, 
coupling constants, as well as the non-thermal propagator denominators. 
The first one (the third line of \eqref{eq:trssnlo}) can be identified 
with the interference of zero-temperature emission amplitude, namely
$|\mathcal{M}_{CA}(q,k_1,k_2)|^2 = \mathcal{M}_{C}\lp\mathcal{M}_{A}\rp^*$.
By using the crossing symmetry one can identify the remaining ones 
with the parts of the amplitudes for absorption processes:
\begin{eqnarray}
 |\mathcal{M}_{CA}(-q,k_1,k_2)|^2 & = & 
|\mathcal{M}_{CA}^{\chi \chi \gamma \rightarrow f \bar f}(q,k_1,k_2)|^2,
\nonumber \\
   -|\mathcal{M}_{CA}(q,-k_1,k_2)|^2& = &
|\mathcal{M}_{CA}^{\chi \chi \bar f \rightarrow \bar f \gamma}(q,k_1,k_2)|^2,
\nonumber \\
 -|\mathcal{M}_{CA}(q,k_1,-k_2)|^2& = &
|\mathcal{M}_{CA}^{\chi \chi f \rightarrow f \gamma}(q,k_1,k_2)|^2,
\label{eq:crossing}
\end{eqnarray}
where the minus sign comes from interchanging the fermions between 
initial and final states.

The example shows that the surviving terms from the three-loop 
CTP self-energy correspond precisely to the collision term in the 
form of (\ref{eq: collision_term_NLO_pre_1}).
In the following we discuss the computation 
of the IR divergent and leading IR finite thermal correction, separately 
for the real and virtual cuts. The results are summarized and discussed in 
the following section~\ref{sec:results}.

\subsubsection{Thermal emission and absorption}

The computation of the emission and absorption processes at non-zero 
temperature follows the same procedure as is well known from the $T=0$ case, 
simply because at NLO all the contributions from the thermal part of 
internal propagators are either exponentially suppressed or kinematically 
forbidden at this order, as explained for the example above.
The only difference comes from the fact that the external particles 
can be thermal, in which case the corresponding external leg is multiplied 
by the phase-space distribution function. 
From (\ref{eq: collision_term_NLO_pre_1}), the thermal emission and 
absorption contributions to the annihilation process are given by
\begin{eqnarray} 
\label{eq: collision_term_NLO_real}
C_{\textrm{NLO}\, T\neq 0\textrm{, real}}^\textrm{ann} = 
&-& \int d\Pi_{\c \bar \c } \, f_\c f_{\bar\c} \,
\int d\Pi_{f\bar{f}\gamma} \Big[
 f_\g \lp |\mathcal{M}_{\c \bar{\c} \rightarrow f\bar{f} \g}|^2 + 
|\mathcal{M}_{\c \bar{\c} \g \rightarrow f\bar{f}}|^2 \rp
\nonumber\\
&& \hspace{-1.75cm} 
+ \,f_{\bar{f}} \lp |\mathcal{M}_{\c \bar{\c}\bar{f} 
\rightarrow \bar{f} \g}|^2\pm|\mathcal{M}_{\c \bar{\c} 
\rightarrow f\bar{f} \g}|^2  \rp
+ f_f \lp |\mathcal{M}_{\c \bar{\c} f 
\rightarrow f \g}|^2\pm|\mathcal{M}_{\c \bar{\c} 
\rightarrow f\bar{f} \g}|^2 \rp
\Big].\qquad
\end{eqnarray}

Let us focus first on photon emission and absorption in $\chi\chi$ 
annihilation as given by the first line of 
(\ref{eq: collision_term_NLO_real}).
In the freeze-out situation the phase-space distribution of the 
photons is always close to equilibrium, and therefore 
emission and absorption of hard photons with energies $\omega$ of order 
of $m_\chi \gg\ T\sim T_f$ is exponentially suppressed by the 
distribution function $f_\gamma$. The 
scattering matrix elements can therefore be evaluated in an expansion 
in $\omega \sim T_f\ll m_\chi$, that is, in the soft-photon 
regime. In particular, the leading 
IR divergence could be obtained from the amplitudes in the eikonal 
approximation. However, since we are interested also in the leading finite 
thermal correction, which turns out to be of order 
$\mathcal{O}(\tau^2)$, we compute the full amplitude.  
After performing the integration over all phase-space variables 
except the energy $\omega$ of the emitted or absorbed particle, 
we are left with an expression of the form 
\be
\label{eq:Sdef}
\int d\Pi_{f\bar{f} \g} f_\g |\mathcal{M}_{\c \bar{\c} \rightarrow f\bar{f} \g}|^2 = 
\int_0^{\omega_\textrm{max}} d\omega f_\g(\omega) 
S_{\c\bar\c\rightarrow f\bar{f} \g}(\omega),
\ee
where the range of the integration for $\omega$ is determined by the 
phase space delta function contained in $d\Pi_{ij\g}$. For absorption 
$\omega_\textrm{max}=\infty$, while for emission there is a kinematic 
limit $\omega_\textrm{max}$. Since $\omega_\textrm{max}= {\cal O}(m_\chi)$, 
the upper limit is not relevant, however, because $f_\gamma$ 
is already exponentially suppressed, and the limit may be extended to 
infinity. At this point we can perform an expansion of $S(\omega)$ 
retaining terms up to linear order in $\omega$,\footnote{For $S_{\c\bar\c\rightarrow i j \g}(\omega)$ 
this corresponds to the 
first {\em two} terms in the expansion, since $S\sim 1/\omega$ for 
small $\omega$.}
and the final integral over $\omega$ can be expressed in terms of 
\begin{eqnarray} 
\label{eq: thermal_integral}
T^{n+1} J_n &\equiv& \int_0^\infty d\omega \, \omega^{n} \, f_B(\omega)   = 
\Biggl\lbrace
\begin{matrix}
{\rm divergent}\quad & n \leq 0 \\[0.2cm]
{\cal O}(T^{n+1}) & n>0 \, .
\end{matrix}
\end{eqnarray}
The divergence for $n=0$ follows from the Bose enhancement 
$f_B(\omega) \sim 1/\omega$ of soft photons and implies a stronger 
divergence than at zero temperature, where the soft IR divergence is 
only logarithmic. There is no such enhancement for fermions, hence 
when the same considerations are applied to the SM fermion 
emission and absorption terms in the last line of 
(\ref{eq: collision_term_NLO_real}), the relevant integrals are 
\begin{eqnarray} 
\label{eq: thermal_integralfermion}
T^{n+1} I_n &\equiv& \int_0^\infty d\omega \, \omega^{n} \, f_F(\omega)   = 
\Biggl\lbrace
\begin{matrix}
{\rm divergent}\quad & n \leq -1 \\[0.2cm]
{\cal O}(T^{n+1}) & n>-1 \, .
\end{matrix}
\end{eqnarray}
Of particular relevance will be the integrals 
\be \label{eq: J1}
J_1 = 2 I_1 = \frac{\pi^2}{6}.
\ee
In order to obtain the above analytic expression $I_1$ 
for the leading thermal 
correction in the case of thermal fermion emission and absorption, we 
have to assume that the fermions are massless. 
We briefly discuss the size of the corrections 
from finite fermion masses in section~\ref{sec:finiteTfermions}.

Returning to photon emission, the amplitude for the annihilation process 
$\chi (p_A)\chi (p_B)\rightarrow f(k_1)\bar f(k_2) \gamma (q)$ 
can be written as\footnote{For the amplitudes we follow the notation 
of \cite{Ciafaloni:2011oq}.}
\begin{equation}
\mathcal{M}_{\rm em} \equiv \mathcal{M}_{\c \c \rightarrow f \bar f \g}
=\frac{e \lambda^2}{2}\left[\left(\mathcal{M}_A-\mathcal{M}_A^{\textrm{exc}} \right)
+\left(\mathcal{M}_B-\mathcal{M}_B^{\textrm{exc}} \right)
+\left(\mathcal{M}_C-\mathcal{M}_C^{\textrm{exc}} \right) \right],
\end{equation}
where the letters $A,B,C$ refer to the amplitudes as given in 
table~\ref{tab:seA}. After the Fierz rearrangement the three terms 
are
\begin{eqnarray}
\mathcal{M}_A-\mathcal{M}_A^{\textrm{exc}}&=&
\frac{\bar u(k_1)\sl\epsilon^*(q)(\sl k_1+\sl q+m_f)P_R\g^\mu v(k_2)}
{2k_1\cdot q}
\nonumber\\
&&\times\,
\left( \frac{\bar v(p_B)P_L\g_\mu u(p_A)}{(p_B-k_2)^2-m_{\phi}^2}
-\frac{\bar v(p_B)P_R\g_\mu u(p_A)}{(p_A-k_2)^2-m_{\phi}^2} \right),
\\
\mathcal{M}_B-\mathcal{M}_B^{\textrm{exc}}&=&\bar u(k_1)P_R\g^\mu v(k_2)
\times \Bigg(\frac{\left(p_B-p_A+k_1-k_2\right)\cdot\epsilon^*(q) 
\bar v(p_B)P_L\g_\mu u(p_A) }{[(p_A-k_1)^2-m_{\phi}^2][(p_B-k_2)^2-
m_{\phi}^2]} \nonumber\\
&&-\,\frac{\left(p_A-p_B+k_1-k_2\right)\cdot \epsilon^*(q) 
\bar v(p_B)P_R\g_\mu u(p_A) }{[(p_B-k_1)^2-m_{\phi}^2][(p_A-k_2)^2-
m_{\phi}^2]} \Bigg),\\
\mathcal{M}_C-\mathcal{M}_C^{\textrm{exc}}&=&\frac{\bar u(k_1)P_R
\g^\mu(-\sl k_2-\sl q+m_f)\sl\epsilon^*(q) v(k_2)}{2k_2\cdot q}
\nonumber\\
&&\times\,\left( \frac{\bar v(p_B)P_L\g_\mu u(p_A)}{(p_A-k_1)^2-m_{\phi}^2} 
 -\frac{\bar v(p_B)P_R\g_\mu u(p_A)}{(p_B-k_1)^2-m_{\phi}^2} \right),
\end{eqnarray}

For the absorption process, due the crossing symmetry, the amplitude 
squared summed over polarization can be obtained from the emission 
process by changing the sign of the four momentum of the particle 
emitted and absorbed from the thermal bath, as in \eqref{eq:crossing}. 

Although the emission and absorption 
contributions have different phase-space integration limits, 
we have already seen that this is irrelevant up to exponentially small 
terms in $m_\chi/T$. Thus, when the emission contribution is expanded 
in the form 
\bea
S_{\c\c\rightarrow f \bar f \g}(\omega) = \sum_{n=-1}^\infty S^{(n)} 
\omega^n, 
\eea
eq.~(\ref{eq:crossing}) implies 
\bea
S_{\c\c\g\rightarrow f \bar f}(\omega) = 
\sum_{n=-1}^\infty (-1)^{n+1} S^{(n)} \omega^n ,
\eea
for the corresponding absorption process. Since 
(\ref{eq: collision_term_NLO_real}) always involves the sum of 
emission and absorption, the even terms in the expansion 
in $\omega$ cancel. Eqs.~(\ref{eq: thermal_integral}), 
(\ref{eq: thermal_integralfermion}) then imply that the leading 
finite-temperature correction is of order $\tau^2\sim T^2/m_\chi^2$.

The contributions from thermal photon emission and absoprtion, though 
divergent, can be computed without regularization in four dimensions, 
since the cancellation of the linear IR divergence proportional 
to $J_{-1}$ with the thermal virtual 
correction will be shown algebraically below before integration 
over the photon energy. The integration over the remaining phase-space 
variables that was already done to arrive at the function $S(\omega)$ 
is finite, since the non-vanishing fermion mass plays the role of the 
regulator for collinear divergences. This is no longer the 
case when the thermal fermion emission and aborption processes 
are considered, since the integral over photon energy contained 
in $S(\omega)$ has to be regularized. In this case we perform the 
phase-space integration  
in dimensional regularization with $D = 4 - 2\eta$ and $\eta < 0$.

\subsubsection{Thermal virtual corrections}

Thermal virtual corrections arise from terms in the CTP sum, to which 
the thermal parts of the diagonal 11 or 22 photon and fermion 
propagators contribute. As the sfermion is at least as heavy as the DM 
particle it has a negligible thermal contribution and we do not 
consider the corresponding amplitudes. 
We only need to include the terms when one of the virtual particles 
is thermal. When two are thermal this gives the imaginary part, 
which does not contribute to the real part of the  
interference with the tree diagram (see e.g.~\cite{Altherr:1989jc}), 
while when three are thermal 
at least one of them has to have momentum of order 
$m_\chi$ and is therefore exponentially suppressed by the phase-space 
distribution function.

We denote the relevant amplitudes by $\mathcal{M}_i$ with $i=1,...,4$, 
and the contribution from the thermal part of the photon (SM fermion) 
propagator by  $\mathcal{M}_i^\gamma$ ($\mathcal{M}_i^f$). 
The corresponding diagrams are displayed in table~\ref{tab:seA}.
The general form of every virtual contribution is
\begin{equation}
\mathcal{M}_i^{\g,f}=\int \frac{d^4q}{(2\pi)^4}\,F^{\g,f}_i(q^0,\vec{q}\,)\,
2\pi \delta(q^2-m^2_{\g,f})f_{B,F}(|q^0|),
\end{equation}
where $m_\g=0$ and for thermal photons
\begin{eqnarray}
F_1^\gamma &=&
-i e^2 \lambda^2 \frac{\bar u(k_1)(2\sl p_A-2\sl k_1+\sl q)
(\sl k_1-\sl q+m_f)P_R u(p_A)\bar v(p_B)P_L v(k_2)}{[(k_1-q)^2-m_f^2]
[(p_A-k_1+q)^2-m_{\phi}^2][(p_A-k_1)^2-m_{\phi}^2]} -(p_A \leftrightarrow p_B),
\nonumber \\[-0.3cm]
\\
F_2^\gamma &=&F_1^\g,
\\[0.3cm]
F_3^\gamma&=&\frac{i e^2 \lambda^2}{2}\frac{(2p_A-2k_1-q)^2 \,
\bar u(k_1) P_R u(p_A)\bar v(p_B) P_L v(k_2)}
{[(p_A-k_1)^2-m_{\phi}^2]^2[(p_A-k_1-q)^2-m_{\phi}^2]} -
(p_A \leftrightarrow p_B),\\[0.1cm]
F_4^\gamma&=&i e^2 \lambda^2 \frac{\bar u(k_1)
\g^\mu (\sl k_1+\sl q+m_f)P_R u(p_A) \bar v(p_B)P_L 
(-\sl k_2+\sl q+m_f)\g_\mu v(k_2)}{[(k_1+q)^2-m_f^2]
[(p_A-k_1-q)^2-m_{\phi}^2][(k_2-q)^2-m_{f}^2]} -(p_A \leftrightarrow p_B),
\nonumber \\[-0.2cm]
\end{eqnarray}
and for thermal fermions, 
\begin{eqnarray}
F_{1}^f&=&i e^2 \lambda^2 \frac{\bar u(k_1)(2\sl p_A-\sl k_1-\sl q) 
(\sl q + m_f) 
P_R u(p_A)\bar v(p_B)P_L v(k_2)}{(q-k_1)^2[(p_A-q)^2-m_{\phi}^2]
[(p_A-k_1)^2-m_{\phi}^2]} -(p_A \leftrightarrow p_B),\\
F_2^f &=& F_1^f,\\[0.3cm]
F_{4}^f&=&-i e^2 \lambda^2 \frac{\bar u(k_1)\g^\mu (\sl q \!+\! m_f) 
P_R u(p_A) 
\bar v(p_B)P_L (\sl q-\!\sl k_1 \!-\sl k_2\!+m_f)\g_\mu v(k_2)}
{[(q-k_1-k_2)^2-m_f^2][(p_A-q)^2-m_{\phi}^2](q-k_1)^2} 
\nonumber\\
&&- \,(p_A \leftrightarrow p_B).\quad\qquad
\end{eqnarray}
Note that $F_{3}^f = 0$, since there are no internal fermion lines in 
this diagram, and that $F_{4}^f$ has to be counted twice, since any one of 
the two fermion lines in the loop can be thermal.

Given these expressions we first perform the integral over $q^0$, 
which leads to
\begin{equation}
\label{eq:generalMv}
\mathcal{M}_i^{\g,f}=\int \frac{d^3\vec{q}}{(2\pi)^3 2\omega}
\left[F_i^{\g,f}(\omega,\vec{q})+F_i^{\g,f}(-\omega,\vec{q})\right]f_{B,F}(\omega), 
\end{equation}
with
\begin{equation}
\omega\equiv 
\Biggl\lbrace
\begin{matrix}
|\vec{q}\,|\quad & \,\textrm{for\ photons} \\[0.2cm]
\sqrt{\vec{q}^{\; 2} + m_{f}^2} & \quad\!\! \textrm{for\ fermions} \, .
\end{matrix}
\end{equation}
Changing the integration variable $\vec{q}\rightarrow -\vec{q}$ 
in the second integral gives $F_i^{\g,f}(q)+F_i^{\g,f}(-q)$ in the bracket. 
Then we compute 
the interference of the resulting expression with the tree-level 
amplitude and perform the integration over the two-body phase space 
together with the angles of $\vec{q}$. 
We are left with an integral over the $\omega$ similar to (\ref{eq:Sdef}), 
which can be computed in expansion in $T/m_\chi$ by expanding 
the integrand in $\omega$. The result involves the same integrals 
$J_n$, $I_n$ as was the case for the emission and absorption terms.

\subsubsection{Thermal corrections to external legs}

The remaining part of the virtual correction can be interpreted as a 
thermal correction to the mass and wave-function renormalization of the 
external SM fermion lines. Due to the universality of the renormalization 
factor, we can follow the standard procedure 
(see e.g.~\cite{Altherr:1989jc,Czarnecki:2011mr}) of computing the 
one-loop corrected thermal propagator
\begin{equation}
S_F^T(p)=\frac{i}{\sl p-m_f-\Rea \Sigma^T}.
\end{equation}
When the result for the self-energy at one loop is written as 
\begin{equation}
\Sigma^T(p)= \sl p c_B - 2 m_f (c_B+c_F) + (\sl K_B + \sl K_F) \,,
\end{equation}
with quantities $c_{B,F}$, $K^\mu_{B,F}$ to be defined shortly, the 
propagator is expressed as 
\begin{equation}
S_F^T(p)=i\ \frac{\sl p \lp 1 - c_B \rp + m_f \lc 1 - 2 (c_B+c_F) \rc - 
\lp \sl K_B + \sl K_F \rp}{p^2 \lp 1 - 2 c_B \rp - m_f^2 
\lc 1 - 4(c_B+c_F) \rc - 2 p\cdot (K_B+K_F)+\mathcal{O}(\alpha^2)}.
\label{eq:SFT}
\end{equation}
The subscript $B$ refers to the contribution when the photon propagator 
in the one-loop self-energy diagram is thermal and the SM fermion 
propagator is not. Vice-versa for the quantities with subscript $F$.  
Then 
\begin{equation}\label{eq: CB-KB}
c_B=2e^2\int \frac{d^4 q}{(2\pi)^3}
\frac{\delta(q^2)f_B(|q^0|)}{(p+q)^2-m_f^2}, 
\qquad K_B^\mu=2e^2\int \frac{d^4 q}{(2\pi)^3}\,q^\mu 
\frac{\delta(q^2)f_B(|q^0|)}{(p+q)^2-m_f^2},
\end{equation}
for the thermal photon contribution, and 
\begin{equation} \label{eq: CF-KF}
c_F=-2e^2\int \frac{d^4 q}{(2\pi)^3}\frac{\delta(q^2-m_f^2)f_F(|q^0|)}
{(p+q)^2}, \qquad K_F^\mu= 2e^2\int \frac{d^4 q}{(2\pi)^3} \,q^\mu 
\frac{\delta(q^2-m_f^2)f_F(|q^0|)}{(p+q)^2}
\end{equation}
for fermions.

The wave-function renormalization factor is derived from the expansion 
of the propagator around the particle pole. Let $\hat p^\mu = (
\hat{p}^0,\vec{p}\,)$ with $\hat{p}^0 = (m_f^2+\vec{p}^{\,2})^{1/2}$
be the on-shell limit of the external momentum $p$, and let $\hat f$ 
denote the value $f(\hat{p})$ of a function $f(p^0,\vec{p}\,)$. 
Then one finds that $c_B$ vanishes on-shell by antisymmetry of the integrand
under $q\rightarrow -q$, i.e. $\hat c_B = 0$, 
so that its expansion around the on-shell value reads 
\begin{equation}
c_B= (p^2-m_f^2)\hat c_B'+\mathcal{O}((p^2-m_f^2)^2) \qquad {\rm with} 
\qquad \hat c_B'=-\frac{\alpha}{\pi}\frac{J_{-1}}{m_f^2},
\end{equation}
and $J_{-1}$ the divergent integral defined in~(\ref{eq: thermal_integral}).
The explicit calculation of the integral defining $c_F$ 
in~(\ref{eq: CF-KF}) shows that the thermal fermion contribution 
$\hat c_F$ is only vanishing in the $m_f=0$ limit, so that in general
\begin{equation}
c_F= \hat c_F + (p^2-m_f^2)\hat c_F'+\mathcal{O}((p^2-m_f^2)^2).
\end{equation}
The coefficients $\hat c_F$ and $\hat c_F'$ can be obtained in 
general by solving the integrals numerically. 
In the massless case they simplify to
\be
\hat c_F|_{m_f=0} = 0,
\qquad
\hat c_F'|_{m_f=0} = \frac{4\alpha}{3\pi|\vec{p}\,|^2}I_{-1},
\ee
with $I_{-1}$ defined in~(\ref{eq: thermal_integralfermion}).
The vector contribution from the photon $\hat K_B^\mu$ in the 
on-shell limit reads
\begin{equation} \label{eq: KB_OS}
\hat K_B^\mu=\frac{\alpha}{\pi} J_1 \frac{T^2}{|\vec{p}\,|}
\left(L_p, \frac{\vec{p}}{|\vec{p}\,|} 
\left[\frac{\hat{p}^0}{|\vec{p}\,|}L_p-2 \right]\right) \qquad {\rm with} 
\qquad L_p=\log \frac{\hat{p}^0+|\vec{p}\,|}{\hat{p}^0-|\vec{p}\,|},
\end{equation}
and $J_1$ given by~(\ref{eq: J1}).
The fermion contribution $\hat K_F^\mu$ in the massless limit
 is divergent on-shell. We use dimensional regularization
  ($D=4-2\eta$  and the $\overline{\rm MS}$ scheme), which gives 
  \begin{equation}\label{eq: KF_OS}
\hat K_F^\mu|_{m_f=0}=
\frac{\alpha}{\pi} I_1^D \frac{T^2}{|\vec{p}\,|} \lp
\mathcal{I}(\eta),
\frac{\vec p}{|\vec p\,|} \lc \frac{\hat p^0}{|\vec p\,|}
\mathcal{I}(\eta) - 2 \rc
\rp
\hspace{0.4cm} \textrm{with} \hspace{0.4cm}  
\mathcal{I}(\eta) = 
\frac{\sqrt{\pi}\,e^{\eta \gamma_E}}
{(-\eta)\Gamma(\frac{1}{2}-\eta)},
\end{equation}
where
\begin{equation}
T^2 I_1^D =  
\mu^{2\eta} \int_0^\infty d\omega \, \omega^{1-2\eta} \, f_F(\omega)
\end{equation}
is the $D$-dimensional generalization 
of~(\ref{eq: thermal_integralfermion}).
The quantities $2 p\cdot K_B$, $2 p\cdot K_F$ appearing in the 
denominator of (\ref{eq:SFT}) can be related to $c_B$, $c_F$ 
by 
\begin{equation}
2 p\cdot K_B = \delta m_B^2 -(p^2-m_f^2) \,c_B,
\qquad 
2 p\cdot K_F = \delta m_F^2  + (p^2+ m_f^2) \,c_F\,,
\end{equation}
as follows immediately from the defining expressions 
(\ref{eq: CB-KB}), (\ref{eq: CF-KF}). Here 
\bea \label{eq:massshift}
\delta m_B^2 &=& \frac{4\alpha}{\pi} J_1 = \frac{2\pi}{3}\alpha T^2,
\nonumber \\
\delta m_F^2 &=& \frac{4\alpha}{\pi}\int_{m_f}^\infty d\omega\ 
\sqrt{\omega^2-m_f^2}\ f_F(\omega) \xrightarrow{m_f= 0}
\;\frac{4\alpha}{\pi} I_1 = \frac{\pi}{3}\alpha T^2,
\eea
contribute to the thermal corrections to the fermion mass.

Expanding the fermion propagator (\ref{eq:SFT}) around the 
corrected mass-shell, we obtain, up to non-singular terms,  
\begin{eqnarray}
S_F^T(p) & = & i \ (1-2m_f^2 (\hat c_B'+\hat c_F') + \hat c_F)\ 
\frac{\sl p+m_f (1-2\hat c_F) - \hat{\slashed K}_B-\hat{\slashed K}_F}
{p^2-m_f^2-\delta m_B^2-\delta m_F^2 + 2m_f^2 \hat c_F} 
+ {\cal O}(\alpha^2).\qquad
\end{eqnarray}
In the massless case $m_f=0$, this simplifies to
\begin{equation}
S_F^T(p) =  i \ \left(1+ \frac{2\alpha}{\pi} J_{-1}\right) 
\frac{\sl p - \hat{\slashed K}_B-\hat{\slashed K}_F}
{p^2-\delta m_B^2-\delta m_F^2} 
+ {\cal O}(\alpha^2)\,.
\end{equation}
In summary, the thermal plasma affects the external SM fermion lines, and 
therefore the annihilation cross section computation, in three ways:
\begin{enumerate}
\item Modification of the spinor orthogonality relations
\be \label{eq: spinsummod}
\sum_s u(p,s)\bar u(p,s) = \sl p+m_f (1-2\hat c_F) - 
\hat{\slashed K}_B-\hat{\slashed K}_F.
\ee
This contribution to the annihilation cross-section 
$\sigma_{\chi\chi \rightarrow f \bar f\ } v_{\rm rel}$ is simply obtained
by computing the tree-level diagrams with the modified 
relation~(\ref{eq: spinsummod}) and taking the $\mathcal{O}(\alpha)$ term.
We note from (\ref{eq: KB_OS}) and (\ref{eq: KF_OS}) that this contribution 
is finite for thermal photon, while it contains a $1/\eta$ pole for thermal 
fermion in the massless limit.
This pole cancels when adding the corresponding real correction ``cut''.

\item Temperature-dependent wave function renormalization
\be
Z_2^T = 1-2m_f^2 (\hat c_B'+\hat c_F') + \hat c_F  
\xrightarrow{m_f= 0}\;
1+ \frac{2\alpha}{\pi} J_{-1}.
\ee
The contribution is simply the $\mathcal{O}(\alpha)$ term in 
$\lc(Z_2^T)^2 - 1\rc (\sigma^\textrm{tree}_{\chi\chi \rightarrow f \bar f\ } v_{\rm rel})$.
We note that this contribution is divergent only for the thermal photon case, 
and it vanishes for the thermal fermion in the massless limit. 

\item Shift of the fermion pole mass by the thermal contributions 
\be
\Delta m_f^2 \equiv \delta m_{B}^2 + \delta m_{F}^2 - 2m_f^2 \hat c_F + \mathcal{O}(\alpha^2),
\ee
which leads to a change in the phase-space integration. This results 
in a contribution to cross section that can be written as 
\bea
\Delta\sigma_{ph} 
&=& \sigma_{\rm tree}(m_f^2 + \Delta m_f^2) - \sigma_{\rm tree}(m_f^2)
\nonumber \\
&=& \frac{d\sigma_{\rm tree}}{dm_f^2}\Delta m_f^2 + \mathcal{O}\lp(\Delta m_f^2)^2\rp,
\eea
where we used the short notation 
$\sigma_{\rm tree} \equiv (\sigma^{\rm tree}_{\chi\chi \rightarrow f \bar f\ } 
v_{\rm rel})$. 
This contribution is finite for both thermal photon and fermion. 
In the massless limit this is ensured by the fact that $\sigma_{\rm tree}$
is analytic in $m_f^2$.
\end{enumerate}


\section{Results}
\label{sec:results}

In this section we summarize the results of the calculation of the 
thermal correction. We first note that we are interested in the 
situation $T\ll m_\chi$ ($\tau\ll 1$). The thermal correction arises from soft 
thermal propagators, $\omega\sim T$ up to exponentially small terms, 
since larger energies are suppressed by the thermal 
distribution functions. After expansion 
in $\omega/m_\chi$, all the thermal real and virtual corrections 
can be expressed in terms of the integrals $J_n$ for photons, 
and $I_n$ for massless fermions, as defined in (\ref{eq: thermal_integral}), 
(\ref{eq: thermal_integralfermion}). Since we are interested in the infrared 
divergence cancellation and the leading thermal correction, we only 
keep terms to order ${\cal O}(\tau^2)$. 

The scattering processes depend further on the masses $m_\c$, $m_f$, 
$m_\phi$, and the DM energy $E_\c$, or the corresponding 
dimensionless variables $\epsilon$, $\xi$ and $e_\c = E_\c/m_\c$. 
Freeze-out occurs when the DM particles are non-relativistic, 
so that we can expand in $e_\c\ll 1$. We performed the calculation for 
the first two terms of this expansion, which correspond to the 
s- and p-wave terms, respectively, keeping the full dependence on 
the scalar and SM fermion mass parameters $\xi$ and $\epsilon$. 
All computations were done in Feynman gauge. 
In addition, we also computed the result without an expansion 
in the non-relativistic DM kinetic energy, keeping the full  
dependence on $e_\chi$. However, in this case we performed an expansion 
for large scalar mediator mass $\xi\gg 1$, up to the order 
$\mathcal{O}(\xi^{-10})$. The large-mass expansion may be physically 
motivated, since often (but not necessarily) the scalar particle 
in the DM model is significantly heavier than the DM particle itself. 
Going to high order in $1/\xi$ is motivated by the observation 
that one needs to retain terms up to $\mathcal{O}(\xi^{-8})$ to 
see the lifting of helicity suppression of the non-thermal NLO 
contribution.


\subsection{IR divergence cancellation}
\label{sec:IRdivresults}

As we have seen above, the basic quantities from which the inclusive 
annihilation cross section is derived are the CTP self-energies.
It is well known that at zero temperature, off-shell self-energy diagrams 
are IR finite. The cancellation of divergences between 
virtual and real corrections to an inclusive process occurs after 
summing all possible cuts of the self-energy diagram. 
Our first important result is that we find that this is also true at 
finite temperature. That is, in the sum of all additional contributions from 
the thermal part of the propagators, the IR divergences cancel. Moreover, 
as at zero temperature, this happens at the level of individual CTP 
self-energy diagrams. This ensures that the 
collision term in the Boltzmann equation is IR finite, since it is 
directly built out of the self-energies $\Sigma^{<,>}$.

In order to show how the cancellation takes place, we discuss in 
detail the correction from thermal photons to the collision term for 
s-wave annihilation. We also verified the cancellation for the 
p-wave term, for the contribution from thermal fermions, and 
without partial wave expansion in the large-scalar mass expansion, 
as discussed above.

At one-loop the amplitude can have singular terms of the 
order $\mathcal{O}(\omega^{-1})$ at small $\omega$, which at $T=0$ 
lead to the usual logarithmic soft divergence. At finite temperature, 
the enhancement of the Bose distribution function $f_B(\omega)$ 
for small energies results in linear and logarithmic divergences, 
encoded in the singular integrals $J_{-1}$ and $J_0$, respectively. 
As already pointed out, the latter vanishes when both the emission 
and absorption of thermal photons are included, due to the different 
sign of these contributions for even orders in $\omega$. The results 
for the remaining part proportional to $J_{-1}$ are given separately 
for all self-energy diagrams in table~\ref{tab:div},
where the prefactor $\alpha/(\pi\epsilon^2)\times a_{\rm tree}$
involving the tree-level s-wave annihilation cross section 
(\ref{eq:abtree}) has been factored out.\footnote{The 
divergence can be factorized from the tree cross section, because it comes 
from the soft region. The same structure of 
the divergence was found for the hard photon scattering in the thermal plasma 
\cite{Indumathi:1996ec}.} We immediately note the aforementioned 
fact that the sum of all contributions cancels for every self-energy 
diagram separately. The logarithm present in the last row is defined in 
table~\ref{tab:def} and contains a collinear divergence 
$L \stackrel{\epsilon\rightarrow 0}{\to} \log\epsilon$ in the limit 
of small SM fermion mass ($\epsilon=m_f/(2 m_\chi)$). This collinear 
divergence also cancels in the sum.

The tree annihilation cross section is helicity-suppressed, 
$a_{\rm tree} \propto \epsilon^2$. The appearance of terms 
in table~\ref{tab:div} and tables \ref{tab:finA} and \ref{tab:finB} below, 
which are not ${\cal O}(\epsilon^2)$ 
for small $\epsilon$, implies that individual terms are not 
helicity-suppressed.

\newcolumntype{V}{>{\centering\arraybackslash} m{.12\linewidth} }
\setlength{\tabcolsep}{3.5pt}
\begin{table}[t]
\centering
\begin{tabular}{V|cc|V|cc}
\hline
\multicolumn{6}{ c}{The divergent part $J_{-1}$} \\
\hline
Type A & Real & Virtual   & Type B & Real & Virtual \\
\hline 
\includegraphics[scale=0.25]{./\figurefolder/3} 
& $1-2 \epsilon ^2$ 
& \hspace{-0.5cm}$-1+2 \epsilon ^2$
& \includegraphics[scale=0.25]{./\figurefolder/3p} 
&$-1 $
& \hspace{-0.3cm}$1$
\\ 

\includegraphics[scale=0.25]{./\figurefolder/4} 
& $1-2 \epsilon ^2$ 
& \hspace{-0.5cm}$-1+2 \epsilon ^2$
& \includegraphics[scale=0.25]{./\figurefolder/4p} 
&$-1 $
& \hspace{-0.3cm}$1$
\\

\includegraphics[scale=0.25]{./\figurefolder/9} 
&0 
& 
& \includegraphics[scale=0.25]{./\figurefolder/9p} 
&0
&
\\

\includegraphics[scale=0.25]{./\figurefolder/8} 
&  0 
& 0 
&  \includegraphics[scale=0.25]{./\figurefolder/8p} 
&0
&0
\\

\includegraphics[scale=0.25]{./\figurefolder/6} 
&  0 
& 0 
&  \includegraphics[scale=0.25]{./\figurefolder/6p} 
&0
&0
\\

\includegraphics[scale=0.25]{./\figurefolder/7} 
&  0 
& 0 
&  \includegraphics[scale=0.25]{./\figurefolder/7p} 
&0
&0
\\

\includegraphics[scale=0.25]{./\figurefolder/5} 
&  0 
& 0 
&  \includegraphics[scale=0.25]{./\figurefolder/5p} 
&0
&0
\\

\includegraphics[scale=0.25]{./\figurefolder/1} 
&   
& 0   
&  \includegraphics[scale=0.25]{./\figurefolder/1p} 
&
&0
\\

\includegraphics[scale=0.25]{./\figurefolder/2} 
&   
& 0  
&  \includegraphics[scale=0.25]{./\figurefolder/2p} 
&
&0
\\

\includegraphics[scale=0.25]{./\figurefolder/10} 
&  $\frac{2   \left(1-2 \epsilon ^2\right)^2}{   \sqrt{1-4 \epsilon ^2}}L$ 
& $-\frac{2   \left(1-2 \epsilon ^2\right)^2 }{ \sqrt{1-4 \epsilon ^2}}L$ 
&  \includegraphics[scale=0.25]{./\figurefolder/10p} 
&  $-\frac{2   \left(1-2 \epsilon ^2\right) }{  \sqrt{1-4 \epsilon ^2}}L$ 
& $\frac{2   \left(1-2 \epsilon ^2\right) }{ \sqrt{1-4 \epsilon ^2}}L$ 
\\

\end{tabular}
\caption{
Coefficients of the divergent integral $J_{-1}$ omitting the 
overall factor $\alpha/(\pi\epsilon^2)\times a_{\rm tree}$. 
Here ``Real'' includes both, emission and absorption, 
while ``Virtual" comprises vertex and external leg corrections. 
An empty space means that the corresponding contribution does not 
exist, while 0 implies that the diagram exists, but is finite. 
$L$ denotes the logarithm as defined in table \ref{tab:def}.}
\label{tab:div}
\end{table}


\subsection{Finite-temperature correction from thermal photons}
\label{sec:finiteT}

Once the divergent $J_{-1}$ and $J_0$ contributions are cancelled, the 
remaining finite correction is necessarily of $\mathcal{O}(\tau^2)$, 
proportional to the integral $J_1$. 
Again, we first show the diagram-by-diagram results for 
the s-wave contributions, which can be found in tables 
\ref{tab:finA} and \ref{tab:finB} for the diagrams of type A and B, 
respectively. A common factor 
$\pi \alpha/(6\epsilon^2)\times a_{\rm tree}$ is left out.

\newcolumntype{V}{>{\centering\arraybackslash} m{.09\linewidth} }
\setlength{\tabcolsep}{15pt}
\begin{table}[t]
\centering
\begin{tabular}{V|cc}
\hline
\multicolumn{3}{ c}{The finite part $J_{1}$} \\
\hline
Type A & Real & Virtual \\
\hline 
\includegraphics[scale=0.2]{./\figurefolder/3} 
& $\frac{2(1-\xi^2)}{D^2 D_\xi^2}+
\frac{(1-2 \epsilon ^2) p_1(\epsilon,\xi)}{2D^2 D_\xi^2}+\frac{1}{2\sqrt{D}}L$
&  \hspace{-1.2cm} $\frac{(1-2 \epsilon ^2 )
(\xi ^2-3D)}{2D D_\xi} \!-\! \frac{1}{2\sqrt{D}} L$
\\ 

\includegraphics[scale=0.2]{./\figurefolder/4} 
& \dittoclosing
&\dittoclosing
\\

\includegraphics[scale=0.2]{./\figurefolder/9} 
& $ -\frac{ 4 (1-2 \epsilon ^2)D}{D_\xi^2} $
&  
\\

\includegraphics[scale=0.2]{./\figurefolder/8} 
& $ -\frac{2  (1-2 \epsilon ^2)\xi ^2 }{D_\xi^2} -\frac{f_1(\epsilon,\xi)}{\sqrt{D}D_\xi^2}L$
& $ \frac{2  (1-2 \epsilon ^2)(D-\xi ^2) }{D_\xi^2} +\frac{ f_1(\epsilon,\xi)}{\sqrt{D}D_\xi^2}L$
\\

\includegraphics[scale=0.2]{./\figurefolder/6} 
& \dittoclosing 
& \dittoclosing 
\\

\includegraphics[scale=0.2]{./\figurefolder/7} 
& \dittoclosing 
& \dittoclosing 
\\

\includegraphics[scale=0.2]{./\figurefolder/5} 
&  \dittoclosing  
& \dittoclosing 
\\

\includegraphics[scale=0.2]{./\figurefolder/1} 
&   
&  $ -\frac{ 4 (1-2 \epsilon ^2)D}{D_\xi^2} $ 
\\

\includegraphics[scale=0.2]{./\figurefolder/2} 
&   
&  \dittoclosing
\\

\includegraphics[scale=0.2]{./\figurefolder/10} 
& $\frac{2  (1-2 \epsilon ^2)p_2(\epsilon,\xi) +(1-\xi^2)^2}{D^2D_\xi^2}\! +\! \frac{ 4f_2(\epsilon,\xi)}{\sqrt{D}D_\xi^2}L$
& $\frac{16\epsilon^2 (2-3\epsilon^2) -(3-\xi^2)^2}{D_\xi^2} \! -\! \frac{ 4f_2(\epsilon,\xi)}{\sqrt{D}D_\xi^2}L$
\\
\hline
\multicolumn{1}{r}{Total:} & \multicolumn{2}{l}{$-\frac{8(1-2\epsilon^2)}{D_\xi}$}\\
\hline
\end{tabular}
\caption{Coefficients of the finite $\mathcal{O}(\tau^2)$ correction 
 for the type A diagrams. An overall factor 
$\pi \alpha/(6\epsilon^2)\times a_{\rm tree}$ is left out. 
$D$, $D_\xi$ and polynomials $p_i$ and $f_i$ are defined in 
table~\ref{tab:def}.}
\label{tab:finA}
\end{table}

\newcolumntype{V}{>{\centering\arraybackslash} m{.09\linewidth} }
\setlength{\tabcolsep}{15pt}
\begin{table}[t]
\centering
\begin{tabular}{V|cc}
\hline
\multicolumn{3}{ c}{The finite part $J_{1}$} \\
\hline
Type B & Real & Virtual  \\
\hline 
\includegraphics[scale=0.2]{./\figurefolder/3p} 
&  $\epsilon^2\left(1\!-\!\frac{12}{D_\xi^2}\!+\!\frac{4}{D D_\xi}\!+\!\frac{2\epsilon^2}{D^2}\right)\!-\!\frac{1}{2\sqrt{D}}L$
&  $\frac{2}{D_\xi}\!-\!\frac{1}{2D}\!+\!\frac{1}{2\sqrt{D}}L$
\\ 
   
\includegraphics[scale=0.2]{./\figurefolder/4p} 
& \dittoclosing
&\dittoclosing
\\

\includegraphics[scale=0.2]{./\figurefolder/9p} 
& $ \frac{ 4 D}{D_\xi^2} $
& 
\\

\includegraphics[scale=0.2]{./\figurefolder/8p} 
& $ \frac{2\xi ^2 }{D_\xi^2} +\frac{f_3(\epsilon,\xi)}{\sqrt{D}D_\xi^2}L$
& $ \frac{2(\xi ^2 -D)}{D_\xi^2} -\frac{f_3(\epsilon,\xi)}{\sqrt{D}D_\xi^2}L$
\\

\includegraphics[scale=0.2]{./\figurefolder/6p} 
& \dittoclosing 
& \dittoclosing 
\\

\includegraphics[scale=0.2]{./\figurefolder/7p} 
& \dittoclosing 
& \dittoclosing 
\\

\includegraphics[scale=0.2]{./\figurefolder/5p} 
&  \dittoclosing  
& \dittoclosing 
\\

\includegraphics[scale=0.2]{./\figurefolder/1p} 
&   
&   $ \frac{ 4 D}{D_\xi^2} $

\\

\includegraphics[scale=0.2]{./\figurefolder/2p} 
&   
&   \dittoclosing

\\

\includegraphics[scale=0.2]{./\figurefolder/10p} 
& $\frac{2  (1-2 \epsilon ^2)p_3(\epsilon,\xi) -4(1-\xi^2)^2}{D^2D_\xi^2} \!-\! \frac{ 2f_4(\epsilon,\xi)}{\sqrt{D}D_\xi^2}L$
& $\frac{(3-\xi^2)^2-8\epsilon^2(1-2\epsilon^2+\xi^2) }{D_\xi^2} \!+\! \frac{ 2f_4(\epsilon,\xi)}{\sqrt{D}D_\xi^2}L$
\\
\hline
\multicolumn{1}{r}{Total:} & \multicolumn{2}{l}{$\frac{8}{D_\xi}$}\\
\hline
\end{tabular}
\caption{Coefficients of the finite $\mathcal{O}(\tau^2)$ correction 
for the type B diagrams. An overall factor 
$\pi \alpha/(6\epsilon^2)\times a_{\rm tree}$ is left out. 
$D$, $D_\xi$ and polynomials $p_i$ and $f_i$ are defined in 
table~\ref{tab:def}.}
\label{tab:finB}
\end{table}

\setlength\extrarowheight{3pt}
\begin{table}[ht]
\centering
\small{
\begin{tabular}{l}

 $D=1-4\epsilon^2$ \\
 $D_\xi=1-4\epsilon^2+\xi^2$ \\
  $L= \log \!\frac{1-2 \epsilon ^2-\sqrt{1-4 \epsilon ^2}}{2 \epsilon ^2}$ \\ \vspace*{-0.3cm}
  \\
\vspace*{0.05cm}
$f_1(\epsilon,\xi) = (1-\epsilon^2)(D-\xi^2)+2\epsilon^2 \xi^2$ \qquad\qquad\qquad  \qquad $f_2(\epsilon,\xi) = (1-\epsilon^2)(D-\xi^2)+2\epsilon^2 $\\  $f_3(\epsilon,\xi) = D(1+2\epsilon^2)-(1-2\epsilon^2)\xi^2 $ \qquad\qquad\qquad  \qquad $f_4(\epsilon,\xi) = (2+D_\xi^2 \epsilon^2-2\xi^2) $ \\
\vspace*{-0.3cm}
\\
 $p_1(\epsilon,\xi)= -3+\xi ^4 \left(1-4 \epsilon ^2-4 \epsilon ^4\right)+\xi ^2 \left(6-24 \epsilon ^2+120 \epsilon ^4+32
   \epsilon ^6\right) -12 \epsilon ^2-20 \epsilon ^4-32 \epsilon ^6-64 \epsilon ^8 $ \\
$p_2(\epsilon,\xi)= 3+\xi ^4 \left(-1+2 \epsilon ^4\right)+\xi ^2 \left(2-4 \epsilon ^2+20
   \epsilon ^4-16 \epsilon ^6\right) -36 \epsilon ^2+114 \epsilon ^4-144 \epsilon ^6+32 \epsilon ^8$ \\
$p_3(\epsilon,\xi)= -2 +\xi ^4 \left(2+5 \epsilon ^2+8 \epsilon ^4\right)+\xi ^2 \left(-6+2 \epsilon ^2-24 \epsilon ^4-64
   \epsilon ^6\right) +37\epsilon ^2-64 \epsilon ^4+16 \epsilon ^6+ 128 \epsilon ^8 $  \\

\end{tabular}
}
\caption{The definitions used in the results tables \ref{tab:div}, \ref{tab:finA} and \ref{tab:finB}.}
\label{tab:def}
\end{table}

We see that the separate contributions are significantly 
more complex than was the case for the divergent parts. 
The first simplification 
occurs when summing over the different cuts of 
a given self-energy diagram. At this stage all the logarithms $L$ 
cancel, which is a sign of cancellation of the collinear divergence 
on a diagram-by-diagram basis. An even more remarkable simplification 
occurs upon adding separately all diagrams of type A and B, respectively 
(given as ``Total'' at the bottom of the tables). 
Finally, helicity-suppression is recovered after summing over A and B. 
The thermal correction to the s-wave annihilation cross section 
(times velocity, see (\ref{eq:sigmaab}))
can be written as
\begin{equation}
a=a_{\rm tree} \lp 1+\Delta_a \rp + \mathcal{O}(\tau^4) \quad {\rm with } 
\quad \Delta_a = \frac{8\pi}{3}\alpha \tau^2\frac{1}{1+\xi^2-4\epsilon^2}.
\label{eq:Deltaa}
\end{equation}
It is worth noting that the leading thermal correction is suppressed not 
only by $\alpha \tau^2$ but also one power of $\xi^2$, if the 
mediator mass is large. This is true not only for the s-wave contributions, 
but also for all partial waves.

Beyond the s-wave case displayed explicitly in the tables, we computed 
the thermal correction to the p-wave cross section; further without the 
partial wave expansion in the limit $\xi\gg 1$, up to the order  
$\mathcal{O}(\tau^2,\xi^{-10})$, retaining full dependence on 
$e_\c$ and $\epsilon$. We find that all cases are covered by the 
remarkably simple expression 
\begin{equation}
\label{eq:resum}
\sigma v = [\sigma v]_{\rm{tree}} 
-\frac{4}{3}\pi\alpha\tau^2 \frac{\partial }{\partial \xi^2}
[\sigma v]_{\rm{tree}}  +\mathcal{O}(\tau^4),
\end{equation}
which implies that the total $\mathcal{O}(\tau^2)$ correction from 
thermal photons can be obtained directly from the tree cross section without 
any explicit calculation. To appreciate the simplicity of this expression, 
note the complicated dependence on $\xi$ and $\epsilon$ of the tree-level 
p-wave cross section given in (\ref{eq:abtree}). The fact that this 
formula holds in all limits that we investigated leads us to 
conjecture that it is generally valid beyond the 
non-relativistic approximation (partial-wave expansion).

The structure of (\ref{eq:resum}) is certainly not accidental, and 
it is not the only example of ``universality''  of a finite-temperature 
correction. In charged particle decay \citep{Czarnecki:2011mr} 
the finite-temperature correction 
was found to be related to the tree-level 
decay width by the simple factor $-\frac{\pi}{3}\alpha\tau^2$, while in 
neutral Higgs decay to two fermions it vanishes~\citep{Donoghue:1983qx}. 
This suggests that the leading thermal photon correction is related to 
the coupling to the electric charge multipoles of the initial or 
final state. In DM pair annihilation the total charge is zero, but 
higher moments are not, which may be the reason for the $\xi$ 
suppression. Further work in this direction is in progress.

The leading ${\cal O}(\tau^2)$ thermal correction does not lift the 
helicity suppression of the s-wave cross section, 
even though the NLO $T=0$ radiative correction does. 
This is easy to understand, since it is the hard photon emission from 
internal bremsstrahlung from the scalar mediator 
that lifts the helicity suppression in the $T=0$ case, while here such 
contributions are strongly suppressed. Helicity-suppression is 
absent in the first sub-leading ${\cal O}(\tau^4)$ temperature 
correction. Explicitly, in the limit of massless SM fermions 
$\epsilon \rightarrow 0$, we find
\begin{equation}
\Delta a_{\tau^4}^{\epsilon=0} = 
\frac{8\pi^2\lambda^4 \alpha\tau^4}{45} \frac{1}{(1+\xi^2)^4}=
\frac{4\pi}{45}\alpha\tau^4\frac{1}{(1+\xi^2)^2} 
\frac{ a_{\rm{tree}}}{\epsilon^2}\Bigl|_{\epsilon=0} .
\end{equation}
This thermal correction can be larger than the tree-level s-wave 
cross section $a_{\rm{tree}}$ when $\epsilon$ is very small 
(e.g., for SM leptons). Nevertheless, it is always parametrically 
smaller than the thermal correction to the p-wave cross section, 
because $\tau \sim v^2$ around freeze-out. It is also smaller than the 
zero-temperature $\mathcal{O}(\alpha)$ NLO correction, which has 
no $\tau^4$ suppression, while both come from internal 
bremsstrahlung, and therefore have the same $\xi^{-8}$ suppression.

Finally, we note that (\ref{eq:resum}) holds also when the DM particle 
is a Dirac fermion, as can be expected from the structure of the total 
in table~\ref{tab:finB}. The difference between the Majorana and Dirac 
cases is that for Dirac fermions the diagrams of type A are absent 
altogether, which also implies the absence of helicity suppression.


\subsection{The finite-$T$ correction from thermal fermions}
\label{sec:finiteTfermions}

Like photons the light SM fermions are very abundant in the 
plasma around freeze-out and also contribute to the finite-temperature 
correction, see (\ref{eq: collision_term_NLO_pre_1}). The method of 
computation of these contributions follows the same steps as for 
thermal photons, and has been described in section \ref{sec:collision_term}. 
However, the results differ considerably between these two cases. 
This is, because at zero temperature soft fermion radiation does 
not cause IR divergences. Furthermore, the Fermi-Dirac distribution is 
finite in the soft limit, hence the degree of divergence is not 
changed at finite temperature. As a consequence the thermal fermion 
contributions have no IR divergences from soft fermions. However, 
for massless fermions there is a divergence from hard-collinear 
photons, which has the same origin as the corresponding $T=0$ 
divergence. When working in the massless limit, we use dimensional 
regularization to regulate this divergence. The poles in $1/\eta$ 
cancel in the sum over all cuts for a given CTP self-energy diagram.

We first discuss the leading finite-temperature correction for the 
case when DM is a Dirac fermion, in which case only diagrams of type B 
are present. Since the 
light SM fermion masses satisfy the condition $m_f \ll eT$, 
they can effectively be treated as massless.\footnote{The top 
quark on the other hand is too heavy to be present in the plasma, 
unless the DM particle mass is 
significantly above 1 TeV. The effects of massive thermal particles 
lead to significantly more complicated integrals, which
can be solved only numerically.} 
We therefore consider the correction for $\epsilon=0$. 
Once again the total result turns out to be simple once all the cuts 
from a given CTP self-energy diagram are summed up. 
The s-wave contribution vanishes for each self-energy diagram separately, 
due to an exact cancellation between real and virtual corrections. 
The s- and p-wave corrections (for $\epsilon =0$) that need to be 
added to the tree cross section are
\begin{equation}
\Delta a_f = 0 \qquad 
\Delta b_f = \frac{16}{9}\alpha\tau^2 \,\frac{\lambda^4}{(1+\xi^2)^3}.
\label{eq:Deltaaf}
\end{equation}

Turning now to the Majorana case, we find the the s-wave and p-wave 
${\cal O}(\tau^2)$ 
contributions from the type A diagrams actually vanish for 
$\epsilon=0$. This means that (\ref{eq:Deltaaf}) also holds for 
Majorana DM. 
However, since $a_{{\rm tree}} \propto \epsilon^2$ the 
limit $\epsilon=0$ is not sufficient for the s-wave contribution, 
in which case the above only shows that there is no lifting 
of helicity suppression from the thermal fermion correction. 
To analyse the result for finite fermion mass we computed 
the thermal correction numerically in the region 
$\epsilon,\tau\ll 1$. We find that there is no $\epsilon^2\tau^2$ 
term in $a$, which means that indeed there is no leading finite thermal 
correction to the s-wave cross section, different from 
the case of thermal photons. On the other hand, for the p-wave 
cross section the thermal correction from fermions is of the 
same order as the one from photons.


\section{Conclusions}
\label{sec:conclusions}

The present work was motivated by the observation that existing 
approaches to calculating the DM relic density at NLO are 
based on zero-temperature calculations of the annihilation cross 
section in the standard freeze-out equation. This procedure has never 
been truly justified. In particular, it ignores the potential presence 
of IR divergences, which, in individual terms, are more severe than 
at zero temperature. In this paper we showed, using a realistic 
example model and the CTP formulation of non-equilibrium quantum 
field theory, that for relic density computations at NLO it is 
indeed sufficient to treat the \textit{annihilation process} at the 
thermal level, leaving the Boltzmann equation intact. That is, 
under the usual assumptions, the freeze-out equation
\begin{equation}
\frac{dn_\c}{dt}+3Hn_\c = 
\langle \sigma^{\rm NLO} v_{\rm rel} 
\rangle_{T\not=0} 
\lp n_\c^{\rm eq}n_{\bar \c}^{\rm eq} - n_\c n_{\bar \c}\rp ,
\label{eq:BoltzmannEqthermal}
\end{equation}
retains its form, and only the annihilation cross 
section receives a finite-temperature correction. 

By computing the thermal contributions to the NLO collision term, 
which includes emission and absorption as well as thermal 
virtual corrections,  
we showed that all soft and collinear divergences cancel, a prerequisite 
for $\langle \sigma^{\rm NLO} v_{\rm rel} 
\rangle_{T\not=0}$ to be well-defined. The cancellation 
was demonstrated in the non-relativistic expansion for the s- and p-wave 
cross sections and, additionally, for the full velocity-dependent 
process but in an expansion in the mass of the t-channel mediator 
up to order $1/\xi^{10}$. We find that the cancellations occur at 
the level of individual CTP dark-matter self-energy diagrams. To 
our knowledge, this is the first time that the IR finiteness of 
relic density computations at NLO has been demonstrated.

Finiteness assured, we investigated the leading finite-temperature 
correction to the annihilation cross section from thermal photons 
and light SM fermions in the plasma. The result can be summarized 
as follows:
\begin{itemize}
\item The leading correction is of order $\alpha\times (T/m_\chi)^2$. 
Since $T\ll m_\chi$ near freeze-out this is parameterically smaller 
than the zero-temperature NLO correction, thus justifying the naive 
zero-temperature radiative correction calculations.
\item Helicity suppression, present for Majorana fermion s-wave 
annihilation, is not lifted by the $T\neq 0$ correction at order 
$\tau^2 = (T/m_\chi)^2$, but only at ${\cal O}(\tau^4)$.
\item The structure of the $\mathcal{O}(\tau^2)$ correction is 
remarkably simple.  The contribution from thermal photons can be 
inferred directly from the tree level annihilation cross section, and is  
given by 
\begin{equation}
-\frac{4}{3}\pi\alpha\tau^2 \frac{\partial }{\partial \xi^2}
[\sigma v]_{\rm{tree}}.
\end{equation}
\end{itemize}
The simplicity of this result and the amount of cancellations 
required to arrive at it call for a deeper explanation. Work in this 
direction is in progress.


\acknowledgments
We would like to thank Marco Drewes for useful discussions and Bj\"{o}rn Garbrecht
for careful reading of the manuscript.
This work was supported by the Gottfried Wilhelm Leibniz programme 
of the Deutsche Forschungsgemeinschaft (DFG)
and the DFG cluster of excellence ``Origin and Structure of the Universe".

\appendix
\section{Feynman rules at finite temperature}  
\label{sec:FeynRules}
We summarize our conventions for the Feynman rules for the scalar and 
fermion propagators in the CTP formalism.
For a given particle species with distribution function $f$, we denote here by $\bar{f}$ 
the distribution of the corresponding antiparticle. 
\bea
i\Delta^{11}(p)&=&\frac{i}{p^2-m^2+i\eta}
+2\pi\delta\lp p^2-m^2\rp \lc \Theta (p^0) f(\vec{p}\,) 
+ \Theta (-p^0) \bar{f}(-\vec{p}\,)\rc,
\label{eq: scalprop11} \\[0.15cm]
i\Delta^{22}(p)&=&\frac{-i}{p^2-m^2-i\eta}
+2\pi\delta\lp p^2-m^2\rp \lc \Theta (p^0) f(\vec{p}\,) 
+ \Theta (-p^0) \bar{f}(-\vec{p}\,)\rc,\\[0.15cm]
i\Delta^{12}(p)&=&2\pi\delta\lp p^2-m^2\rp\lc \Theta (p^0) 
f(\vec{p}\,) + \Theta (-p^0) (1+\bar{f}(-\vec{p}\,)) \rc,\\[0.15cm]
i\Delta^{21}(p)&=&2\pi\delta\lp p^2-m^2\rp \lc \Theta (p^0) 
(1+f(\vec{p}\,)) + \Theta (-p^0) \bar{f}(-\vec{p}\,)\rc,\\[0.15cm]
iS^{11}(p)&=&\frac{i\lp\sl p+m\rp}{p^2-m^2+i\eta}
\!-\!2\pi\lp\sl p+m\rp\!\delta\!\lp p^2-m^2\rp\! \lc \Theta (p^0) 
f(\vec{p}\,) \!+\! \Theta (-p^0) \bar{f}(-\vec{p}\,) \rc,
\label{eq: fermprop11}\\[0.15cm]
iS^{22}(p)&=&\frac{-i\lp\sl p+m\rp}{p^2-m^2-i\eta}
\!-\!2\pi\lp\sl p+m\rp\!\delta\!\lp p^2-m^2\rp\! \lc \Theta (p^0) 
f(\vec{p}\,) \!+\! \Theta (-p^0) \bar{f}(-\vec{p}\,) \rc,
\qquad
\\[0.15cm]
iS^{12}(p)&=&-2\pi\lp\sl p+m\rp\delta\lp p^2-m^2\rp \lc \Theta (p^0) 
f(\vec{p}\,) - \Theta (-p^0) (1-\bar{f}(-\vec{p}\,)) \rc,
\label{eq: fermprop12}\\[0.15cm]
iS^{21}(p)&=& - 2\pi\lp\sl p+m\rp\delta\lp p^2-m^2\rp \lc - \Theta (p^0) 
(1-f(\vec{p}\,)) + \Theta (-p^0) \bar{f}(-\vec{p}\,) \rc,
\label{eq: fermprop21}
\eea
The photon propagator in the Feynman gauge is
\be
iD^{ab}_{\mu\nu}\lp p \rp = -g_{\mu\nu}\ i\Delta^{ab}(p)\rvert_{m=0}.
\ee

\setcounter{section}{5}
\setcounter{equation}{0}
\vskip0.5cm
\section*{\Large Erratum}
\addcontentsline{toc}{section}{Erratum}

In the above a systematic approach starting from 
non-equilibrium quantum field theory to relic density computations at 
next-to-leading order (NLO) was presented. The primary purpose of this work 
was to demonstrate the cancellation of the temperature-dependent infrared 
divergences. In addition, the leading finite temperature-dependent correction 
in a model, where dark matter annihilation into Standard Model (SM) fermions 
is mediated by an electrically charged scalar, was computed and found to be 
of $\op(T^2)\,$. It was also noted that this correction exhibited 
a surprisingly simple structure. This result is incorrect, and the  
$\op(T^2)\,$ correction actually vanishes altogether. Below we list 
the corrections to the original manuscript and provide the leading 
finite-temperature contribution, which is of order $\op(T^4)\,$. 
An explanation of the temperature-dependence of the correction and the 
absence of the $\op(T^2)\,$ term can be given in terms of an operator product 
expansion (M.~Beneke, F.~Dighera and A.~Hryczuk, {\it 
Finite-temperature modification of heavy particle 
decay and dark matter annihilation}, July 2016).

\newcolumntype{V}{>{\centering\arraybackslash} m{.09\linewidth}}
\setlength{\tabcolsep}{15pt}
\begin{table}[b]
\centering
\begin{tabular}{|V|cc|}
\hline
\multicolumn{3}{ |c|}{The finite part $J_{1}$} \\
\hline \hline
Type $A$ & Real & Virtual  \\[0.0cm]
\hline 
\vskip0.2cm
\includegraphics[scale=0.2]{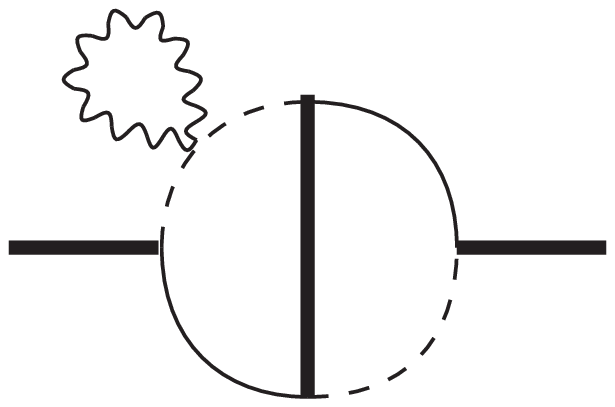} 
&   
&  $ \frac{ 4 (1-2 \epsilon ^2)}{D_\xi^2} $
\\[0.0cm]
\includegraphics[scale=0.2]{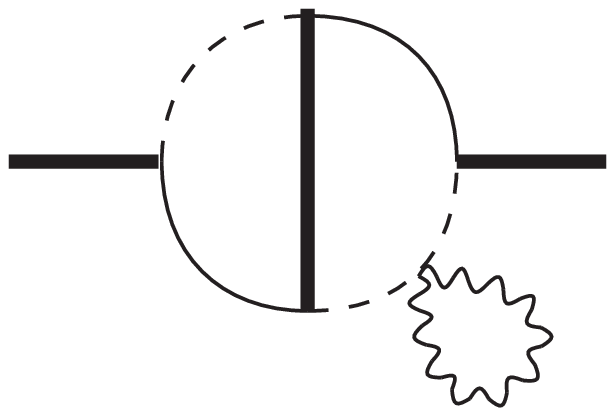} 
&   
&   \dittoclosing
\\
\hline
\end{tabular}
\caption{The self-energy diagrams of type~A, with corresponding coefficients of the finite ${\cal O}(\tau^2)$ correction, to be added to Table~4.}
\label{tab:diags_A}
\end{table}
%
\setlength{\tabcolsep}{15pt}
\begin{table}[t]
\centering
\begin{tabular}{|V|cc|}
\hline
\multicolumn{3}{ |c|}{The finite part $J_{1}$} \\
\hline \hline
Type $B$ & Real & Virtual  \\
\hline 
\vskip0.2cm
\includegraphics[scale=0.2]{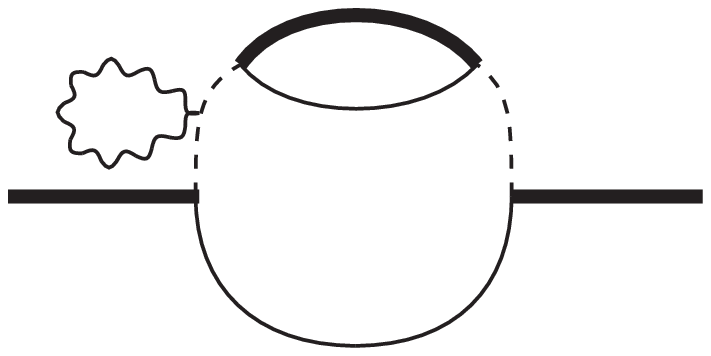} 
&   
&   $ -\frac{ 4 }{D_\xi^2} $
\\
\includegraphics[scale=0.2]{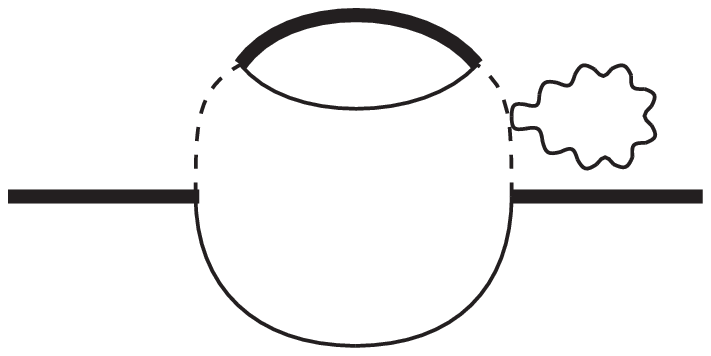} 
&   
&   \dittoclosing
\\
\hline
\end{tabular}
\caption{The self-energy diagrams of type~B, with corresponding coefficients of the finite ${\cal O}(\tau^2)$ correction, to be added to Table~5.}
\label{tab:diags_B}
\end{table}

\vskip0.2cm
1. The diagram from the photon-tadpole contribution to the scalar self-energy 
was missed. This diagram is infrared-finite 
and vanishes at $T=0$, but contributes at finite temperature at order 
$\tau^2=T^2/m_\chi^2$. Tables 4 and 5 must 
be amended by Tables~\ref{tab:diags_A} and \ref{tab:diags_B}, respectively, 
below. (An analogous amendment is necessary for Tables 1 and 2.) 
Once the (purely virtual) terms from these 
diagrams are added, the ``Total''  $\op(\tau^2)\,$ correction given at the 
bottom of Tables 4 and 5 is exactly zero.

\vskip0.2cm
2. It follows that Eqs.~(5.1) and (5.2) must be 
corrected. $\Delta_a$ in Eq.~(5.1) is zero, and the $\op(\tau^2)\,$ 
term in Eq.~(5.2) is absent. The surprisingly simple form of the 
finite-temperature correction in Eq.~(5.2) is a consequence of the 
simplicity of the missed photon self-energy tadpole. The diagrams 
shown in Tables~\ref{tab:diags_A} and \ref{tab:diags_B} factorize 
into the tadpole times the tree diagram with an additional scalar propagator, 
which gives rise to the derivative in $\xi^2$ in Eq.~(5.2). Since 
the  $\op(\tau^2)\,$ correction must vanish on general grounds as follows 
from the operator product expansion, the incorrect $\op(\tau^2)\,$ 
correction found 
in the original text is simply the negative of the contribution 
from the tadpole diagram.
 
\vskip0.2cm
3. The leading temperature-dependent correction from the thermal bath of 
photons is of order~${\cal O}(\tau^4)$. Because the photon tadpole diagrams 
do not contribute at this order, the expression in the s-wave limit and 
for massless SM fermions given in Eq.~(5.3), 
\begin{equation}
\label{eq:T4_phot_swave}
\Delta a_{\tau^4}^{\epsilon=0} = 
\frac{8\pi^2\lambda^4 \alpha\tau^4}{45} \frac{1}{(1+\xi^2)^4}=
\frac{4\pi}{45}\alpha\tau^4\frac{1}{(1+\xi^2)^2} 
\frac{ a_{\rm{tree}}}{\epsilon^2}\Bigl|_{\epsilon=0}\,, 
\end{equation}
is correct. Since this is now the leading correction, we also give 
the full result without restriction to the s-wave limit ($e_\chi=1$), 
and for finite SM fermion mass, but expanded in the mass of the 
mediator ($\xi=m_\phi/m_\chi\gg 1$):
\bea
\label{eq:T4_phot_xi4}
s\, \sigma_{\rm ann}v_{\rm rel}\,|_{\tau^4,\ {\rm thermal \ photons}} &=& 
\frac{4\pi^2\lambda^4 \alpha\tau^4}
{135\, e_\chi^3\, (e_\chi^2 - 4\epsilon^2)^{5/2} \, \xi^4}\,
\Big( - 2 e_\chi^6 (e_\chi^2 -1 ) + \epsilon^2 e_\chi^4 (22 e_\chi^2 - 25) 
  \nonumber \\[0.1cm]
 && \hspace{-3cm}
-  \epsilon^4 e_\chi^2 (80 e_\chi^2 - 101) + 3\epsilon^6 (38 e_\chi^2 - 47)\
\Big) +\op(\xi^{-6})\ . \qquad\qquad
\eea

\vskip0.2cm
4. The finite-temperature correction from the SM fermions in the 
thermal bath given in Eq.~(5.4) is also incorrect. 
There is no $\op(\tau^2)$ contribution in the massless fermion 
limit. (Here the error arose due to an inconsistent Fierz 
transformation in dimensional regularization applied to the collinear 
divergences.) The leading $\op(\tau^4)$ correction for massless fermions 
is 
\begin{equation}
\label{eq:T4_ferm_swave}
\Delta a_{\tau^4,\ {\rm thermal \ fermions}}^{\epsilon=0} = 
\frac{7\pi^2\lambda^4 \alpha\tau^4}{45}\ \frac{3\xi^4+4\xi^2+5}
{(\xi^4-1)^3}\ ,
\end{equation}
in the s-wave limit for arbitrary mediator mass, and 
\be
\label{eq:T4_ferm_xi4}
s\, \sigma_{\rm ann}v_{\rm rel}\,
|_{\tau^4,\ {\rm thermal \ fermions}}^{\epsilon=0} = 
\frac{28 \pi^2 \lambda^4 \alpha \tau^4 \left(e_\chi^2-1\right)}
{135\, e_\chi^2\, \xi^4} \ .
\ee
without restriction to the s-wave limit, but at leading order in 
an expansion in the heavy mediator mass.



\begin{thebibliography}{10}

\bibitem{Baro:2007em}
N.~Baro, F.~Boudjema and A.~Semenov, {\it {Full one-loop corrections to the
  relic density in the MSSM: A Few examples}},  {\em Phys.Lett.} {\bf B660}
  (2008) 550--560 [\href{http://arXiv.org/abs/0710.1821}{{\tt 0710.1821}}].

\bibitem{Baro:2009na}
N.~Baro, F.~Boudjema, G.~Chalons and S.~Hao, {\it {Relic density at one-loop
  with gauge boson pair production}},  {\em Phys.Rev.} {\bf D81} (2010) 015005
  [\href{http://arXiv.org/abs/0910.3293}{{\tt 0910.3293}}].

\bibitem{Herrmann:2009wk}
B.~Herrmann, M.~Klasen and K.~Kovarik, {\it {Neutralino Annihilation into
  Massive Quarks with SUSY-QCD Corrections}},  {\em Phys.Rev.} {\bf D79} (2009)
  061701 [\href{http://arXiv.org/abs/0901.0481}{{\tt 0901.0481}}].

\bibitem{Herrmann:2009mp}
B.~Herrmann, M.~Klasen and K.~Kovarik, {\it {SUSY-QCD effects on neutralino
  dark matter annihilation beyond scalar or gaugino mass unification}},  {\em
  Phys.Rev.} {\bf D80} (2009) 085025
  [\href{http://arXiv.org/abs/0907.0030}{{\tt 0907.0030}}].

\bibitem{Harz:2012fz}
J.~Harz, B.~Herrmann, M.~Klasen, K.~Kovarik and Q.~L. Boulc'h, {\it
  {Neutralino-stop co-annihilation into electroweak gauge and Higgs bosons at
  one loop}},  {\em Phys.Rev.} {\bf D87} (2013) 054031
  [\href{http://arXiv.org/abs/1212.5241}{{\tt 1212.5241}}].

\bibitem{Chatterjee:2012db}
  A.~Chatterjee, M.~Drees and S.~Kulkarni,
 {\it Radiative Corrections to the Neutralino Dark Matter Relic Density - an Effective Coupling Approach},
  Phys.\ Rev.\ D {\bf 86} (2012) 105025
  [\href{http://arXiv.org/abs/1209.2328}{{\tt 1209.2328}}].

\bibitem{Ciafaloni:2013hya}
P.~Ciafaloni, D.~Comelli, A.~De~Simone, E.~Morgante, A.~Riotto {\em et.~al.},
  {\it {The Role of Electroweak Corrections for the Dark Matter Relic
  Abundance}},  {\em JCAP} {\bf 1310} (2013) 031
  [\href{http://arXiv.org/abs/1305.6391}{{\tt 1305.6391}}].

\bibitem{Herrmann:2014kma}
B.~Herrmann, M.~Klasen, K.~Kovarik, M.~Meinecke and P.~Steppeler, {\it
  {One-loop corrections to gaugino (co-)annihilation into quarks in the MSSM}},
    Phys.\ Rev.\ D {\bf 89} (2014) 114012
   [\href{http://arXiv.org/abs/1404.2931}{{\tt 1404.2931}}].

\bibitem{Baro:2008bg}
N.~Baro, F.~Boudjema and A.~Semenov, {\it {Automatised full one-loop
  renormalisation of the MSSM. I. The Higgs sector, the issue of tan(beta) and
  gauge invariance}},  {\em Phys.Rev.} {\bf D78} (2008) 115003
  [\href{http://arXiv.org/abs/0807.4668}{{\tt 0807.4668}}].

\bibitem{Baro:2009gn}
N.~Baro and F.~Boudjema, {\it {Automatised full one-loop renormalisation of the
  MSSM II: The chargino-neutralino sector, the sfermion sector and some
  applications}},  {\em Phys.Rev.} {\bf D80} (2009) 076010
  [\href{http://arXiv.org/abs/0906.1665}{{\tt 0906.1665}}].

\bibitem{SloopS}
``http://code.sloops.free.fr/index.php.''

\bibitem{Herrmann:2007ku}
B.~Herrmann and M.~Klasen, {\it {SUSY-QCD Corrections to Dark Matter
  Annihilation in the Higgs Funnel}},  {\em Phys.Rev.} {\bf D76} (2007) 117704
  [\href{http://arXiv.org/abs/0709.0043}{{\tt 0709.0043}}].

\bibitem{DMatNLO}
``http://dmnlo.hepforge.org.''

\bibitem{Bloch:1937pw}
F.~Bloch and A.~Nordsieck, {\it {Note on the Radiation Field of the electron}},
   {\em Phys.Rev.} {\bf 52} (1937) 54--59.

\bibitem{Kinoshita:1962ur}
T.~Kinoshita, {\it {Mass singularities of Feynman amplitudes}},  {\em
  J.Math.Phys.} {\bf 3} (1962) 650--677.

\bibitem{Lee:1964is}
T.~Lee and M.~Nauenberg, {\it {Degenerate Systems and Mass Singularities}},
  {\em Phys.Rev.} {\bf 133} (1964) B1549--B1562.

\bibitem{Grandou:1991qr}
T.~Grandou, M.~Le~Bellac and D.~Poizat, {\it {Cancellation of infrared and
  collinear singularities in relativistic thermal field theories}},  {\em
  Nucl.Phys.} {\bf B358} (1991) 408--434.

\bibitem{Grandou:1990ir}
T.~Grandou, M.~Le~Bellac and D.~Poizat, {\it {Remarks on infrared singularities
  of relativistic thermal field theories}},  {\em Phys.Lett.} {\bf B249} (1990)
  478--484.

\bibitem{Altherr:1991cq}
T.~Altherr, {\it {Infrared singularities cancellation in reaction rates at
  finite temperature}},  {\em Phys.Lett.} {\bf B262} (1991) 314--319.

\bibitem{Landshoff:1994ud}
P.~Landshoff and J.~Taylor, {\it {Photon radiation in a heat bath}},  {\em
  Nucl.Phys.} {\bf B430} (1994) 683--694
  [\href{http://arXiv.org/abs/hep-ph/9402315}{{\tt hep-ph/9402315}}].

\bibitem{Baier:1989ub}
R.~Baier, E.~Pilon, B.~Pire and D.~Schiff, {\it {Finite Temperature Radiative
  Corrections to Early Universe Neutron - Proton Ratio: Cancellation of
  Infrared and Mass Singularities}},  {\em Nucl.Phys.} {\bf B336} (1990) 157.

\bibitem{Baier:1988xv}
R.~Baier, B.~Pire and D.~Schiff, {\it {Dilepton production at finite
  temperature: Perturbative treatment at order $\alpha_s$}},  {\em Phys.Rev.}
  {\bf D38} (1988) 2814.

\bibitem{Altherr:1989yn}
T.~Altherr and T.~Becherrawy, {\it {Cancellation of Infrared and Mass
  Singularities in the Thermal Dilepton Rate}},  {\em Nucl.Phys.} {\bf B330}
  (1990) 174.

\bibitem{Gabellini:1989yk}
Y.~Gabellini, T.~Grandou and D.~Poizat, {\it {Electron - Positron Annihilation
  in Thermal {QCD}}},  {\em Annals Phys.} {\bf 202} (1990) 436--466.

\bibitem{Altherr:1988bg}
T.~Altherr, P.~Aurenche and T.~Becherrawy, {\it {On Infrared and Mass
  Singularities of Perturbative {QCD} in a Quark - Gluon Plasma}},  {\em
  Nucl.Phys.} {\bf B315} (1989) 436.

\bibitem{Czarnecki:2011mr}
A.~Czarnecki, M.~Kamionkowski, S.~K. Lee and K.~Melnikov, {\it {Charged
  Particle Decay at Finite Temperature}},  {\em Phys.Rev.} {\bf D85} (2012)
  025018 [\href{http://arXiv.org/abs/1110.2171}{{\tt 1110.2171}}].

\bibitem{Besak:2012qm}
D.~Besak and D.~B\"odeker, {\it {Thermal production of ultrarelativistic
  right-handed neutrinos: Complete leading-order results}},  {\em JCAP} {\bf
  1203} (2012) 029 [\href{http://arXiv.org/abs/1202.1288}{{\tt 1202.1288}}].

\bibitem{Anisimov:2010gy}
A.~Anisimov, D.~Besak and D.~B\"odeker, {\it {Thermal production of relativistic
  Majorana neutrinos: Strong enhancement by multiple soft scattering}},  {\em
  JCAP} {\bf 1103} (2011) 042 [\href{http://arXiv.org/abs/1012.3784}{{\tt
  1012.3784}}].

\bibitem{Salvio:2011sf}
A.~Salvio, P.~Lodone and A.~Strumia, {\it {Towards leptogenesis at NLO: the
  right-handed neutrino interaction rate}},  {\em JHEP} {\bf 1108} (2011) 116
  [\href{http://arXiv.org/abs/1106.2814}{{\tt 1106.2814}}].

\bibitem{Laine:2011pq}
M.~Laine and Y.~Schr\"oder, {\it {Thermal right-handed neutrino production rate
  in the non-relativistic regime}},  {\em JHEP} {\bf 1202} (2012) 068
  [\href{http://arXiv.org/abs/1112.1205}{{\tt 1112.1205}}].

\bibitem{Garbrecht:2013gd}
B.~Garbrecht, F.~Glowna and M.~Herranen, {\it {Right-Handed Neutrino Production
  at Finite Temperature: Radiative Corrections, Soft and Collinear
  Divergences}},  {\em JHEP} {\bf 1304} (2013) 099
  [\href{http://arXiv.org/abs/1302.0743}{{\tt 1302.0743}}].

\bibitem{Wizansky:2006fm}
T.~Wizansky, {\it {Finite temperature corrections to relic density
  calculations}},  {\em Phys.Rev.} {\bf D74} (2006) 065007
  [\href{http://arXiv.org/abs/hep-ph/0605179}{{\tt hep-ph/0605179}}].

\bibitem{Weldon:1991eg}
H.~A. Weldon, {\it {Bloch-Nordsieck cancellation of infrared divergences at
  finite temperature}},  {\em Phys.Rev.} {\bf D44} (1991) 3955--3963.

\bibitem{Weldon:1993qh}
H.~A. Weldon, {\it {Cancellation of infrared divergences in thermal QED}},
  {\em Nucl.Phys.} {\bf A566} (1994) 581C--584C
  [\href{http://arXiv.org/abs/hep-ph/9308249}{{\tt hep-ph/9308249}}].

\bibitem{Indumathi:1996ec}
D.~Indumathi, {\it {Cancellation of infrared divergences at finite
  temperature}},  {\em Annals Phys.} {\bf 263} (1998) 310--339
  [\href{http://arXiv.org/abs/hep-ph/9607206}{{\tt hep-ph/9607206}}].

\bibitem{Schwinger:1960qe}
J.~S. Schwinger, {\it {Brownian motion of a quantum oscillator}},  {\em
  J.Math.Phys.} {\bf 2} (1961) 407--432.

\bibitem{Keldysh:1964ud}
L.~Keldysh, {\it {Diagram technique for nonequilibrium processes}},  {\em
  Zh.Eksp.Teor.Fiz.} {\bf 47} (1964) 1515--1527.

\bibitem{Baym:1961zz}
G.~Baym and L.~P. Kadanoff, {\it {Conservation Laws and Correlation
  Functions}},  {\em Phys.Rev.} {\bf 124} (1961) 287--299.

\bibitem{KB}
L.~Kadanoff and G.~Baym, {\em Quantum Statistical Mechanics}.
\newblock Benjamin, New York, 1962.

\bibitem{kainulainen:2001cn}
K.~Kainulainen, T.~Prokopec, M.~G. Schmidt and S.~Weinstock, {\it {First
  principle derivation of semiclassical force for electroweak baryogenesis}},
  {\em JHEP} {\bf 0106} (2001) 031
  [\href{http://arXiv.org/abs/hep-ph/0105295}{{\tt hep-ph/0105295}}].

\bibitem{prokopec:2003pj}
T.~Prokopec, M.~G. Schmidt and S.~Weinstock, {\it {Transport equations for
  chiral fermions to order h bar and electroweak baryogenesis. Part 1}},  {\em
  Annals Phys.} {\bf 314} (2004) 208--265
  [\href{http://arXiv.org/abs/hep-ph/0312110}{{\tt hep-ph/0312110}}].

\bibitem{Calzetta:1986cq}
E.~Calzetta and B.~Hu, {\it {Nonequilibrium Quantum Fields: Closed Time Path
  Effective Action, Wigner Function and Boltzmann Equation}},  {\em Phys.Rev.}
  {\bf D37} (1988) 2878.

\bibitem{Carrington:2004tm}
M.~E. Carrington and S.~Mrowczynski, {\it {Transport theory beyond binary
  collisions}},  {\em Phys.Rev.} {\bf D71} (2005) 065007
  [\href{http://arXiv.org/abs/hep-ph/0406097}{{\tt hep-ph/0406097}}].

\bibitem{LeBellac}
M.~L. Bellac, {\em Thermal Field Thoery}.
\newblock Cambridge University Press, 1996.

\bibitem{Ciafaloni:2011oq}
P.~Ciafaloni, M.~Cirelli, D.~Comelli, A.~De~Simone, A.~Riotto {\em et.~al.},
  {\it {On the Importance of Electroweak Corrections for Majorana Dark Matter
  Indirect Detection}},  {\em JCAP} {\bf 1106} (2011) 018
  [\href{http://arXiv.org/abs/1104.2996}{{\tt 1104.2996}}].

\bibitem{Denner1992}
A.~Denner, H.~Eck, O.~Hahn and J.~Kueblbeck, {\it Feynman rules for
  fermion-number-violating interactions},  {\em Nuclear Physics} {\bf B387}
  (1992) 467--481.

\bibitem{Altherr:1989jc}
T.~Altherr and P.~Aurenche, {\it {Finite Temperature QCD Corrections to Lepton
  Pair Formation in a Quark - Gluon Plasma}},  {\em Z.Phys.} {\bf C45} (1989)
  99.

\bibitem{Donoghue:1983qx}
J.~F. Donoghue and B.~R. Holstein, {\it {Renormalization and Radiative
  Corrections at Finite Temperature}},  {\em Phys.Rev.} {\bf D28} (1983) 340.

\end{thebibliography}
\providecommand{\href}[2]{#2}\begingroup\raggedright\endgroup

\end{document}